\documentclass[fleqn,usenatbib]{mnras}
\pdfoutput=1

\usepackage{newtxtext}
\usepackage[pdftex]{graphicx,color}
\usepackage[T1]{fontenc}
\usepackage[utf8]{inputenc}
\usepackage{lmodern}
\usepackage[normalem]{ulem}

\definecolor{urlblue}{rgb}{0,0,0.9}
\definecolor{linkgreen}{rgb}{0,0.45,0}
\definecolor{linkorange}{rgb}{0.7,0.1,0.0}

\usepackage{nicematrix}
\usepackage{amsmath, amssymb}
\usepackage{cuted}
\usepackage[english]{babel}
\usepackage{enumerate}
\usepackage{multirow}
\usepackage{booktabs}
\usepackage{makecell}

\usepackage{supertabular}
\usepackage{listings}

\graphicspath{{./figs/}}

\AtBeginDocument{\hypersetup{
linkcolor=linkgreen,
citecolor=linkorange,
urlcolor=urlblue}}

\DeclareUnicodeCharacter{0303}{~}

\title[J-PLUS DR3: Galaxy-Star-Quasar classification]{J-PLUS: Galaxy-Star-Quasar classification for DR3}

\author[R.~von~Marttens et al.]{{
R.~von~Marttens,$^{1,2,3}$\thanks{E-mail: rodrigovonmarttens@gmail.com}
V.~Marra,$^{4,5,6}$
M.~Quartin,$^{7,8,9}$
L.~Casarini,$^{10}$
P.~O.~Baqui,$^{11}$}
\newauthor{
A. Alvarez-Candal,$^{12,13,3}$
F. J. Galindo-Guil,$^{14}$
J.A. Fernández-Ontiveros,$^{14}$
Andrés del Pino,$^{14}$}
\newauthor{
L.A. D\'iaz-Garc\'ia,$^{15}$
C. L\'opez-Sanjuan,$^{16}$
J.~Alcaniz,$^{3}$
R.~Angulo,$^{17,18}$
A.~J.~Cenarro,$^{16}$}
\newauthor{
D.~Cristóbal-Hornillos,$^{14}$
R.~Dupke,$^{3,19,20}$
A.~Ederoclite,$^{21}$
C.~Hernández-Monteagudo,$^{16,22,23}$}
\newauthor{
A.~Marín-Franch,$^{16}$
M.~Moles,$^{14}$
L.~Sodré,$^{21}$
J.~Varela,$^{16}$
H.~Vázquez Ramió$^{16}$}\\
$^{1}$Instituto de Física, Universidade Federal da Bahia, 40210-340, Salvador-BA, Brasil\\
$^{2}$PPGCosmo, Universidade Federal do Espírito Santo, 29075-910, Vitória, ES, Brazil\\
$^{3}$Observat\'orio Nacional, Rua General Jos\'e Cristino 77, Rio de Janeiro, RJ, 20921-400, Brazil\\
$^{4}$Departamento de Física, Universidade Federal do Espírito Santo, 29075-910, Vitória, ES, Brazil\\
$^{5}$INAF -- Osservatorio Astronomico di Trieste, via Tiepolo 11, 34131 Trieste, Italy\\
$^{6}$IFPU -- Institute for Fundamental Physics of the Universe, via Beirut 2, 34151, Trieste, Italy\\
$^{7}$Instituto de Física, Universidade Federal do Rio de Janeiro, 21941-972, Rio de Janeiro, RJ, Brazil\\
$^{8}$Observatório do Valongo, Universidade Federal do Rio de Janeiro, 20080-090, Rio de Janeiro, RJ, Brazil\\
$^{9}$Institute of Theoretical Physics, Heidelberg University, Philosophenweg 16, 69120 Heidelberg, Germany\\
$^{10}$Departamento de Física, Universidade Federal de Sergipe, 49100-000, Aracaju, SE, Brazil\\
$^{11}$Núcleo Cosmo-ufes, Universidade Federal do Espírito Santo, 29075-910, Vitória, ES, Brazil\\
$^{12}$Instituto de Astrof\'isica de Andaluc\'ia, CSIC, Apt 3004, E18080 Granada, Spain\\
$^{13}$Instituto de F\'isica Aplicada a las Ciencias y las Tecnolog\'ias, Universidad de Alicante, San Vicent del Raspeig, E03080, Alicante, Spain\\
$^{14}$Centro de Estudios de F\'isica del Cosmos de Arag\'on (CEFCA), Plaza San Juan 1, 44001 Teruel, Spain\\
$^{15}$Instituto de Astrof\'isica de Andaluc\'ia (IAA-CSIC), P.O.~Box 3004, 18080 Granada, Spain\\
$^{16}$Centro de Estudios de F\'isica del Cosmos de Arag\'on (CEFCA), Unidad Asociada al CSIC, Plaza San Juan 1, 44001 Teruel, Spain\\
$^{17}$Donostia International Physics Center (DIPC),  Manuel Lardizabal Ibilbidea, 4, San Sebasti\'an, Spain\\
$^{18}$Ikerbasque, Basque Foundation for Science, E-48013 Bilbao\\
$^{19}$Department of Astronomy, University of Michigan, 311West Hall, 1085 South University Ave., Ann Arbor, USA\\
$^{20}$Department of Physics and Astronomy, University of Alabama, Box 870324, Tuscaloosa, AL, USA\\
$^{21}$Departamento de Astronomia, Instituto de Astronomia, Geofísica e Ciências Atmosféricas, Universidade de São Paulo, 05508-090, São Paulo, SP, Brazil \\
$^{22}$Instituto de Astrof\'isica de Canarias, C/ V\'ia L\'actea, s/n, E-38205, La Laguna, Tenerife, Spain\\
$^{23}$Departamento de Astrof\'isica, Universidad de La Laguna, E-38206, La Laguna, Tenerife, Spain}

\date{Accepted XXX. Received YYY; in original form ZZZ}

\pubyear{202X}

\begin{document}
\label{firstpage}
\pagerange{\pageref{firstpage}--\pageref{lastpage}}
\maketitle

\begin{abstract}

The Javalambre Photometric Local Universe Survey (J-PLUS) is a 12-band photometric survey using the 83-cm JAST telescope. Data Release 3 includes 47.4 million sources. J-PLUS DR3 only provides star-galaxy classification so that quasars are not identified from the other sources. Given the size of the dataset, machine learning methods could provide a valid alternative classification and a solution to the classification of quasars. Our objective is to classify J-PLUS DR3 sources into galaxies, stars and quasars, outperforming the available classifiers in each class. We use an automated machine learning tool called {\tt TPOT} to find an optimized pipeline to perform the classification. The supervised machine learning algorithms are trained on the crossmatch with SDSS DR18, LAMOST DR8 and \textit{Gaia}. We checked that the training set of about 660 thousand galaxies, 1.2 million stars and 270 thousand quasars is both representative and contain a minimal presence of contaminants (less than 1\%). We considered 37 features: the twelve photometric bands with respective errors, six colors, four morphological parameters, galactic extinction with its error and the PSF relative to the corresponding pointing. With {\tt TPOT} genetic algorithm, we found that XGBoost provides the best performance: the AUC for galaxies, stars and quasars is above 0.99 and the average precision is above 0.99 for galaxies and stars and 0.96 for quasars. XGBoost outperforms the  classifiers already provided in J-PLUS DR3 and also classifies quasars.
\end{abstract}


%%%%%%%%%%%%%%%%%%%%%%%%%%%%%%%%%%%%%%
%%%%%%%%%%%%%%%%%%%%%%%%%%%%%%%%%%%%%%
%%%%%%%%%%%%%%%%%%%%%%%%%%%%%%%%%%%%%%
\section{Introduction}

The Javalambre Photometric Local Universe Survey (J-PLUS) is expected to observe 8500 deg$^2$ of the northern sky \citep{Cenarro:2018uoy}. J-PLUS makes use of the Javalambre Auxiliary Survey Telescope (JAST80)  equipped with the panoramic CCD camera T80Cam, which has a field of view  of 2 deg$^2$ sampled by a mosaic of 9500x9500 pixels, and a pixel scale of 0.56"/pixel.
The J-PLUS DR3 Data Release%
\footnote{\href{https://www.j-plus.es/datareleases/data_release_dr3}{www.j-plus.es/datareleases/data\_release\_dr3}}
comprises 1642 J-PLUS fields observed in twelve optical bands, amounting to 3192 deg$^2$ (2881 deg$^2$ after masking).
More precisely, J-PLUS uses the  $ugriz$ broad bands and the intermediate bands J0378, J0395, J0410, J0430, J0515, J0660 and J0861, providing an unprecedented multicolor view of the Universe close to us. 
J-PLUS DR3 is based on images collected from November 2015 to February 2022 and features 47.4 million objects in the $r$ detection band (29.8 million with $r \le 21$), with forced-photometry in all other filters.\footnote{This is the ``dual mode'' catalog, where SExtractor \citep{Bertin:1996fj} is first run on a reference  image for source detection and for the definition of the position and sizes of the apertures. Then it is run on each of the filter images to perform the photometry within the apertures defined by the reference image (forced photometry), independently if the source is detected or not in the other bands. Here, the $r$-band coadded images were chosen as the reference images for constructing the dual-mode catalogs.}
For point-like sources the limiting magnitude is $\sim$21.8~mag within a circular aperture of 3 arcsec for $S/N\ge 5$.
As compared to previous releases, DR3 covers a much larger footprint and benefits from  several important technical improvements concerning background determination and subtraction as well as photometric calibration~\citep{2019A&A...631A.119L}.

The main goal of this work is to provide a value-added catalog (VAC) with a reliable galaxy-star-quasar classification for the full J-PLUS DR3 catalog. In order to achieve our objective we adopt a supervised machine learning (ML) approach through the flexible and automated Tree-based Pipeline Optimization Tool (TPOT)~\citep{le2020scaling,Olson2016EvoBio,OlsonGECCO2016, le2020scaling,Olson2016EvoBio,OlsonGECCO2016}, which, after performing a genetic algorithm optimization, returns the most suitable pipeline to address a given problem. 
The performance of supervised machine learning depends strongly on the adopted training set, which we build using the crossmatch of J-PLUS DR3 with the SDSS DR18%
\footnote{\href{https://www.sdss.org/dr18/}{www.sdss.org/dr18/}}, LAMOST DR8%
\footnote{\href{https://dr8.lamost.org}{www.dr8.lamost.org}}
and {\it Gaia} DR3%
\footnote{\href{https://www.cosmos.esa.int/web/gaia/data-release-3}{www.cosmos.esa.int/web/gaia/data-release-3}}
datasets.

J-PLUS DR3 includes two star-galaxy classifiers:  
the \texttt{CLASS\_STAR} provided by Source Extractor \citep[SExtractor,][]{Bertin:1996fj} and the stellar-galaxy loci classifier ({\tt SGLC}) introduced in \citet{2019A&A...622A.177L}.
\texttt{CLASS\_STAR} is based on a pre-trained neural network 
and uses as input the following quantities: 8 isophotal areas, the maximum pixel value above the sky, and one ``seeing'' control parameter. On the other hand, the {\tt SGLC} classifier takes advantage of the bimodality  in the concentration-magnitude\footnote{The concentration index serves as a metric for quantifying the degree of light concentration within the object of interest.} diagram, which allows one to separate compact from extended sources. The {\tt SGLC} classification procedure first models the distributions and, afterwards, with appropriate priors, performs a Bayesian classification of the sources. In both cases, the output is probabilistic, which means that they return the probability of each object of being a star (compact point-like source) or a galaxy (extended source).
It is worth mentioning that {\tt CLASS\_STAR} and {\tt SGLC} are not expected to identify quasars, but, since they are compact point-like sources, they are likely classified as stars.
We wish to outperform the J-PLUS classifiers regarding the classification of stars and galaxies and to provide a valid quasar classification.
To this end, we will consider 37 observational features taking advantage of the ability of machine learning algorithms to understand the complexity of non-linearly correlated multivariate data.

A galaxy-star-quasar classification has been recently performed in \citet{bailer2019quasar,2020A&A...639A..84C,fu2021finding,2022A&A...659A.144W}.
In particular, \citet{2022A&A...659A.144W} considered J-PLUS DR1 data and classified, out of the 13 million sources, only the 3.5 million objects that had valid photometric information in all the 12 bands.
Here, we wish to classify all the 47.4 million DR3 objects, without restricting to sources detected in all bands. This will substantially increase the legacy value of our catalog.
Indeed, given the large footprint, J-PLUS has the potential to provide quasar candidates up to $r\!\approx\! 21$ that could be targeted by subsequent follow-up surveys. It is also worth stressing that the methodology developed in this work may be applied to the Javalambre Physics of the Accelerating Universe Astrophysical Survey \citep[J-PAS, ][]{Bonoli:2020ciz}, complementing the results of \citet{2021A&A...645A..87B}, who employed machine learning techniques to classify J-PAS objects as either extended or point-like.

This paper is organized as follows.
In Sect.~\ref{sec:ml} we will present  our machine-learning framework and in Sect.~\ref{sec:training} our training set.
Performance will be discussed in Sect.~\ref{sec:results} and representativeness in Sect.~\ref{sec:represa}. The value-added catalog is presented in Sect.~\ref{sec:vac}. We conclude in Sect.~\ref{sec:conclusions}.

%%%%%%%%%%%%%%%%%%%%%%%%%%%%%%%%%%%%%%
%%%%%%%%%%%%%%%%%%%%%%%%%%%%%%%%%%%%%%
%%%%%%%%%%%%%%%%%%%%%%%%%%%%%%%%%%%%%%
\section{(Supervised) Machine learning}
\label{sec:ml}

In this section, we present the details of our classification process. Among others, we detail the prior information we used, the method applied, and the metrics to evaluate our results. Notably, as previously mentioned, we consider supervised ML techniques, which means that we make use of labeled information (objects whose classifications are known) to establish a model that allows us to classify other objects. Further ingredients of our approach are discussed in the following.

%%%%%%%%%%%%%%%%%%%%%%%%%%%%%%%%%%%%%%
\subsection{One-vs-All method}
\label{ssec:onevsall}

The current classification setting identifies sources exclusively as one of three categories: galaxy, star or quasar. Consequently, we are dealing with a multi-class classification problem.%
\footnote{Note that this is different from multi-label (or multi-output) classification, where multiple classes can be assigned to the same entry.}
Here, we adopt the One-vs-All (OVA) method, where a $k$-class classifier is obtained via $k$ binary classifiers which classify a given category against the rest.
We adopt probabilistic classifiers, so that the output comprises three probabilities, one for each binary classifier: the probability of being a galaxy $p_{\rm gal}$, a star $p_{\rm star}$ or a quasar $p_{\rm qso}$.
These probabilities are independent and can be combined in order to obtain normalized probabilities:
\begin{align}
\tilde p_{\rm gal} &= \frac{p_{\rm gal}}{p_{\rm gal}+p_{\rm star}+p_{\rm qso}} \,, \nonumber \\
\tilde p_{\rm star} &= \frac{p_{\rm star}}{p_{\rm gal}+p_{\rm star}+p_{\rm qso}} \,, \nonumber \\
\tilde p_{\rm qso} &= \frac{p_{\rm qso}}{p_{\rm gal}+p_{\rm star}+p_{\rm qso}} \,, \label{ptilde}
\end{align}
so that $\tilde p_{\rm gal}+\tilde p_{\rm star}+\tilde p_{\rm qso}=1$.
From a formal perspective, our ML method is a mapping $f:\mathbb{R}^{n}\rightarrow\{\tilde p_{\rm gal},\tilde p_{\rm star}, \tilde p_{\rm qso}\}$, where $n$ is the dimension of the feature space (in our case, $n=37$). Note that the probabilities are given by real numbers in the range $\left[0,1\right]$.

\begin{figure}
\centering
\includegraphics[width=.8\columnwidth]{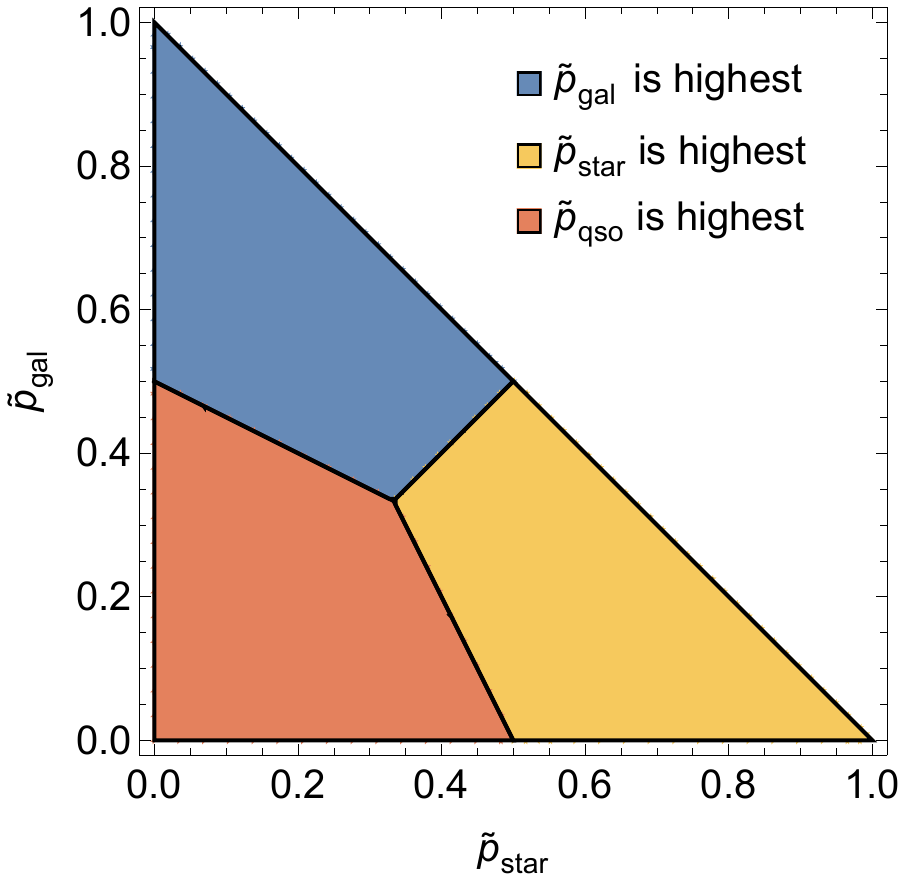}
\caption{Probability space for classifying an object according to the highest probability. Depending on the optimization goal other strategies may be chosen.}
\label{fig:3k}
\end{figure}

Both probabilities $p$ and $\tilde p$ are useful. The former are relative to the binary classifiers and are used to compute, e.g., completeness and purity.
The latter could be used to classify the sources: the simplest choice is to classify an object according to the class with the highest probability. This scenario is shown in Fig.~\ref{fig:3k}. However, one may divide the allowed probability space according to other criteria in order to optimize a given metric. This is similar to the case of a binary classifiers for which one can choose a probability threshold $p_{\rm cut}$ different from 1/2. Note that Fig.~\ref{fig:3k} reduces to the binary classifiers if one of the $\tilde p$ is zero.
 
%%%%%%%%%%%%%%%%%%%%%%%%%%%%%%%%%%%%%%
\subsection{Feature configuration}
\label{ssec:features}

Regarding the ML setup, the first important point to be addressed is which physical quantities will be adopted as  features for the classification process. In this work, we consider 37 observables, measured by J-PLUS, which can be divided into five groups: photometry, colors, morphology, extinction and point spread function. It is worth mentioning that all the features, for all objects within the J-PLUS DR3 catalog, can be downloaded using the ADQL protocol from the J-PLUS webpage.%
\footnote{\url{https://www.j-plus.es/datareleases/data_release_dr3}}
The features used in this work are the following:
\begin{itemize}

\item {\bf Photometry:} We consider the twelve photometric bands observed in J-PLUS with their respective errors. The rationale behind including the errors is to characterize the statistical distribution associated with a  measurement. Observations may be inhomogeneous because, for example, observing conditions vary, and the measurement errors should be able to account at least in part for this potential bias. \textit{A posteriori} one finds that, for some bands, the error may be more important than the actual measurement \citep{2021A&A...645A..87B}.
All the features of this group can be found in the {\tt jplus.MagABDualObj} catalog. More precisely, whereas the magnitudes are contained in the entry {\tt MAG\_AUTO}, their respective errors are contained in the entry {\tt MAG\_ERR\_AUTO}. It is important to mention that the objects are detected in the $r$ band. This means that some objects are possibly non-detected with the other filters and, in this case, the magnitude in the corresponding filter is given the value 99, a label for non-detection.
In other words, the requirement from the pipeline of not having empty values in input data is fulfilled by setting the non-detections to a flag value 99.
The idea is that magnitude information has no physical meaning in this case.
Future work could study the impact of sources with non-detection on the performance of the classifier.
\vspace{2mm}

\item {\bf Colors:} We  consider the following six colors (difference between the magnitudes of a given object in two different bands): the four adjacent broad-band colors $u-g$, $g-r$, $r-i$ and $i-z$ \citep{2015ApJ...811...95P}, and also the colors $u$JAVA$-$J0395 and $u$JAVA$-$J0410, which were found useful to separate quasars from stars \citep{2022A&A...657A..35G}.
\vspace{2mm}

\item {\bf Morphology: }Concerning structural features, we consider the light concentration within the object, quantified by the concentration $c_r$, the ellipticity $A/B$, the full width half maximum (FWHM) assuming a Gaussian kernel, and the ratio $\mu$ between {\tt MU\_MAX} and {\tt MAG\_APER\_3\_0}, which correspond to the peak surface brightness above background and the magnitude within a fixed circular aperture of 3.0", respectively.
All the features are also from the {\tt jplus.MagABDualObj} catalog. The ellipticity is obtained from the ratio {\tt A\_WORLD}/{\tt B\_WORLD}, where {\tt A\_WORLD} and {\tt B\_WORLD} correspond to the RMS along the major and minor axes, respectively. The concentration is given by the difference between {\tt MAG\_APER\_1\_5} and {\tt MAG\_APER\_3\_0}, where the former is the magnitude within 1.5". 
\vspace{2mm}

\item {\bf Extinction: }We consider the E(B-V) color excess dust map at infinity~\citep{Schlegel:1997yv} with its respective error. Both quantities are obtained from the {\tt jplus.MWExtinction} catalog, in the entries {\tt ebv} and {\tt ebv\_err}, respectively.
\vspace{2mm}

\item {\bf PSF: }We also consider the mean point spread function (PSF) in the $r$ band.%
\footnote{Available at \url{http://archive.cefca.es/catalogues/jplus-dr3/image_search.html}}
In this case, the considered PSF value is constant within a given pointing, for a total of 1642 values.

\end{itemize}

Table~\ref{tab:features} presents a comprehensive summary of the features used in this study.

\begin{table}
\centering
\begin{tabular}{l|l}
\hline\hline
 & \multicolumn{1}{c}{Features}                               \\ \hline
\multirow{4}{*}{Photometry}   & Broad bands magnitudes: $ugriz$                \\
                              & Intermediate bands magnitudes: J0378, J0395, J0410,        \\
                              & J0430, J0515, J0660 and J0861                    \\
                              & Magnitude errors in all bands                  \\ \hline
\multirow{2}{*}{Colors}       & $u-g$, $g-r$, $r-i$, $i-z$,\\
&$u$JAVA$-$J0395 and $u$JAVA$-$J0410 \\ \hline
\multirow{5}{*}{Morphology}   & Concentration $c_r$                                        \\
                              & Ellipticity $A/B$                                          \\
                              & Full width half maximum (FWHM)                             \\
                              & Ratio between peak surface brightness above background     \\
                              & and the magnitude within a fixed circular aperture of 3.0" \\ \hline
\multirow{2}{*}{Extinction}   & E(B-V) color excess                                        \\
                              & E(B-V) color excess error                                  \\ \hline
PSF                           & Point spread function in the $r$ band                      \\ \hline\hline
\end{tabular}
\caption{The 37 features used in our ML pipeline. Additional details regarding the features and instructions for accessing J-PLUS data can be found on \href{https://archive.cefca.es/doc/manuals/catalogues_portal_users_manual.pdf}{https://archive.cefca.es/doc/manuals/catalogues\_portal\_users\_manual.pdf}.}
\label{tab:features}
\end{table}

%%%%%%%%%%%%%%%%%%%%%%%%%%%%%%%%%%%%%%
\subsection{TPOT Pipeline}
\label{ssec:tpotxgb}

Given the extensive number of possibilities, the optimal choice of the ML method or pipeline for a specific classification problem remains an important inquiry in the realm of supervised classification. Considering the absence of a predefined answer, we exploit the TPOT framework to address this challenge. TPOT works using genetic programming to search through a large space of possible machine learning pipelines, that is, combinations of data preprocessing steps, feature selection techniques, and machine learning algorithms, to find the best-performing pipeline for our classification problem. While the current version of TPOT does not provide all the pipelines scrutinized in our specific analysis, it leverages the methods available in Scikit-Learn\footnote{\href{https://scikit-learn.org/stable/index.html}{https://scikit-learn.org}}, including Random Forest, Gradient Boosting, and Nearest Neighbors. This approach has already been used in the astrophysical context, for example, in~\citet{vonMarttens:2021fmj} and~\citet{VargasdosSantos:2019ovq}. TPOT explores numerous pipeline possibilities through a genetic algorithm on three stages:
\begin{enumerate}[i)]

\item First, TPOT tests changes of variables in the training set. At this stage, it can re-scale the data (feature transformation), remove useless data (feature selection) or build new features from the combination of old ones (feature construction);

\item Then, TPOT selects the best ML method by maximizing a given performance metric; here we will adopt the area under the curve of the receiver operating characteristic curve, see Section~\ref{ssec:metrics}. This includes the possibility of stacking different methods to produce a powerful combined pipeline. It is worth mentioning that, in the standard version, TPOT tests all methods available in \texttt{scikit-learn};

\item Finally, TPOT optimizes the hyper-parameters of the chosen method.

\end{enumerate}

The number of pipelines that are analyzed depends on the choice of three TPOT hyper-parameters: {\it Generations}, {\it Population size} and {\it Offspring size}. The first defines the number of iterations in the process of optimizing the pipeline; the second indicates the number of options that are retained to the next generation; and the third designates the number of new options to be produced in each generation. At the end of the TPOT computation, the total number of pipelines analyzed is given by {\it Population Size $+$ Generations $\times$ Offspring Size}. In this work, following the TPOT documentation, we set all these hyper-parameters to 100, which means that 10100 pipelines are analyzed.

Our analysis returns eXtreme Gradient Boosting \citep[XGBoost,][]{chen2016xgboost} as the most suitable pipeline, without combining with other methods and changing the feature vector.
In other words, the model uses the initial 37 features.
More details on the pipeline can be found in Appendix~\ref{ap:pipeline}.
We use 80\% of the labeled training set for training and 20\% for testing.

%%%%%%%%%%%%%%%%%%%%%%%%%%%%%%%%%%%%%%
\subsection{Extreme Gradient Boosting (XGBoost)}

Even though TPOT has the ability of stacking different methods in order to find the more suitable pipeline for a given problem, our analysis delivered the Extreme Gradient Boosting  as the most suitable ML method for our classification.  XGBoost is an ensemble learning algorithm that combines the strengths of gradient boosting and regularization techniques to build a predictive model. It operates by building a series of weak prediction models, typically decision trees, and then combines them to form a strong ensemble model.

The XGBoost process works by optimizing an objective function, which contains two components: the loss function and the regularization term. Whereas the loss function quantifies the error between the predicted labels and the true labels, the regularization term penalizes the complexity of the model, in order to avoid overfitting and encourages the model to select simple and more robust weak learners. XGBoost employs a boosting process to iteratively improve the performance of the model. At each iteration, a new weak learner (decision tree) is added to the ensemble model to correct the errors made by the existing ensemble. After the boosting process is complete, the final model is used to make predictions on new instances by summing the predictions of all weak learners in the ensemble, weighted by their corresponding learning rates. For more technical details, we refer to~\citep{chen2016xgboost}.

%%%%%%%%%%%%%%%%%%%%%%%%%%%%%%%%%%%%%%
\subsection{Performance metrics}
\label{ssec:metrics}

Performance can be evaluated using the confusion matrix, which thoroughly compares predicted and true values.
For the $k$ binary classifiers the confusion matrix is $2\times 2$ with four entries: true positives (TP), true negatives (TN), false positives (FP) and false negatives (FN). 
If instead we consider the  multi-class classifier, the confusion matrix is $3 \times 3$ with 9 entries, as the false positive and negatives are split into the remaining two classes.
The confusion matrices are defined according to a criterion to decide how a source is classified.
In the case of $2\times 2$ confusion matrices, one has to specify a probability threshold $p_{\rm cut}$ that separates the two classes, that is, TP, TN, FP and FN depend on $p_{\rm cut}$.
In the case of $3\times 3$ confusion matrices one usually simply classifies according to the highest probability, as discussed in Fig.~\ref{fig:3k}, but alternative schemes may be more appropriate for specific cases.

In the following, we will adopt the $3 \times 3$ confusion matrix to judge the performance of the best classifier.
We will also consider the $2\times 2$ confusion matrices in order to obtain the receiver operating characteristic (ROC) curve and the purity-completeness curve, which are commonly used graphical representations to assess the performance of classification models.
The ROC curve is a parametric plot of the true positive rate (TPR) and false positive rate (FPR) as a function of $p_{\rm cut}$:
\begin{align}
\text{TPR}(p_{\rm cut})=\frac{\rm TP}{\rm  TP+FN}\,, \quad \quad
\text{FPR}(p_{\rm cut})=\frac{\rm FP}{\rm FP+TN} \,, 
\end{align}
with $0\le p_{\rm cut} \le 1$.
TPR is also called recall and, in astronomy, is the completeness. The performance of a classifier can then be summarized with the area under the ROC curve (AUC). The AUC can assume values between 0 and 1. A perfect classifier has a value of 1, while a random classifier, on average, a value of 1/2.

The purity-completeness curve is a useful method to assess the performance of an unbalanced classifier---the training set does not feature the same number of galaxies, stars and quasars.
It is a parametric plot of the completeness (or recall) and the purity (or precision) as a function of $p_{\rm cut}$:
\begin{align}
\text{Purity}(p_{\rm cut}) = \frac{\rm TP}{\rm TP+FP}\,, \quad
\text{Completeness}(p_{\rm cut}) = \frac{\rm TP}{\rm TP+FN} \,, 
\end{align}

In order to summarize the purity-completeness curve, we consider the average precision (AP), which is the area under the purity-completeness curve and takes values between 0 and 1.

Finally, the $F$-score is often used as an overall performance metric.  If one assigns equal importance to both completeness and purity, one has the particular case of the ``$F_1$'' score, the harmonic mean of completeness and purity:
\begin{align}
F_1 (p_{\rm cut}) \!=\! \frac{2}{1/\text{Purity}\!+\!1/\text{Completeness}} \!=\! \frac{2 \rm TP}{\rm 2 TP \!+\! FP \!+\! FN} .
\end{align}

%%%%%%%%%%%%%%%%%%%%%%%%%%%%%%%%%%%%%%
%%%%%%%%%%%%%%%%%%%%%%%%%%%%%%%%%%%%%%
%%%%%%%%%%%%%%%%%%%%%%%%%%%%%%%%%%%%%%
\section{Training set}
\label{sec:training}

The performance of a supervised ML classification depends crucially on three aspects: the choice of the algorithm, the quality of the data, and the trustworthiness and breadth of the training set.
The first aspect is discussed in Sec.~\ref{sec:ml}. The last two aspects are interconnected: for a given quality of data, improving the training set beyond a certain point is inconsequential. This can be quantified using, for example, the learning curve approach \citep[Sec.~D]{vonMarttens:2018bvz}.
As we use only about 5\% of the full catalog for training (see below), we aim to maximize the number of objects in the training set without neglecting its purity.
Specifically, we wish to increase the number of quasars, which are the most difficult to classify.
To this end, we crossmatched the J-PLUS DR3 dataset with SDSS DR18, LAMOST DR8 and {\it Gaia} DR3. All these catalogs have a high quality and a large overlap with the area observed by J-PLUS. This crossmatch yields the spectral classification of the objects, with the features obtained from the J-PLUS data.

%%%%%%%%%%%%%%%%%%%%%%%%%%%%%%%%%%%%%%
\subsection{Crosmatch between J-PLUS DR3 and SDSS DR18}
\label{sssec:jplusxsdss}

\begin{figure}
\centering
\includegraphics[trim={0      .3cm 0  0}, clip, width=\columnwidth]{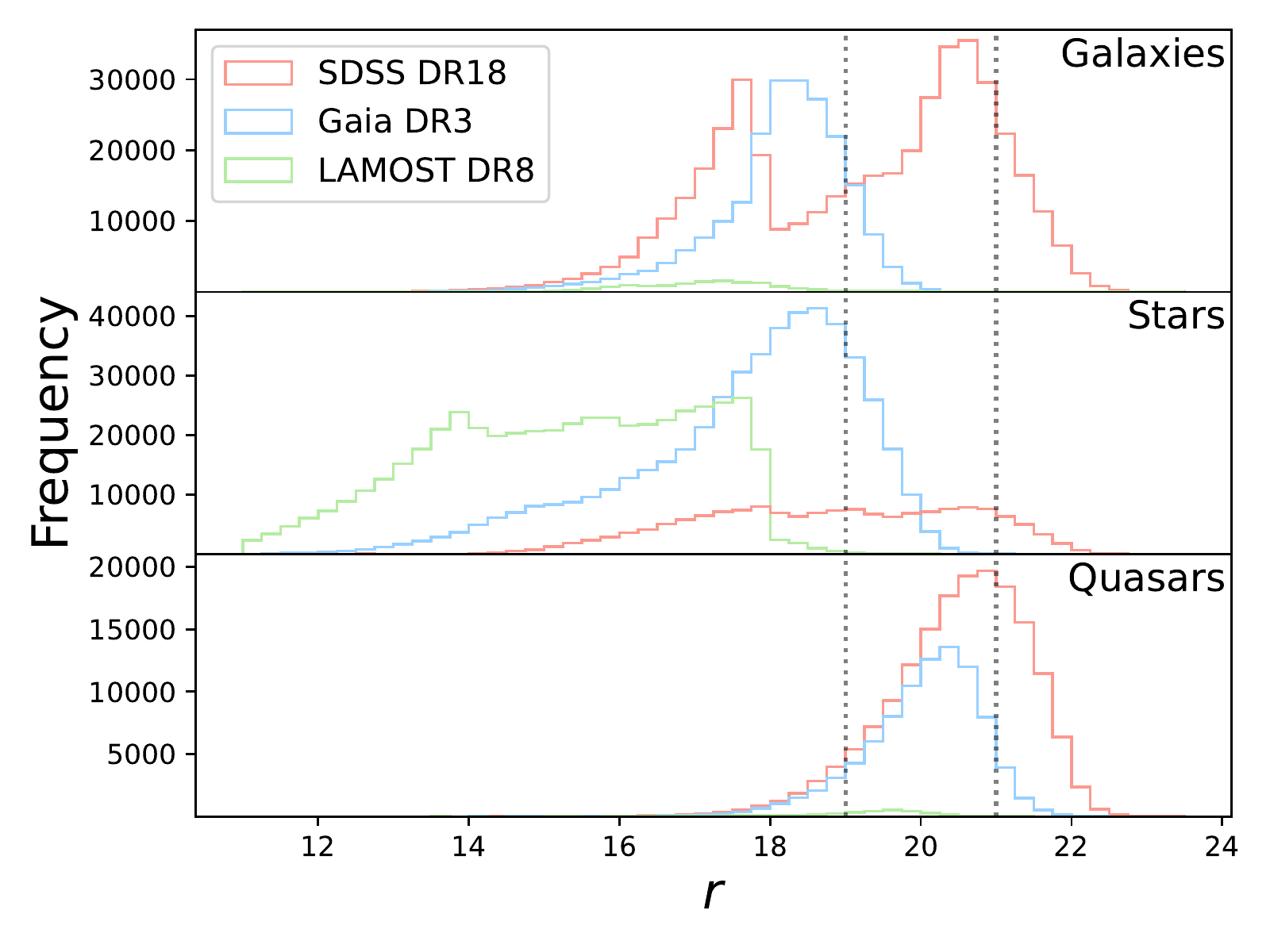}
\caption{Distribution of galaxies (top), stars (middle) and quasars (bottom) according to the $r$ band for the training sets coming from SDSS DR18, LAMOST DR8 and {\it Gaia} DR3.}
\label{fig:hist_trin}
\end{figure}

The first part of the training set consists of eBOSS SDSS DR18 sources that are classified through $R\sim 2000$ spectroscopy \citep{SDSS:2023tbz}.
The crossmatch between J-PLUS DR3 and SDSS DR18 contains 772,323 objects, being 437,976 galaxies, 161,203 stars, and 173,144 quasars. We consider all these objects in our training set.
The distributions of stars, galaxies and quasars are shown in Fig.~\ref{fig:hist_trin}, the  abundances in Table~\ref{tab:set}.

%

%%%%%%%%%%%%%%%%%%%%%%%%%%%%%%%%%%%%%%
\subsection{Crosmatch between J-PLUS DR3 and LAMOST DR8}
\label{sssec:jplusxlamost}

The second part of the training set comes from the Large Sky Area Multi-Object Fiber Spectroscopic Telescope \citep[LAMOST,][]{2022ApJS..259...51W}. More precisely, we use the low-resolution spectral classification from DR8.\footnote{\href{http://www.lamost.org/dr8/}{dr8.lamost.org/v2.0}} It is important to mention that LAMOST DR8 has a $g$-band limiting magnitude of about 20 mag, and a resolution $R \sim 1000$.

In total, the crossmatch between J-PLUS DR3 and LAMOST DR8 has 1,477,218 objects. However, ignoring the crossmatch with SDSS DR18, which has a higher purity,%
\footnote{SDSS DR18 features a higher resolution spectroscopy ($R\sim 2000$) as compared to LAMOST DR8 ($R\sim 1000$).} to avoid double counting, 
leaves 983,502 objects. Among these objects we have 13,995 galaxies and 3,348 quasars. All these galaxies and quasars are included in the training set, but, in order to increase purity and avoid an imbalanced data set, we apply the condition that {\tt CLASS\_STAR} and {\tt SGLC} must be bigger than 0.7 for the stars. In other words, we use the J-PLUS classifiers to improve our training set. After applying this condition the number of stars becomes 966,159, which we still consider an imbalanced number in comparison with galaxies and quasars. Thus, as the last step, we randomly sample 500000 stars from the remaining ones.%
\footnote{We tested sampling different number of stars but found that the best results were obtained sampling 500000 stars.}
The final contribution from LAMOST DR8 to the training set contains 13,995 galaxies, 500,000 stars and 3,348 quasars. The distribution of LAMOST DR8 sources is shown in Fig.~\ref{fig:hist_trin} and Table~\ref{tab:set}.

\setlength{\tabcolsep}{7pt}
\renewcommand{\arraystretch}{1.7}
\begin{table}
\centering
\small
\begin{tabular}{lccc|c}
\hline
\hline
Catalog        & Galaxy  & Star            & Quasar  & Total \\
\hline
SDSS DR18      & 437,976 & 161,203 & 173,144  & 772,323 \\
LAMOST DR8     & 13,995  & 500,000 & 3,348   & 517,343 \\
{\it Gaia} DR3 & 212,716 & 500,000 & 90,548 & 803,264 \\
\hline
Total          & 664,687 & 1,161,203 & 267,040 & 2,092,930 \\
\hline
\hline
\end{tabular}
\caption{Training set composition. See Fig.~\ref{fig:inter} for the diagram summarizing how the training set is built.}
\label{tab:set}
\end{table}

%%%%%%%%%%%%%%%%%%%%%%%%%%%%%%%%%%%%%%
\subsection{Crosmatch between J-PLUS DR3 and {\it Gaia} DR3}
\label{sssec:jplusxgaia}

Finally, the third part comes from   {\it Gaia} DR3 %\citep{2016A&A...595A...1G}
\citep{vallenari2022gaia}, which can be downloaded through the ADQL protocol in the {\it Gaia} webpage.\footnote{\href{https://www.cosmos.esa.int/web/gaia/home}{www.cosmos.esa.int/web/gaia/home}} Specifically for galaxies and quasars, we adopt the extragalactic information  from the {\tt gaiadr3.galaxy\_candidates} and {\tt gaiadr3.qso\_candidates} catalogs. As described in \citet[Section 8]{Gaia:2022vcs}, we use the purer galaxy and quasar candidate sub-samples, which are claimed to have an overall purity of 95\%. Regarding the extragalactic objects, the full crossmatch of  J-PLUS DR3 and {\it Gaia} DR3 contains 297,523 galaxies and 198,602 quasars, and after removing the repeated objects with SDSS DR18 and LAMOST DR8, which have a higher purity,%
\footnote{{\it Gaia} DR3 classification is based on lower-resolution spectra so that $30\leq\lambda/\Delta\lambda\leq 100$, see \citet[][Fig.~3]{carrasco2021internal}.}
one is left with 212,716 galaxies and 90,548 quasars, which are included in the training set.

On the other hand, regarding stars, we have considered the {\tt gaiadr3.gaia\_source} catalog. The full crossmatch between J-PLUS DR3 and  {\it Gaia} DR3 stars contains more than 20 million sources, after ignoring the repeated objects within SDSS DR18 and LAMOST DR8 to avoid double counting. Similarly to the LAMOST case, we also apply a quality cut condition to avoid an imbalanced data set and improve purity. Due to the huge number of stars in the crossmatch, in this case we consider only stars that satisfy the condition {\tt CLASS\_STAR}={\tt SGLC}=1, which results in 966,159 stars. This condition effectively limits the $r$-band magnitude to about $r\approx 20$, which should not be a problem as the SDSS DR18 training set includes fainter stars, till $r\approx 22$, as shown in Fig.~\ref{fig:hist_trin} (middle panel). Also in this case we randomly sample 500,000 objects. The distributions are shown in Fig.~\ref{fig:hist_trin}, the  abundances in Table~\ref{tab:set}.

%%%%%%%%%%%%%%%%%%%%%%%%%%%%%%%%%%%%%%
\subsection{On the quality of the LAMOST DR8 and {\it Gaia} DR3 contributions}
\label{sssec:pur_lamost_gaia}

\begin{figure}
\centering
\includegraphics[trim={0      .3cm 0  0}, clip, width=\columnwidth]{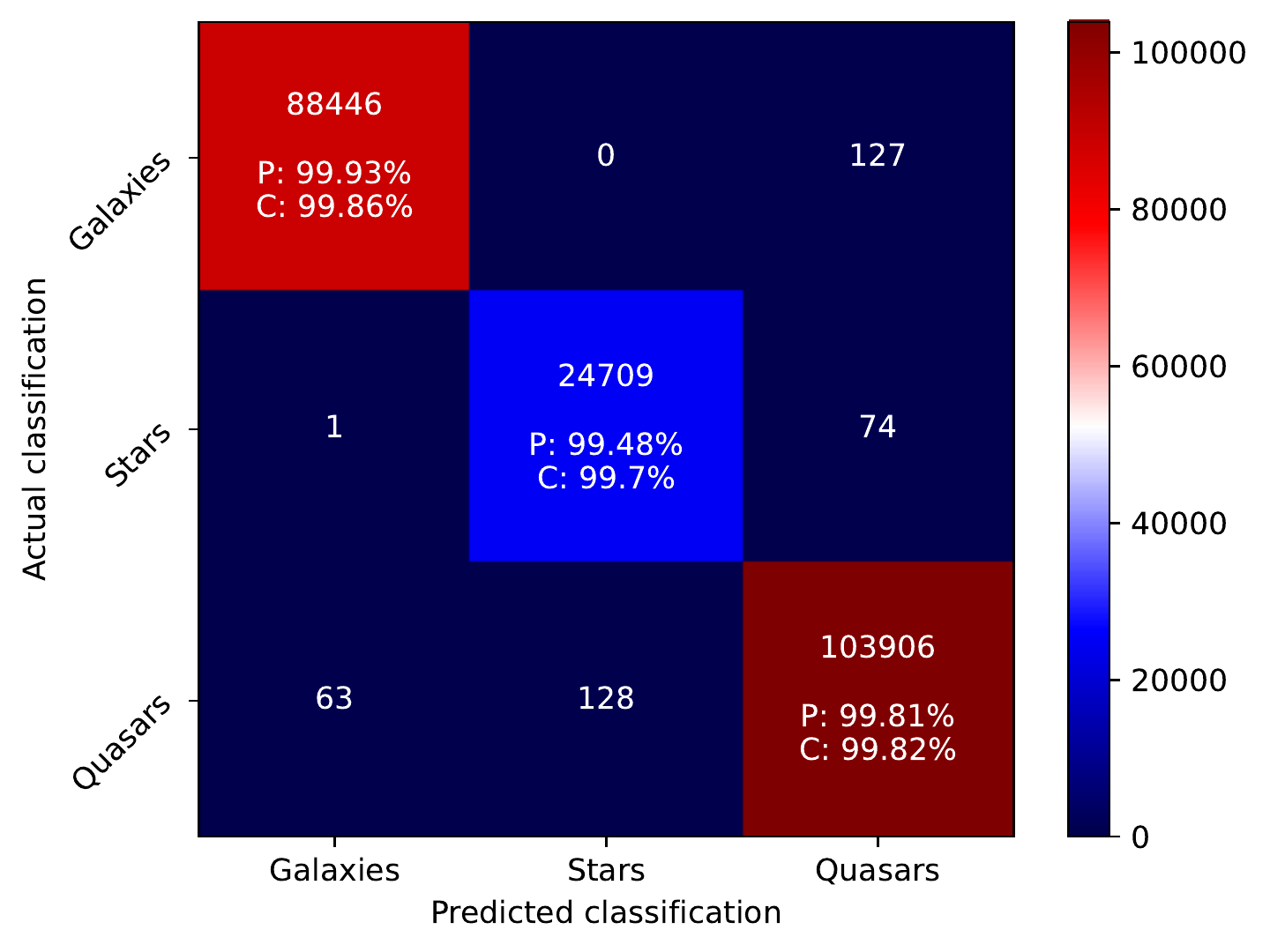}
\caption{Confusion matrix for assessing the quality of classification of LAMOST DR8 + {\it Gaia} DR3 selected sources ($x$-axis) against the classification of SDSS DR18 selected sources ($y$-axis). ``P'' stands for purity and ``C'' for completeness.}
\label{fig:confusion}
\end{figure}

Only SDSS DR18 makes use of high resolution spectra to classify the sources, while LAMOST DR8 and {\it Gaia} DR3 classify the sources combining low resolution spectra with other techniques. For this reason, before performing our classification we need to validate the LAMOST DR8 and {\it Gaia} DR3 contributions. For this purpose, we consider the subset of the training set composed by the intersection between the crossmatch of J-PLUS DR3 and SDSS DR18 with the union of LAMOST DR8 and {\it Gaia} DR3. This subset consists of the sources that are in the training set and are classified by SDSS DR18 and also by LAMOST DR8 and/or {\it Gaia} DR3\footnote{When a given object is classified by both LAMOST DR8 and {\it Gaia} DR3, we consider that the LAMOST DR8 classification is prioritary.}. Note that this subset consists of the objects that were ignored to avoid double counting in the training set. Then, in order to test the quality of our truth table, we build a confusion matrix comparing the SDSS DR18 results, considered as the actual classification, with the LAMOST DR8 and {\it Gaia} DR3 results, considered as the predicted classification. The confusion matrix obtained is shown in Fig.~\ref{fig:confusion} and, assuming that this subset is representative for the LAMOST DR8 and {\it Gaia} DR3 contributions, we conclude that the training set has purity and completeness always above 99\%.

\begin{figure}
\centering
\includegraphics[width=.7 \columnwidth]{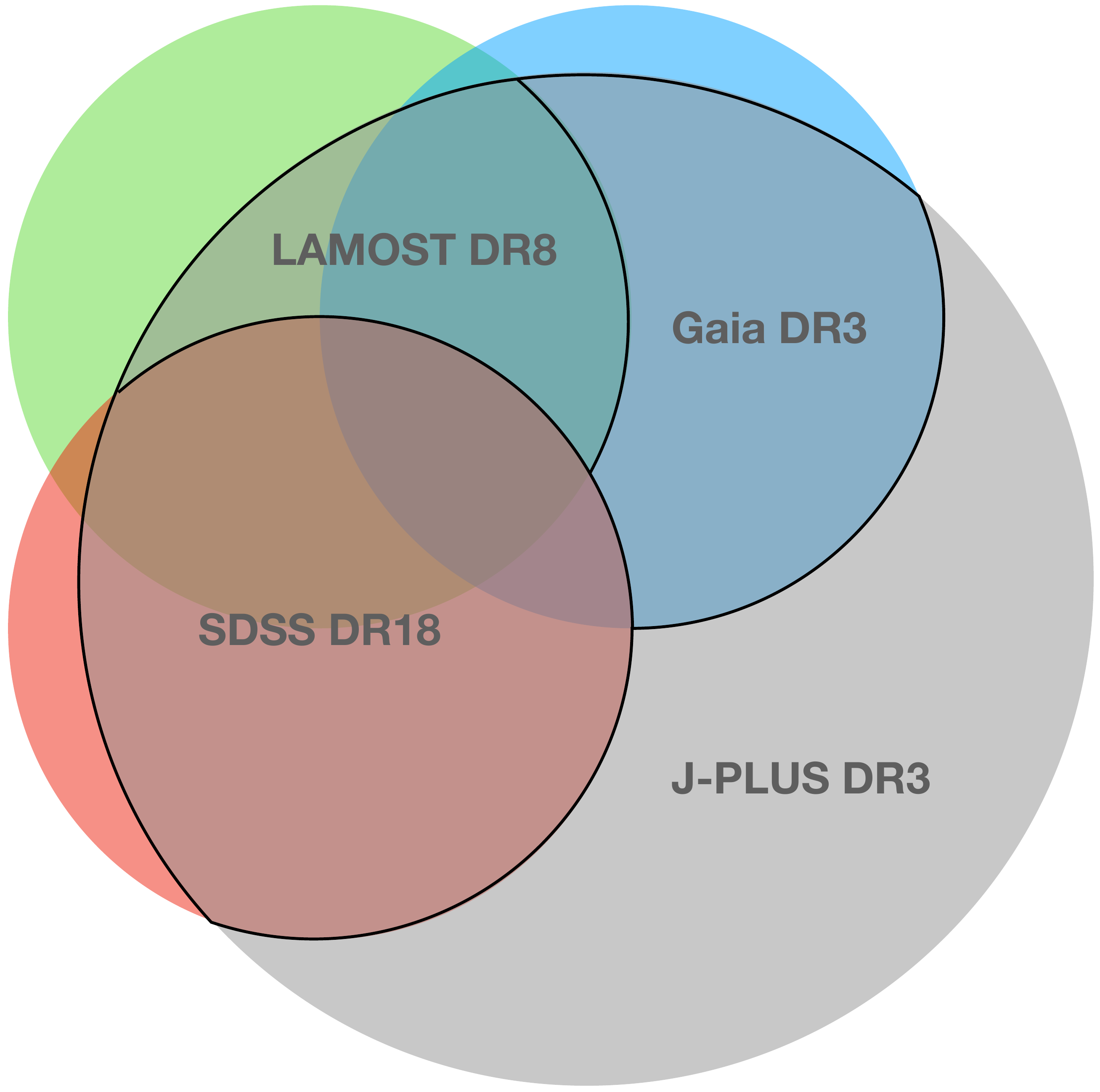}
\caption{Venn-like diagram that summarizes how the training set is built. For common sources, priority is given to the SDSS objects, then LAMOST and finally {\it Gaia}. The red SDSS DR18 (green LAMOST DR8) [blue {\it Gaia} DR3]  area delimited by a solid black line refers to SDSS (LAMOST) [{\it Gaia}] contribution to the training set.  All three cases are quantified in Table~\ref{tab:set}, where the number of sources can also be found.
}
\label{fig:inter}
\end{figure}
%

%%%%%%%%%%%%%%%%%%%%%%%%%%%%%%%%%%%%%%
\subsection{Final set}
\label{sssec:tr}

Our training set is built by incrementally crossmatching the J-PLUS DR3 catalog  with SDSS DR18, LAMOST DR8 and {\it Gaia} DR3.
Table~\ref{tab:set} shows the contribution of the catalogs of the previous sections to the final training set. Fig.~\ref{fig:inter} it illustrates the crossmatch of J-PLUS DR3 with SDSS DR18, LAMOST DR8, and Gaia DR3, prioritized in this specific sequence when handling repeated objects. In other words, it summarizes how the training set was built.

\begin{figure}
\centering
\includegraphics[trim={0      .3cm 0  0}, clip, width=\columnwidth]{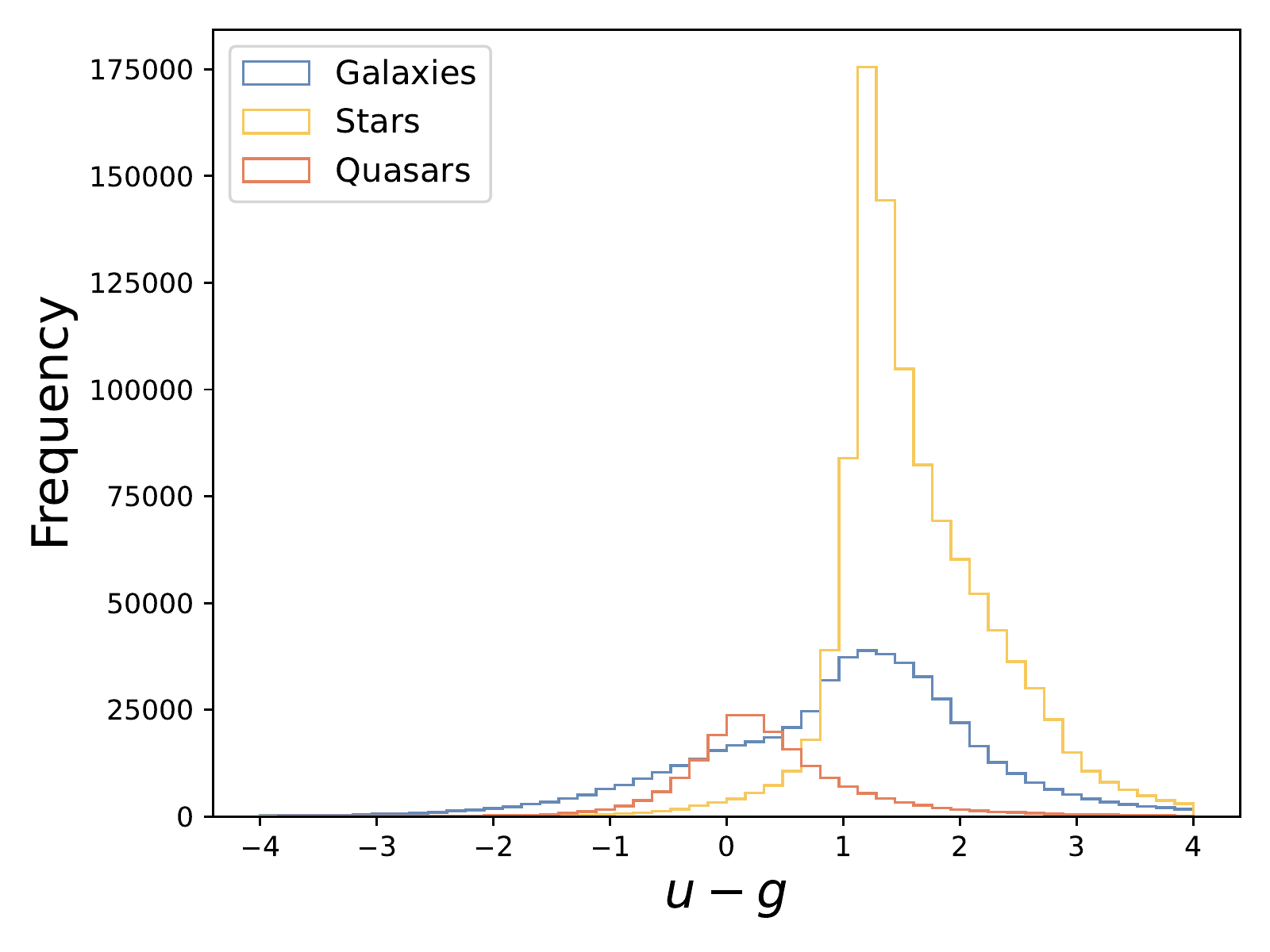}
\includegraphics[trim={0      .3cm 0  0}, clip, width=\columnwidth]{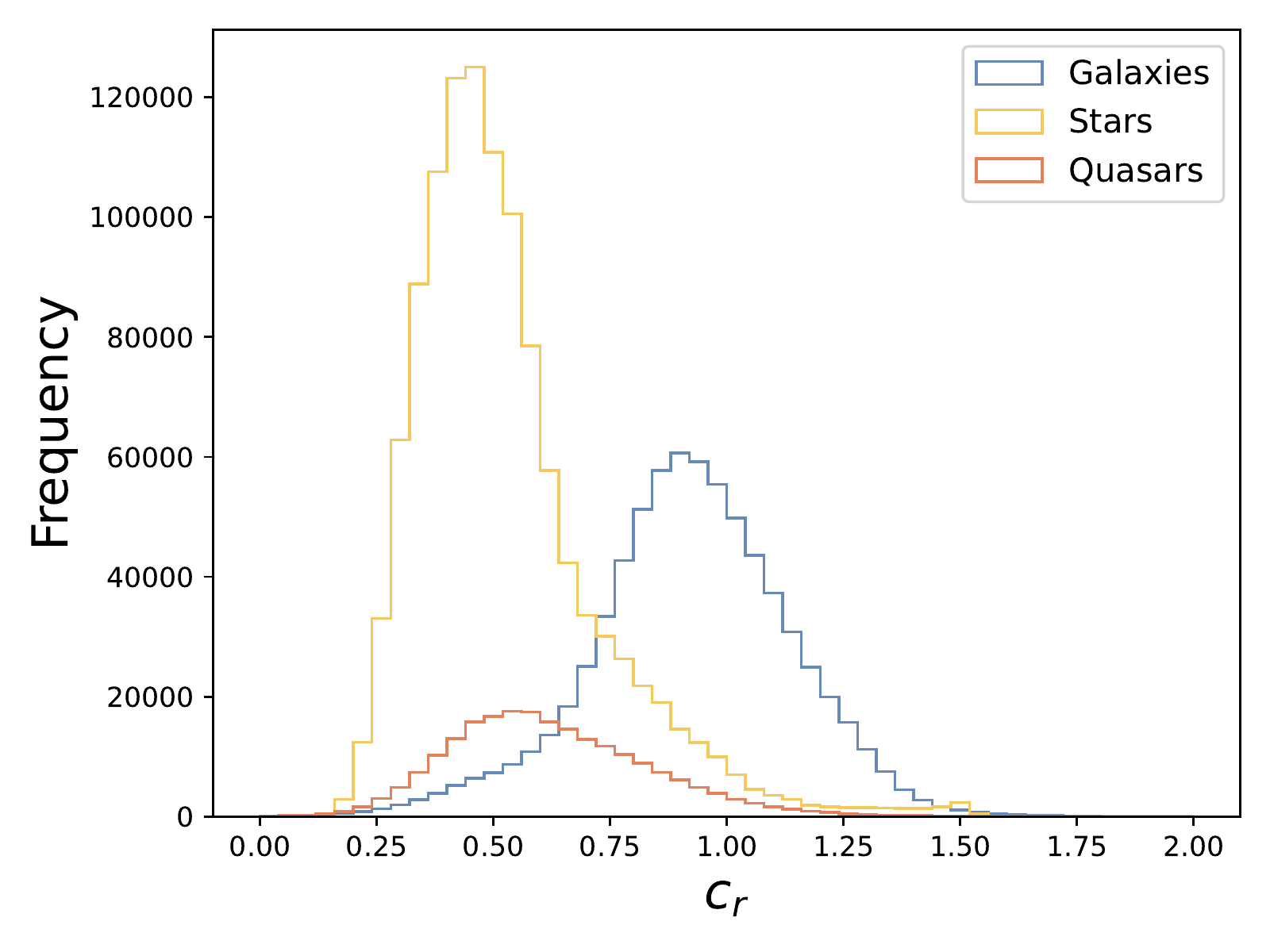}
\caption{Histograms of $u-g$ color (top) and concentration (bottom) for the training set divided by galaxies, stars and quasars.}
\label{fig:hist_train}
\end{figure}

In order to give an intuitive understanding of how ML models can be used to effectively differentiate between galaxies, stars, and quasars within the J-PLUS DR3 data, we show in Fig.~\ref{fig:hist_train} the distribution of the $u-g$ color (top) and the concentration $c_r$ (bottom). By analyzing the top panel, it becomes apparent that the $u-g$ color can be used for distinguish quasars from star and galaxies. On the other hand, the bottom panel showcases that the concentration can be used to distinguish galaxies (extended sources) from stars and quasars (point-like sources).
We have selected the $u-g$ color and the concentration as key parameters to illustrate the informativeness of our training set. These choices are grounded in the feature importance analysis, the details of which will be discussed in Section~\ref{subsec:fi}.
Of course, Fig.~\ref{fig:hist_train} is merely illustrative, and the proposed classification is a multidimensional and complex problem.

As we will see in Section~\ref{sec:represa}, the training set well represents the magnitude range and the angular distribution of the J-PLUS DR3 dataset, as well as the distribution of the most important features.
%Finally, we checked that by considering half of the training set the ML performance degrades.
%This indicates that we did not yet reach saturation as far as the size of the training set is concerned and that further gain in performance could be achieved by adopting a yet larger training set.

%%%%%%%%%%%%%%%%%%%%%%%%%%%%%%%%%%%%%%
%%%%%%%%%%%%%%%%%%%%%%%%%%%%%%%%%%%%%%
%%%%%%%%%%%%%%%%%%%%%%%%%%%%%%%%%%%%%%
\section{Results}
\label{sec:results}

In this section, the classification performances of our model are presented. As  mentioned in Sec.~\ref{ssec:tpotxgb}, the TPOT analysis is performed considering 80\% of the training set presented in Sec.~\ref{sec:training}. On the other hand, the results presented here consist in applying the TPOT-optimized pipeline to the remaining 20\% of the training set that was not used for its construction (the test set). In other words, here we assess whether our pipeline performs well and can be applied to the entire J-PLUS DR3 catalog.
We find convenient to show the results according to different $r$-band bins. We consider the following bins: $r\le 19$, $19<r\le 21$ and $r>21$. From now on, when not explicitly mentioned, all analyses as a function of the probabilities use the probabilities of the binary classifiers, i.e., the unnormalized probabilities $p_{\rm gal}$,  $p_{\rm star}$ and $p_{\rm qso}$.

\subsection{Confusion matrix}
\label{cmf}

\begin{figure}
\centering
% trim={<left> <lower> <right> <upper>}
\includegraphics[trim={0      .3cm 0  0}, clip, width=0.48\textwidth]{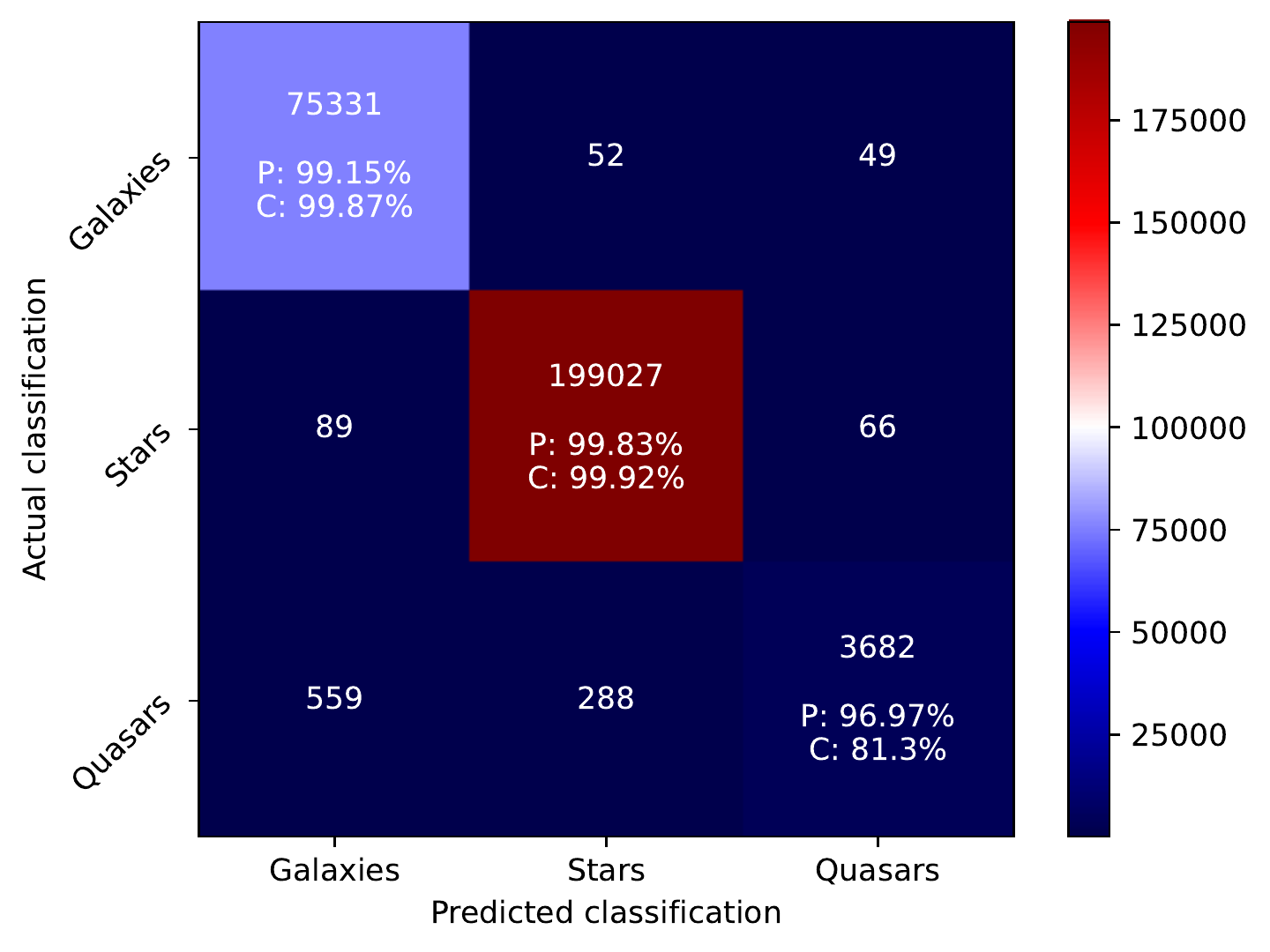}
\includegraphics[trim={0      .3cm 0  0}, clip, width=0.48\textwidth]{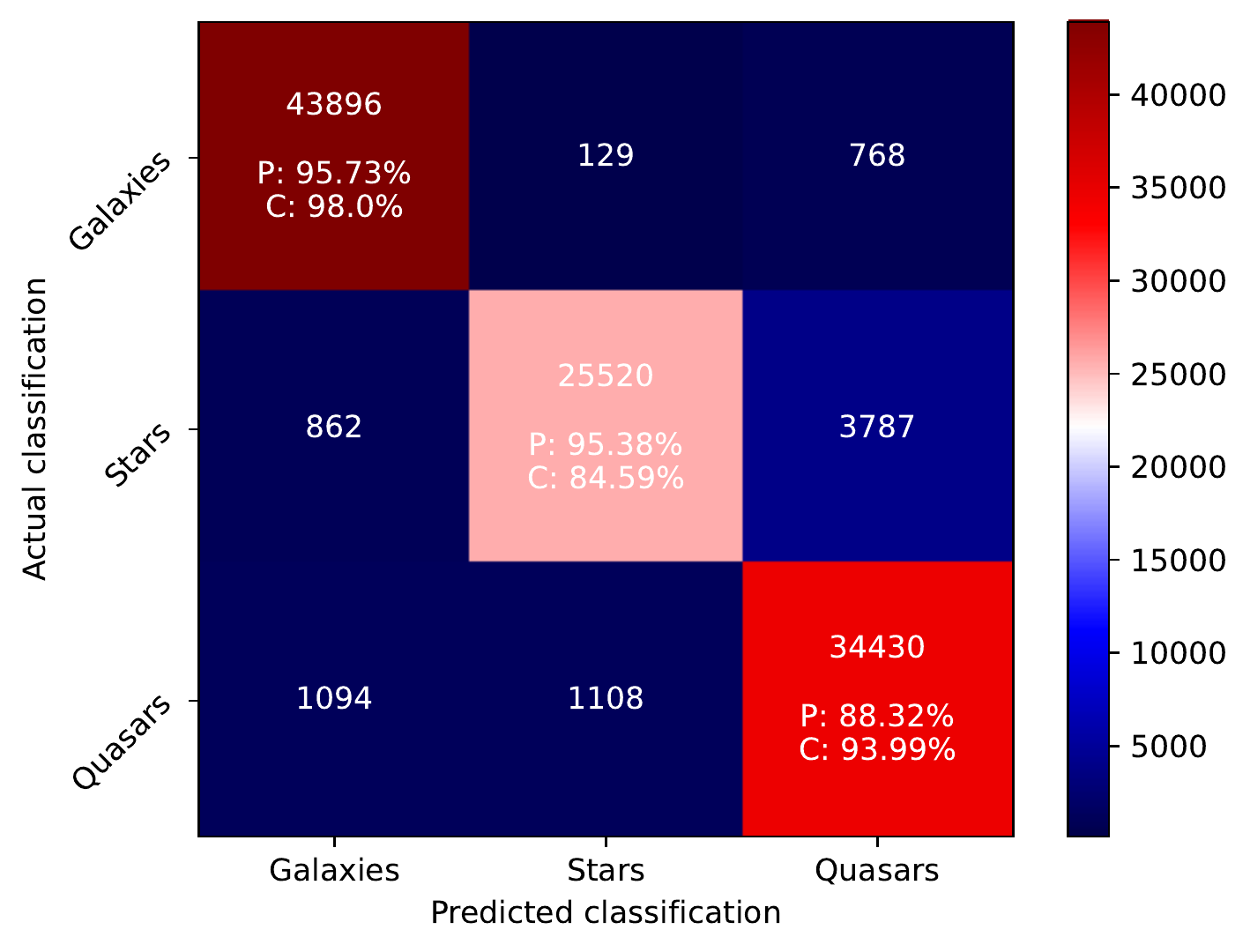}
\includegraphics[trim={0      .3cm 0  0}, clip, width=0.48\textwidth]{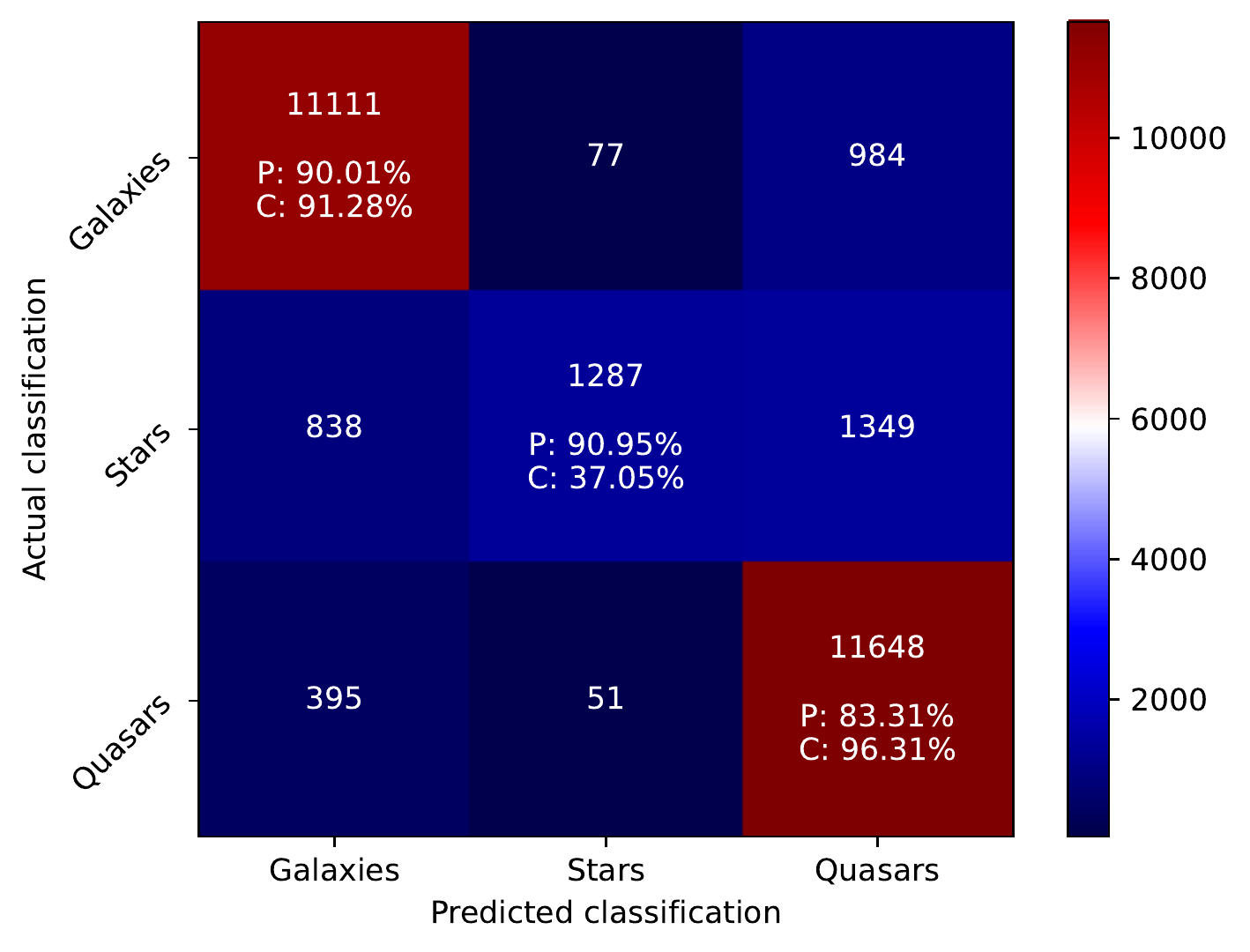}
\caption{Confusion matrices for $r\le 19$ (top), $19<r\le 21$ (middle) and $r>21$ (bottom) when adopting the highest probability criterion. The color code refers to the number of objects of each category.}
\label{fig:conf_r}
\end{figure}
\begin{figure*}
\centering
% trim={<left> <lower> <right> <upper>}
\includegraphics[trim={0      1.8cm .22cm  0}, clip, width=0.358\textwidth]{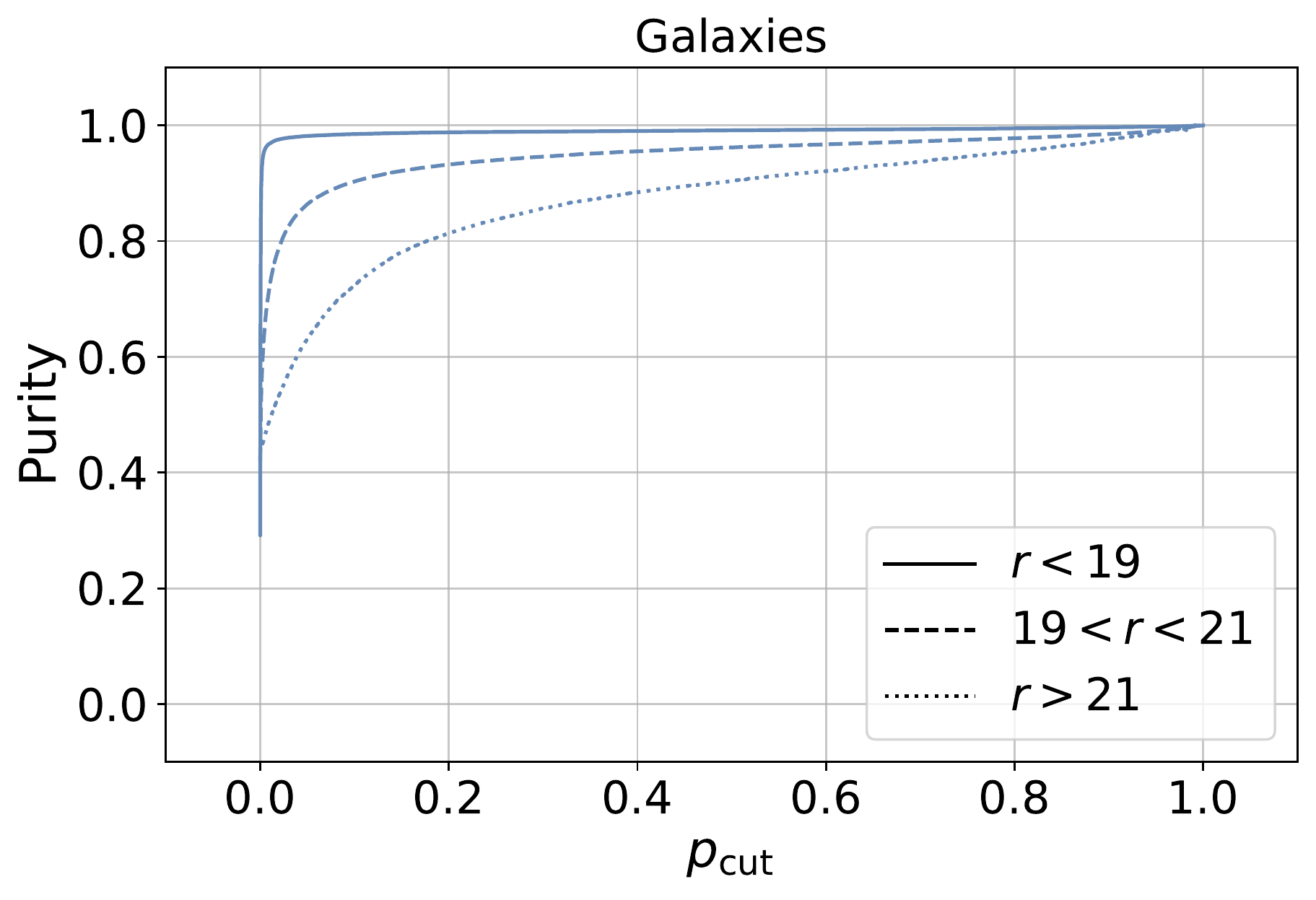}
\includegraphics[trim={2.2cm 1.8cm .22cm  0}, clip,width=0.316\textwidth]{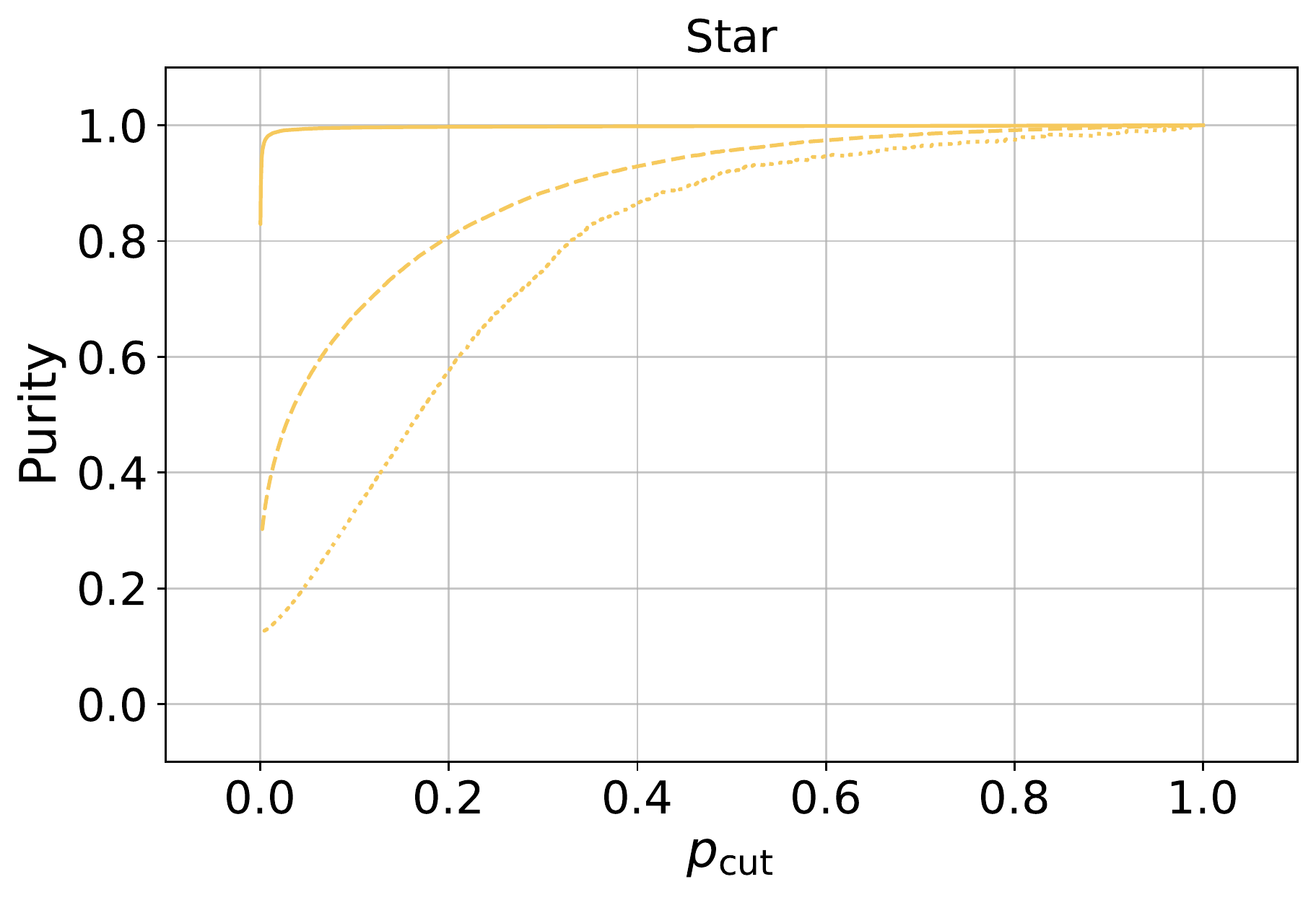}
\includegraphics[trim={2.2cm 1.8cm .22cm  0}, clip,width=0.316\textwidth]{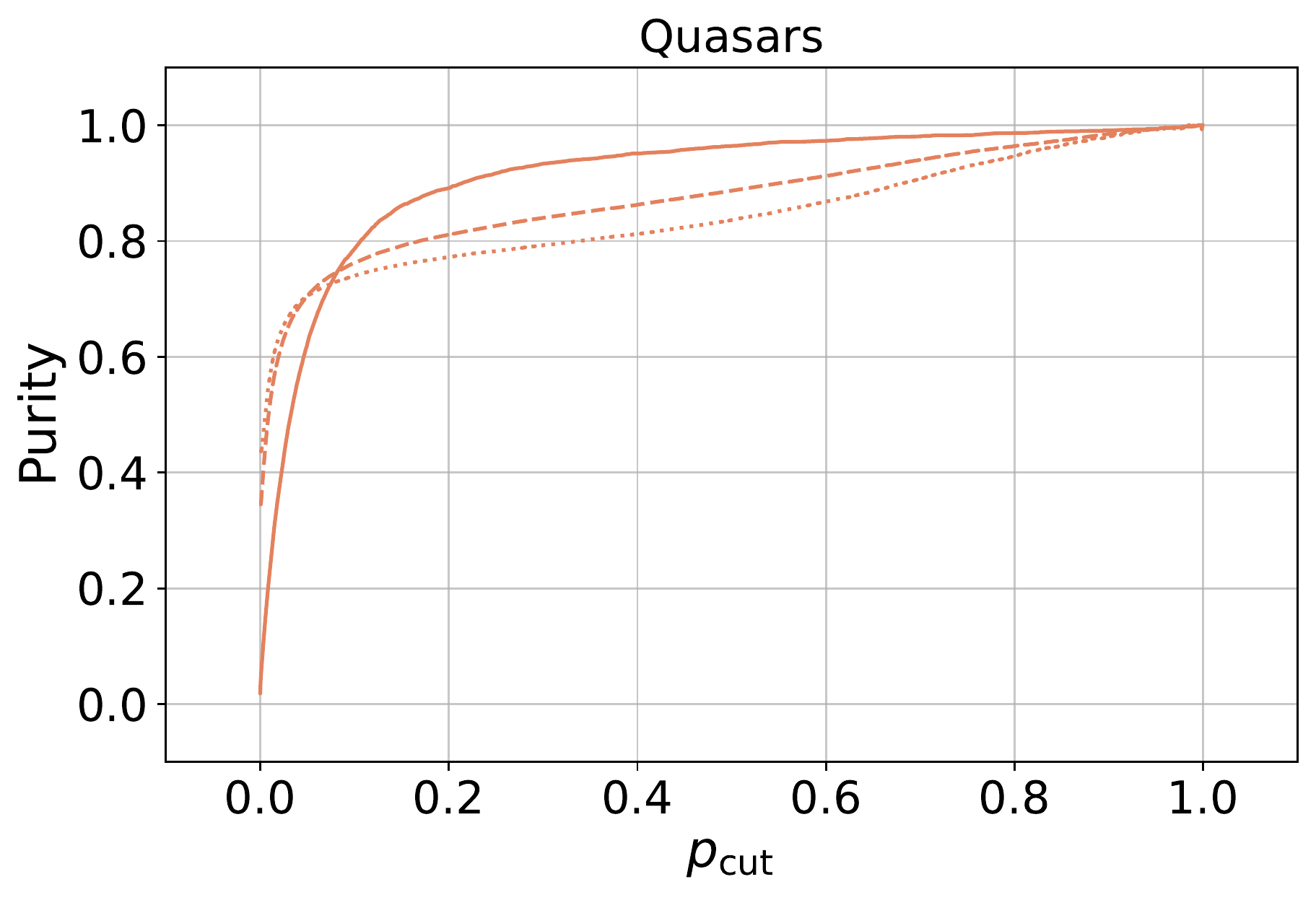}
\includegraphics[trim={0      1.8cm .22cm .75cm}, clip,width=0.358\textwidth]{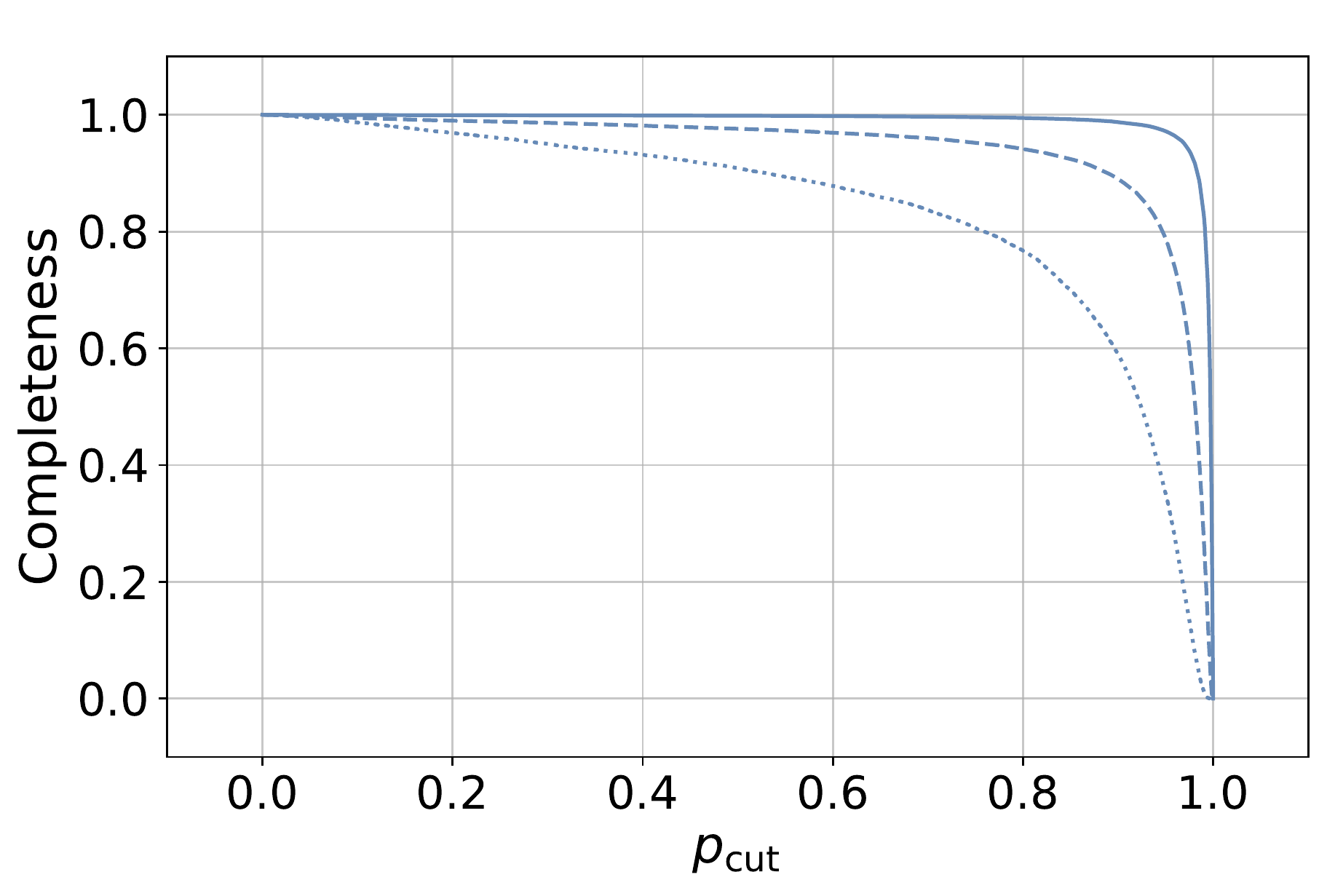}
\includegraphics[trim={2.2cm 1.8cm .22cm .75cm}, clip,width=0.316\textwidth]{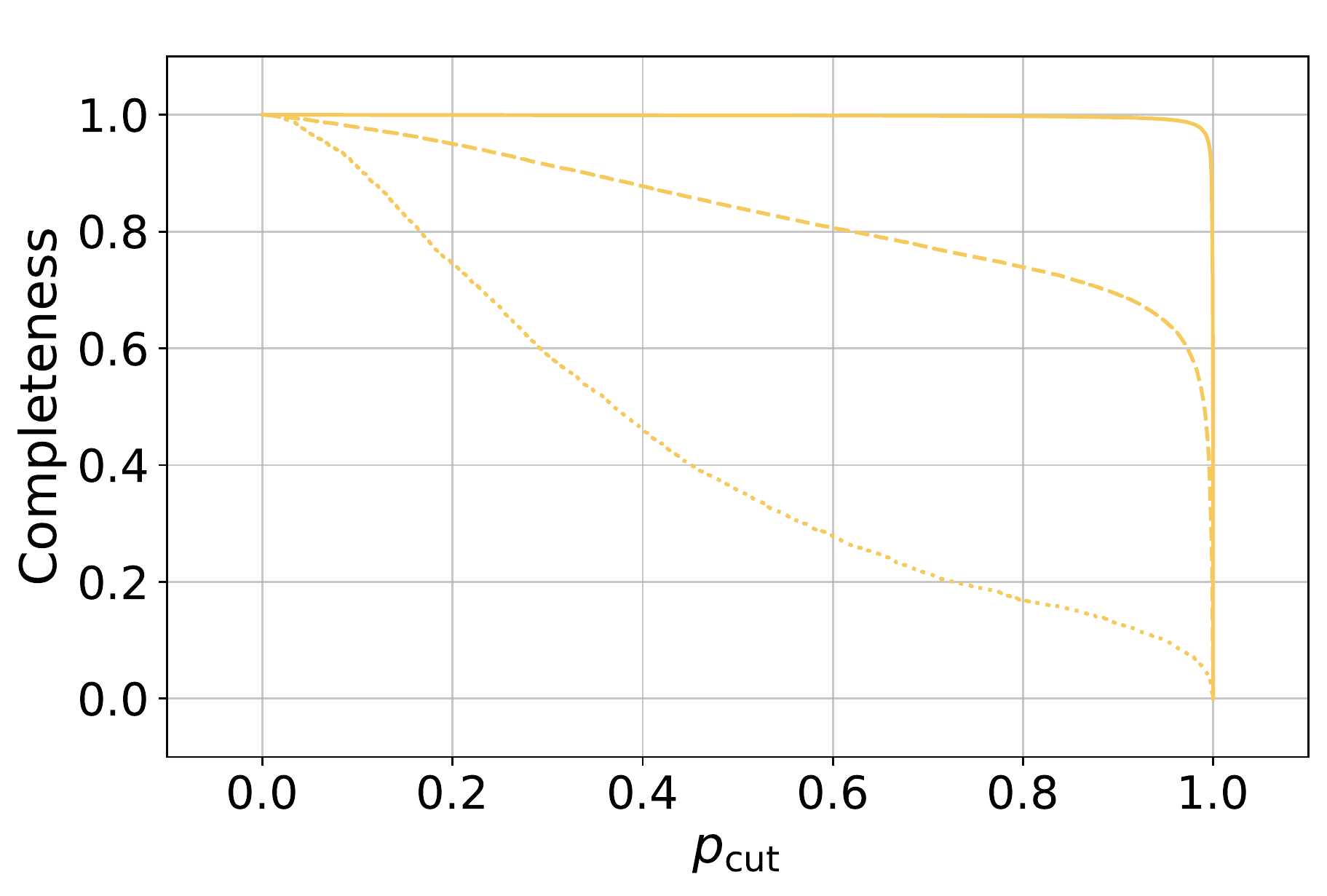}
\includegraphics[trim={2.2cm 1.8cm .22cm .75cm}, clip,width=0.316\textwidth]{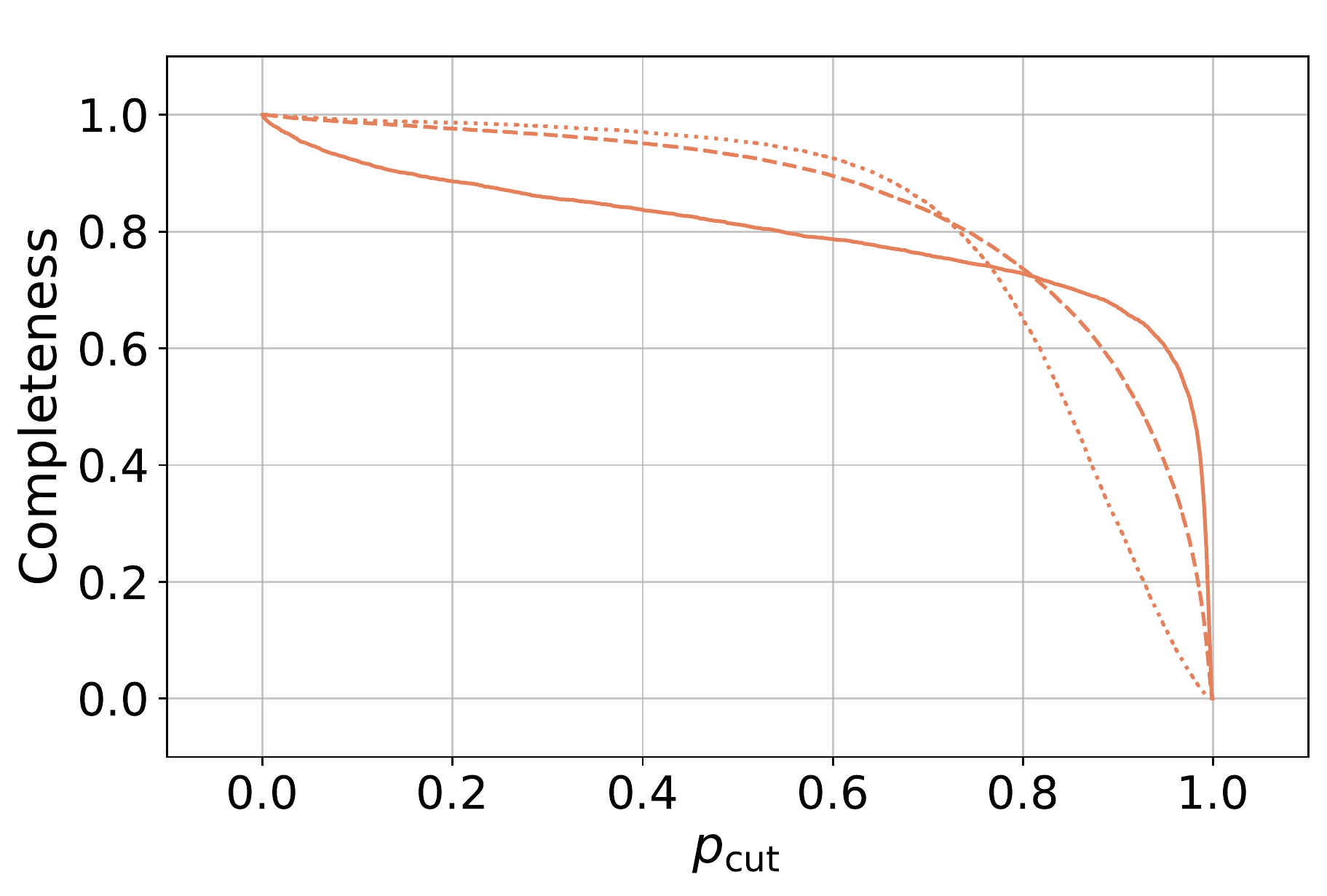}
\includegraphics[trim={0      .3cm     .22cm .75cm}, clip,width=0.358\textwidth]{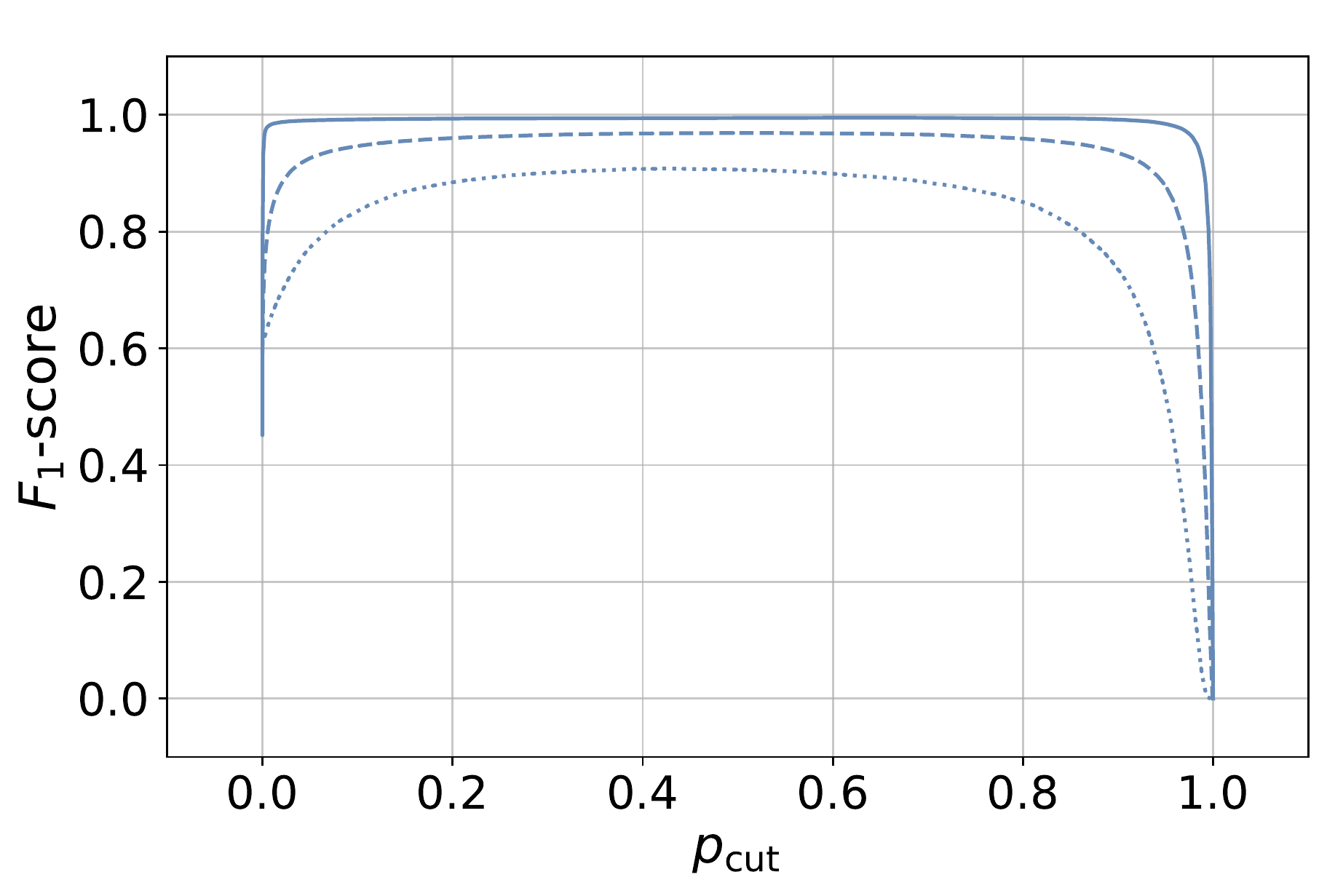}
\includegraphics[trim={2.2cm .3cm     .22cm .75cm}, clip,width=0.316\textwidth]{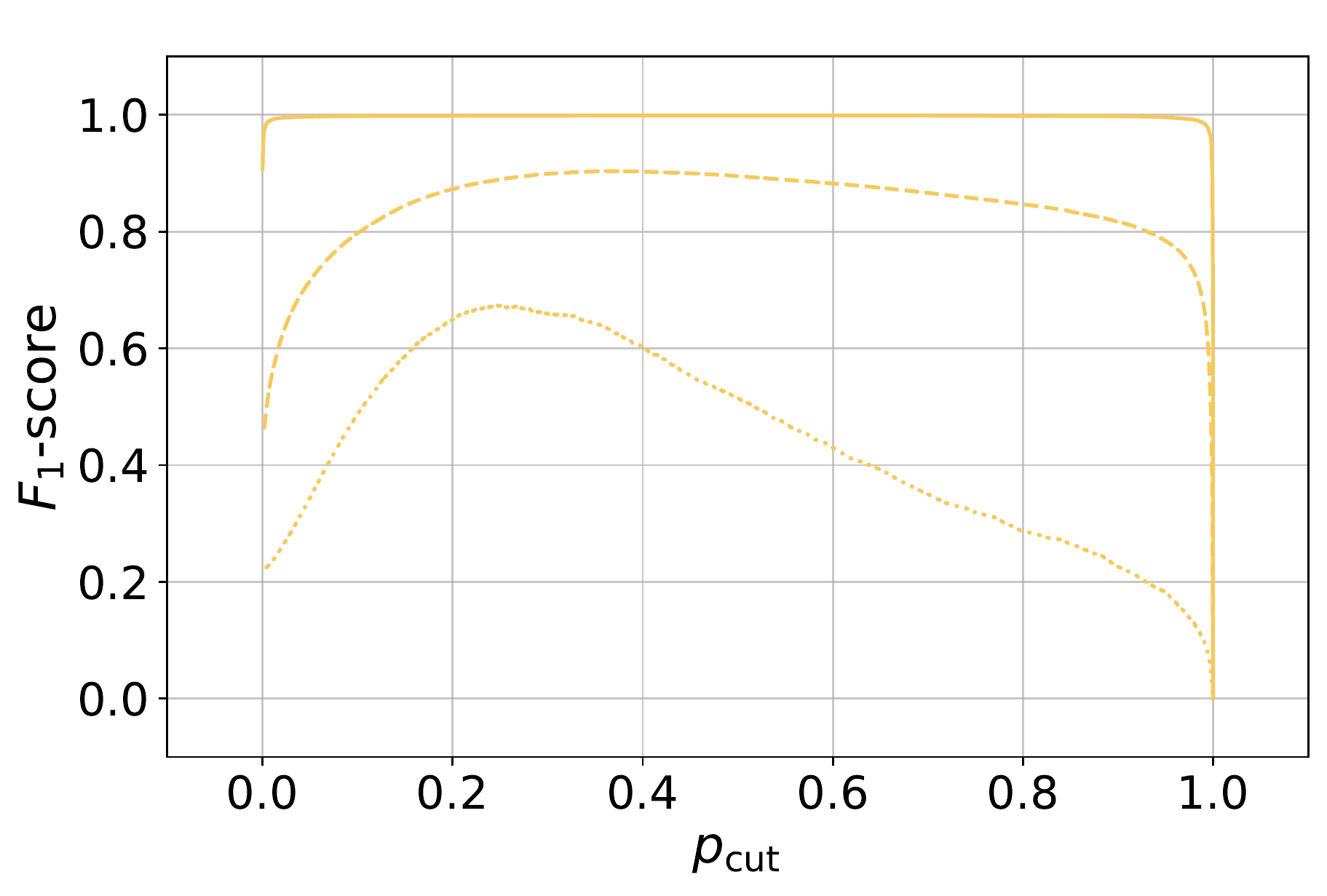}
\includegraphics[trim={2.2cm .3cm     .22cm .75cm}, clip,width=0.316\textwidth]{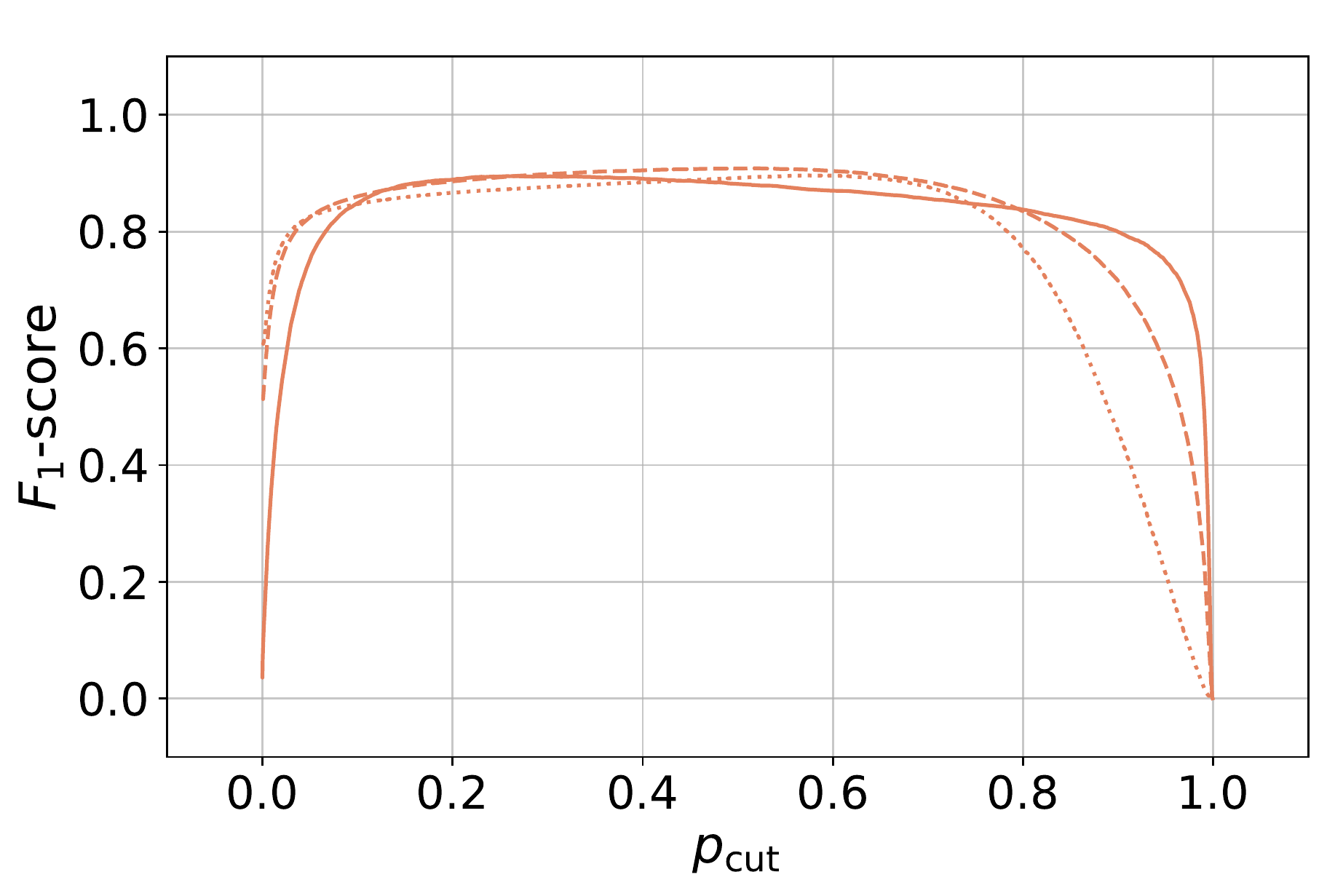}
\caption{Purity (top), completeness (center) and  $F_1$-score (bottom) for various magnitude bins for galaxies (left), stars (middle) and quasars (right). These plots use the three binary classifiers (see Section~\ref{ssec:onevsall}) and it is useful to choose the probability threshold $p_{\rm cut}$ that gives the desired value of purity, completeness or $F_1$-score.}
\label{fig:prob}
\end{figure*}

We show in Fig.~\ref{fig:conf_r}, for the magnitude bins $r\le 19$, $19<r\le 21$  and $r>21$, the $3 \times 3$ confusion matrices using the criterion of Fig.~\ref{fig:3k}, that is, the highest probability determines the classification. We also display purity and completeness. In Table~\ref{tab:miss} we show the overall confusion matrix, that is, for the three magnitude bins together.
As said earlier, given the three probabilities $\tilde p_{\rm gal}$, $\tilde p_{\rm star}$ and $\tilde p_{\rm qso}$, one may adopt an alternative criterion that is best suited to maximize, for instance, the purity of quasars.

When analyzing the brightest bin, $r\le 19$, notable performance was observed for both stars and galaxies, exhibiting purity and completeness rates exceeding 99\%. However, a moderate number of quasars ($\lesssim20\%$) was misclassified as galaxies and stars, resulting in a reduction in overall quasar purity and completeness. This strong performance for galaxies and stars can be attributed to the majority of these objects being encompassed within the brightest bin. On the other hand, the outcome for quasars can be attributed to their relatively small number in this magnitude bin, compared to the total training dataset, see Fig.~\ref{fig:hist_trin}. Additionally, due to their point-like nature, quasars can be erroneously identified as stars, further contributing to the misclassification.

Within the intermediate bin, $19<r\le 21$, all categories are well represented and the sources are well within the limiting depth of J-PLUS, indicating the effective functioning of the classifier across all classes. Galaxies exhibit a slightly superior performance (purity and completeness greater than 95\%), likely owing to their extended nature, which facilitates more distinctive features for classification. It is worth noting that some mismatch classification is observed between stars and quasars within this bin. The proximity in their morphological properties and characteristics contributes to a higher degree of misclassification between these two classes. Nonetheless, the overall performance of the classifier remains good, successfully handling the diverse range of objects within the intermediate bin.

For the faintest bin, $r>21$, the performance of the classifier declines for galaxies and stars in comparison to the intermediate bin, while the quasar classification remained approximately at the same level. Regarding galaxies, there is a reduction of approximately 5\% in both purity and completeness. In contrast, for stars, purity experiences a 5\% decrease, accompanied by a more pronounced drop of about 45\% in completeness. This deterioration in terms of completeness and purity for galaxies and stars can be attributed to the reduction in the number of objects available in the training set, see Fig.~\ref{fig:hist_trin}. Stars suffer the most significant impact, as their total count within this bin is considerably small relative to the overall population of stars. Notably, a substantial portion of misclassified stars are erroneously labeled as quasars. Despite this, all classifications remain relatively robust. Nevertheless, the purity of quasar classification is compromised due to the persistent confusion between quasars and stars. This observation underscores the challenges posed by the limited availability of faint objects and the inherent difficulty in accurately distinguishing between stars and quasars within this context.

In~\citet[Fig.~3]{2022A&A...659A.144W} it is presented a confusion matrix for the galaxy-star-quasar classification of J-PLUS DR1. Since they do not divide the analysis in $r$-band bins, this result would be similar to the sum of the three confusion matrices presented in Fig.~\ref{fig:conf_r}. However, a direct comparison of the confusion matrices is not meaningful because in~\citet{2022A&A...659A.144W} the objects used in the fitting process are included in the confusion matrix, while the confusion matrices of Fig.~\ref{fig:conf_r} are built considering only the objects that are not used in the fitting process.

\subsection{Purity and completeness}

If one is only interested in classifying a specific class, quasars for example, then they may use the binary classifier's probability $p_{\rm qso}$ and choose $p_{\rm cut}$ in order to obtain the desired purity, completeness or $F_1$-score. In Fig.~\ref{fig:prob} we show, for the magnitude bins $r\le 19$, $19<r\le 21$  and $r>21$, the purity, completeness and $F_1$-score in terms of the $p_{\rm cut}$ for all classes: galaxies, stars and quasars.

The first row of Fig.~\ref{fig:prob} presents the purity  curve as a function of $p_{\rm cut}$. In this graph type, the initial point at $p_{\rm cut}=0$ corresponds to the purity of the complete sample. The graph's progression illustrates the purity of the sample considering only objects with $p\geq p_{\rm cut}$. Thus, it is expected that the initial value is the smallest, with the curve gradually increasing until reaching its maximum of unity (perfect purity) when $p_{\rm cut}=1$. A good classifier exhibits a steep curve, quickly approaching values close to 1 for purity. Analyzing galaxies and stars, it becomes evident that the classifier's performance improves as the brightness of the bin increases. Although the performance difference among galaxy bins is modest, the discrepancy is more pronounced for stars, indicating a substantial decline in performance for the faintest bin. Regarding quasars, the classifier's performance is slightly lower, with a slower growth in the curve. Additionally, the classifier's performance remains relatively consistent across all bins for quasars. This analysis aligns with the discussion presented in Fig.~\ref{fig:conf_r}, confirming the observed trends and reinforcing the findings.

The second row of Fig.~\ref{fig:prob} illustrates completeness as a function of $p_{\rm cut}$. In this case, the initial point must be 1 since it represents the completeness of the entire sample. The progression of the curve reflects the decline in completeness when considering only objects with $p\geq p_{\rm cut}$. Hence, a superior classifier exhibits a slower decrease in the curve, indicating better preservation of completeness. Similar to the observations in the purity, a comparable pattern emerges in this analysis as well. Examining galaxies and stars, both exhibit outstanding performance in the brightest bin, with performance gradually deteriorating as the bin becomes fainter. As for quasars, a slightly lower but relatively consistent performance is observed across all magnitude bins. These trends align with the earlier discussion, confirming the consistency in the behavior of the classifier and reinforcing the conclusions drawn from the results.

Finally, the third row in Fig.~\ref{fig:prob} presents the $F_1$-score, which represents the geometric mean between purity and completeness. A higher $F_1$-score indicates a classifier's improved overall performance. Consequently, the closer the curve approaches 1, the better the classifier performs.

The analysis depicted in Fig.~\ref{fig:prob} offers a valuable tool for utilizing our newly-defined VAC in specific applications. Depending on the problem at hand, one can select a particular $p_{\rm cut}$ based on their desired emphasis on purity, completeness, or both ($F_1$-score). This flexibility allows for a tailored approach in choosing an optimal $p_{\rm cut}$ threshold that aligns with the specific requirements and trade-offs of the application. The distribution of probabilities $ p_{\rm gal}$, $ p_{\rm star}$ and~$ p_{\rm qso}$ for actual galaxies, stars and quasars, respectively, are shown in Appendix~\ref{ap:goodies}.

\begin{figure}
\centering
% trim={<left> <lower> <right> <upper>}
\includegraphics[trim={0      1.6cm 0  0}, clip, width=\columnwidth]{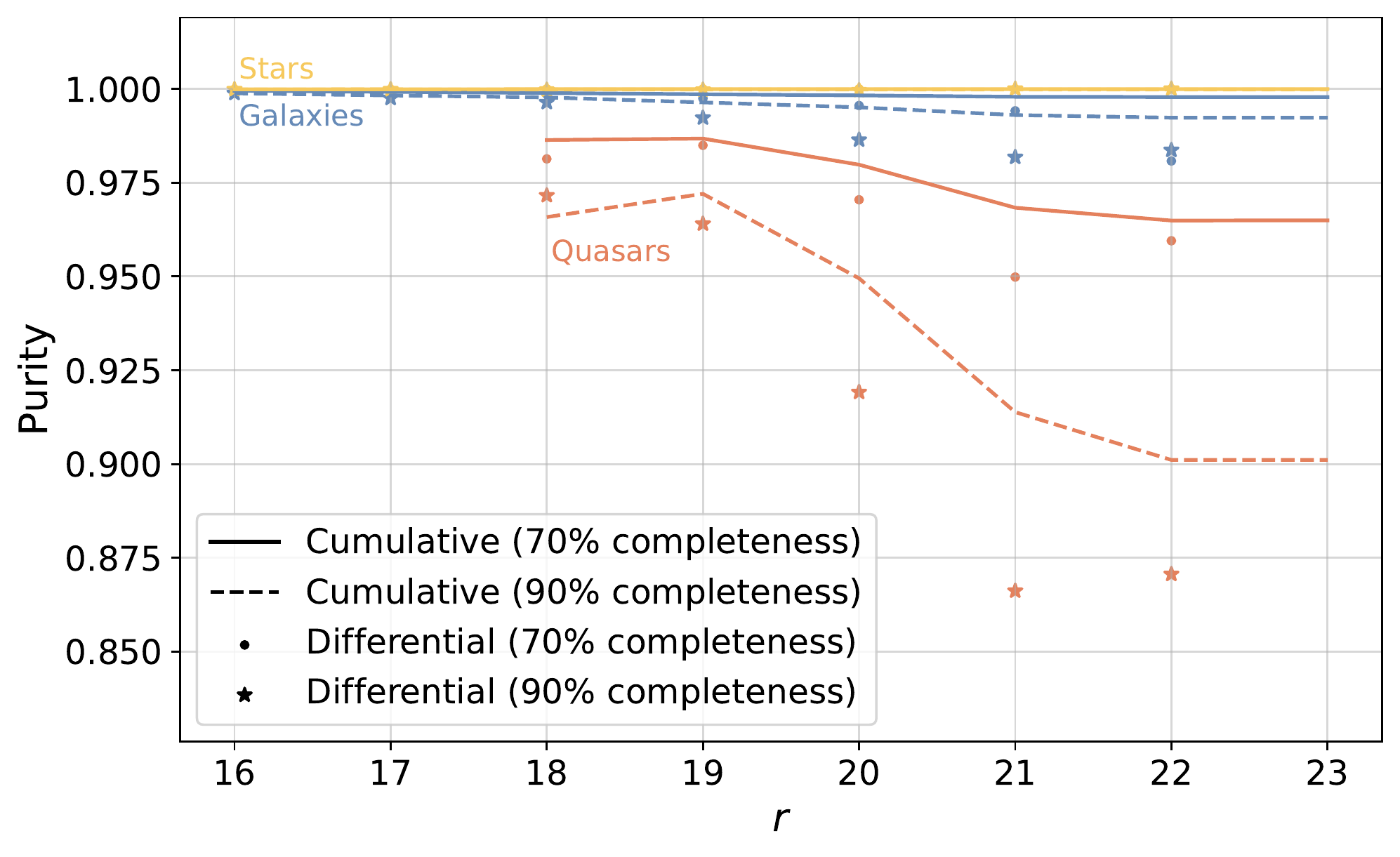}\\
\noindent \hspace{.15cm}\includegraphics[trim={0      .4cm 0 .2cm}, clip, width=.983\columnwidth]{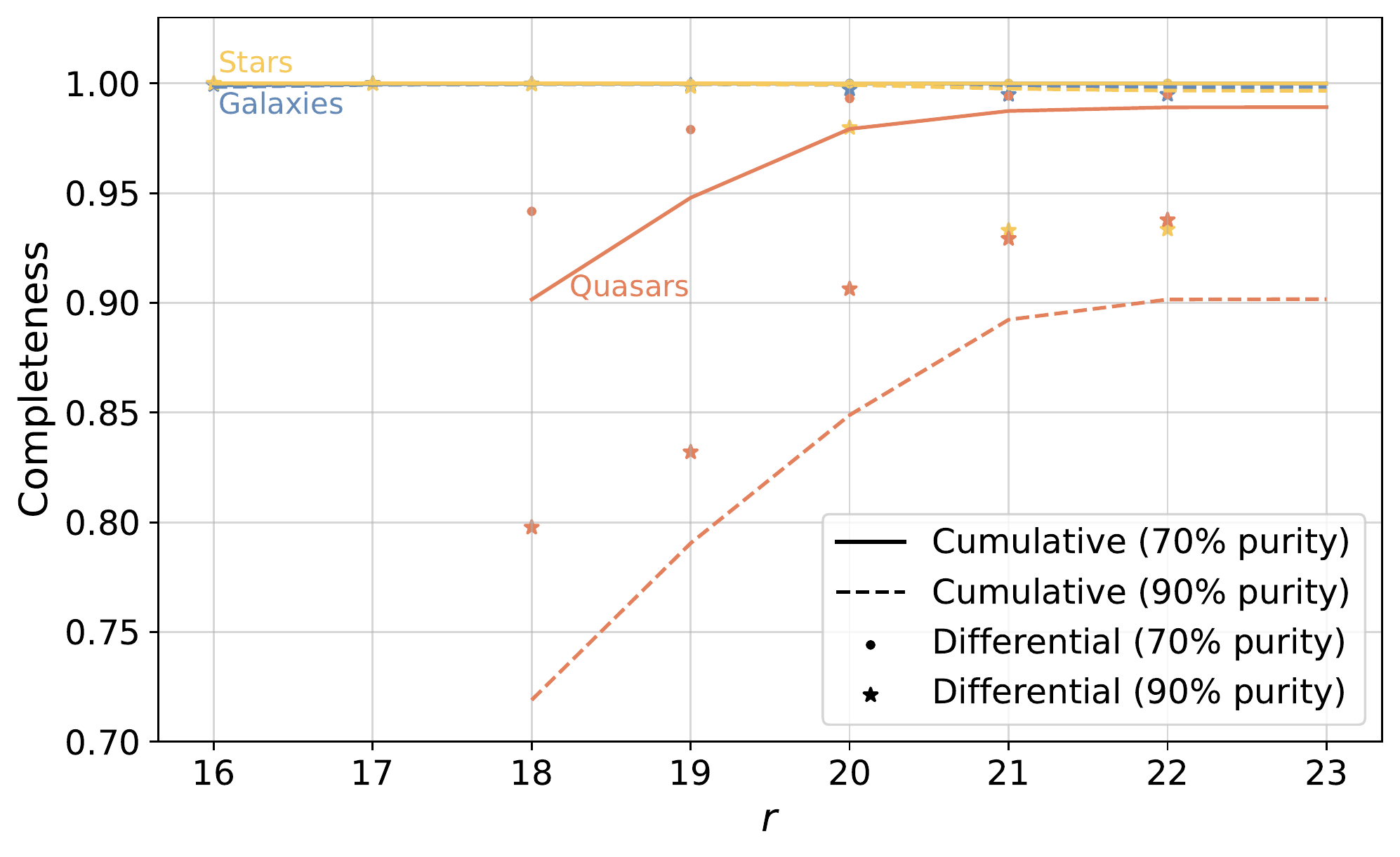}
\caption{Cumulative and differential purity with respect to the $r$ magnitude for a fixed value of completeness (top), and completeness for a fixed value of purity (bottom) for galaxies, stars and quasars.}
\label{fig:pc}
\end{figure}

Fig.~\ref{fig:pc} presents the variations in cumulative and differential purity/completeness for galaxies, stars, and quasars concerning the $r$ magnitude while maintaining a fixed level of completeness/purity. In the cumulative scenario, the purity/completeness for a specific $r$ magnitude is determined by considering all galaxies up to that magnitude threshold. For instance, the purity/completeness at $r=20$ corresponds to the purity/completeness of the subsample containing objects with $r\leq 20$. In the differential scenario, we compute the purity/completeness within magnitude bins of $\Delta r=1$~mag. Whereas we analyze seven magnitude bins centered at ${\rm mag}=[16,17,18,19,20,21,22]$ for galaxies and stars, our examination for quasars is limited to the five last bins. The difference in the number of bins for quasars stems from the scarcity of quasar occurrences in the brightest bins. In order to produce these plots, the value of $p_{\rm cut}$ is computed as a function of $r$ in order to ensure the specified purity/completeness levels, as indicated in the plot legends. Overall, the results obtained demonstrate significant success across all cases, with a marginal deterioration for faint objects, likely attributable to the degradation in the quality of J-PLUS observations for $r>21$~mag.

\subsection{Misclassifications}

Here, we focus on misclassification. The quantitative examination of misclassified instances, stratified by class (galaxies, stars, and quasars), is presented in Fig.~\ref{fig:conf_r} and Table~\ref{tab:miss}. The outcomes reveal that our classifier exhibits a good performance on the test set, albeit slightly diminished in the case of quasars -- a result that underscores the difficulty associated with classifying quasars.

\begin{figure*}
\centering
% trim={<left> <lower> <right> <upper>}
\includegraphics[width=0.33\textwidth]{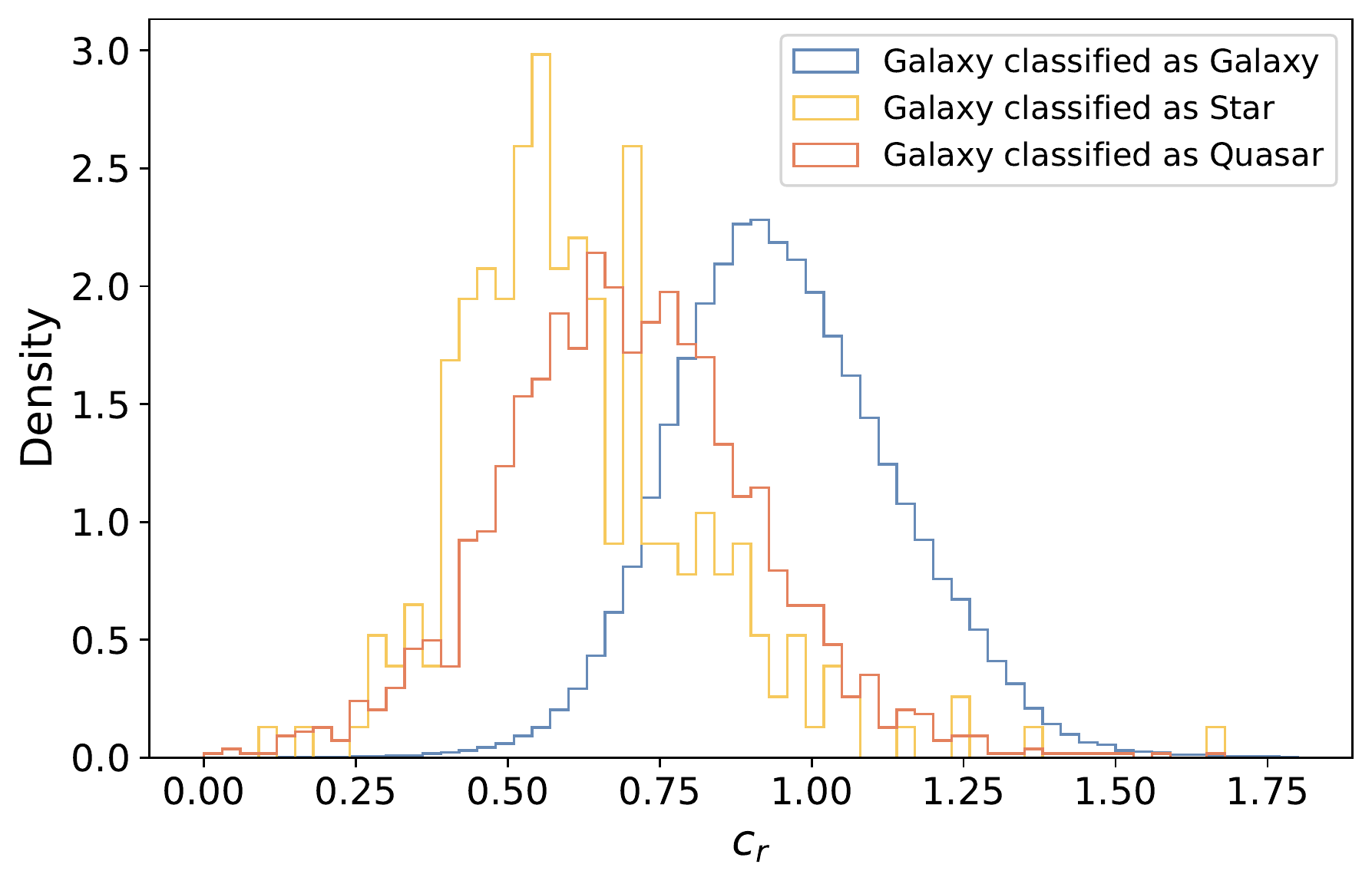}
\includegraphics[width=0.33\textwidth]{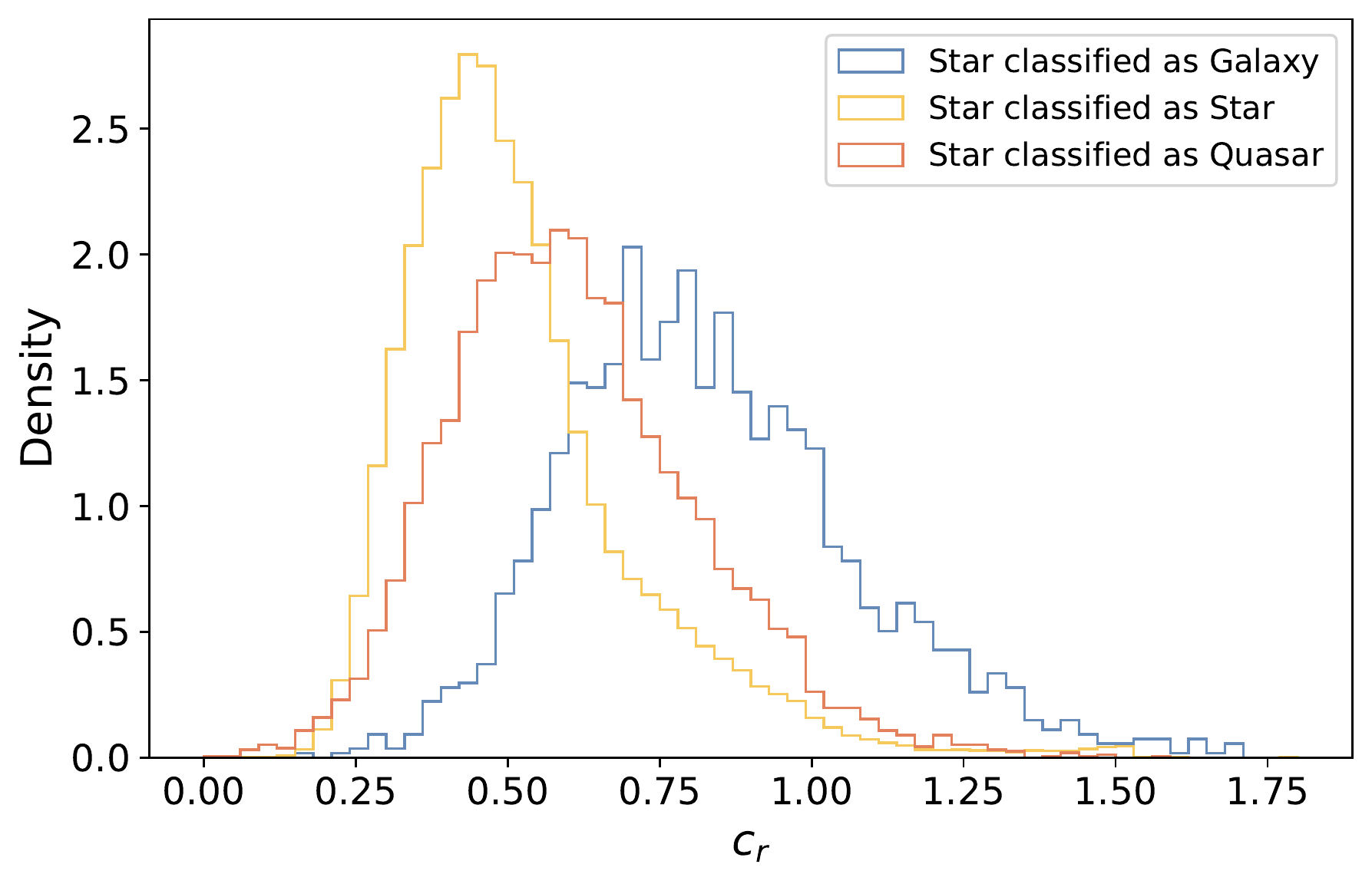}
\includegraphics[width=0.33\textwidth]{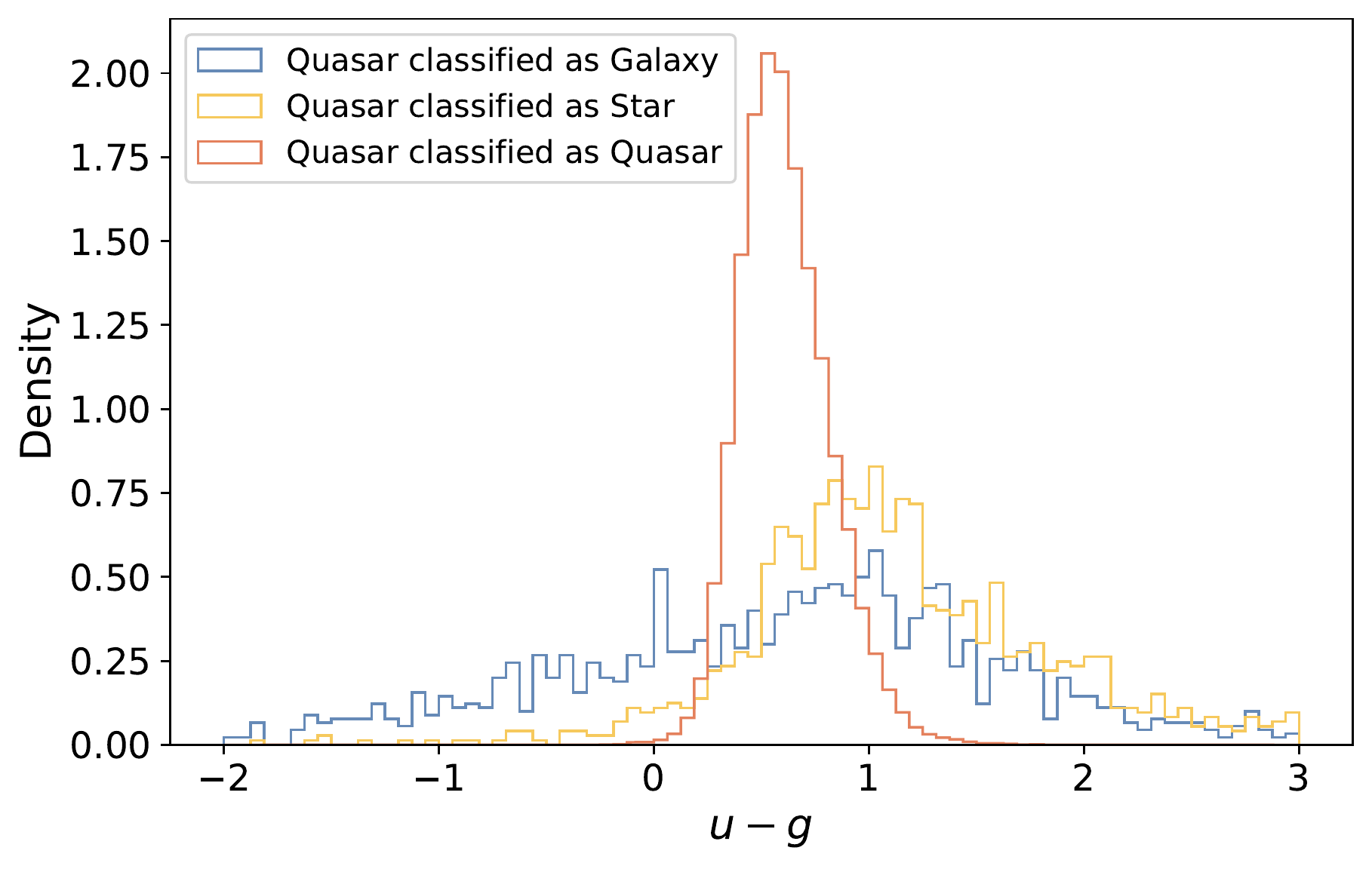}
\caption{Distribution of key features used by the classifier, highlighting trends in misclassification between these categories.}
\label{fig:miss}
\end{figure*}

Figure~\ref{fig:miss} shows key feature distributions used by the classifier, according to our feature importance analysis (see Sec~\ref{subsec:fi}). It presents the data for the entire object cohort, divided by classes and by correct or incorrect classifications. The left panel shows the concentration values for actual galaxies. Notably, those misclassified as stars or quasars have lower concentration, suggesting confusion with point-like entities. The central panel depicts concentration values for stars, highlighting a trend where misclassified stars have higher concentration levels, causing confusion with extended objects. The right panel displays the color index $u-g$ for quasars. Misclassified quasars in this group have a slightly more positive $u-g$ color index, aligning with the galaxy and star distribution shown in Fig.~\ref{fig:hist_train}.

\begin{figure}
\centering
\includegraphics[width=\columnwidth]{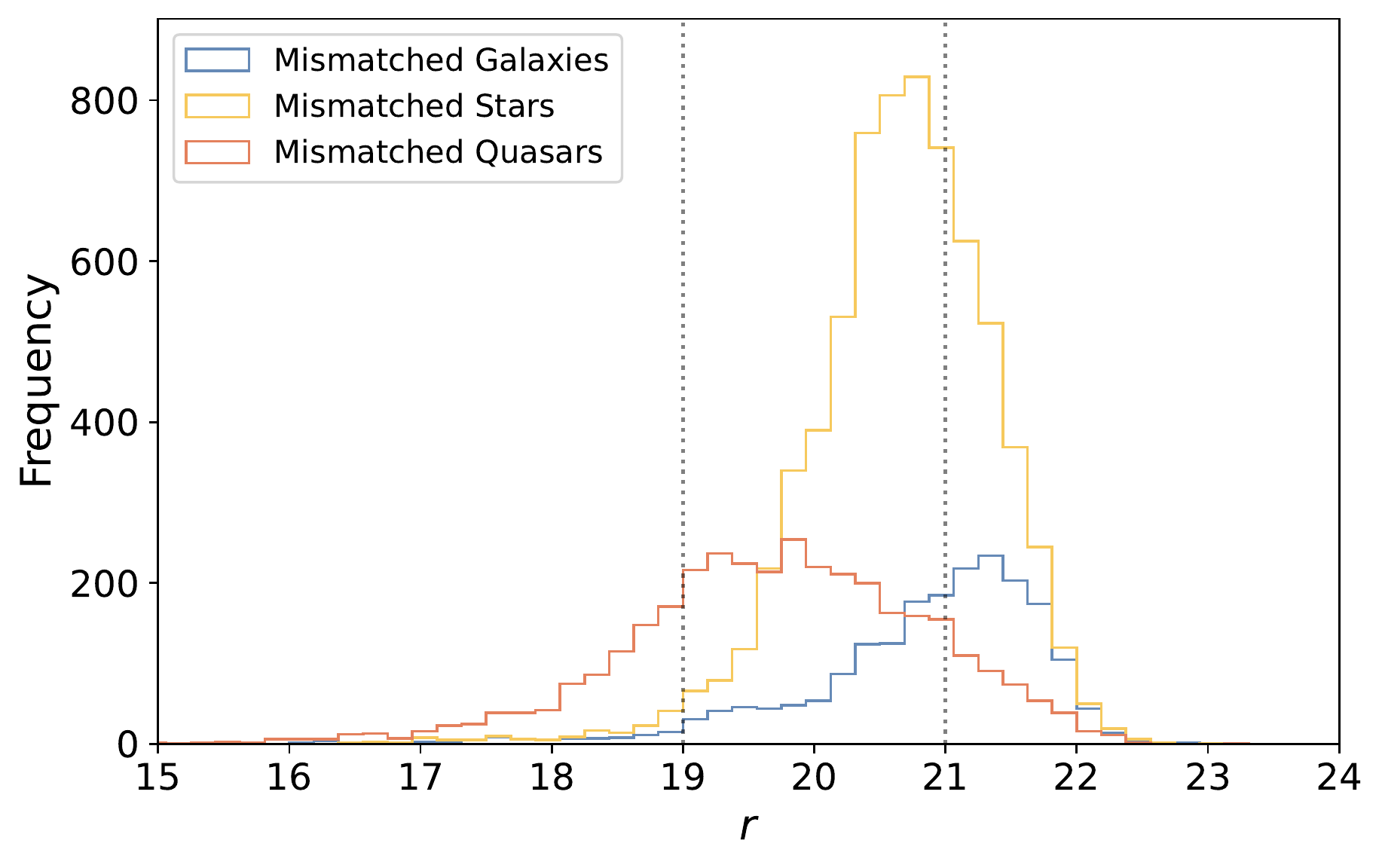}
\caption{Distribution of misclassified objects by magnitude.}
\label{fig:hist_miss}
\end{figure}

In Fig.~\ref{fig:hist_miss}, we show the distribution of misclassified objects by magnitude. The figure reveals that galaxies and stars are more often misclassified when dimmer ($r>21$), while quasars are most frequently misclassified at intermediate magnitudes, reflecting confusion with stars and galaxies.

Finally, we did not find significant misclassification trends with respect to the sky positions of the sources nor crowding in specific sky regions, except an excess of objects that are classified as stars in the region close to the galactic plane, see Appendix~\ref{ap:sky}. This may indicate an effect coming from the galactic plane or from crowding in the photometry.

\begin{table}
\centering
\begin{tabular}{ccc}
\hline\hline
\multicolumn{3}{c}{\textbf{Actual Galaxies (N = 132,397)}}                                                                                                                                                                                 \\  \hline
\multicolumn{1}{c|}{\begin{tabular}[c]{@{}c@{}}Classified as\vspace{-3mm}\\ Galaxy\end{tabular}} & \multicolumn{1}{c|}{\begin{tabular}[c]{@{}c@{}}Classified as\vspace{-3mm}\\ Star\end{tabular}} & \begin{tabular}[c]{@{}c@{}}Classified as\vspace{-3mm}\\ Quasar\end{tabular}  \\ \hline 
\multicolumn{1}{c|}{130,338}                                                        & \multicolumn{1}{c|}{258}                                                           & 1,801                                                     \vspace{-3mm}      \\ 
\multicolumn{1}{c|}{(98,44\%)}                                                      & \multicolumn{1}{c|}{(0,20\%)}                                                      & (1,36\%)                                                        \\ \hline\hline
\multicolumn{3}{c}{\textbf{Actual Stars (N = 232,825)}}                                                                                                                                                                                    \\ \hline
\multicolumn{1}{c|}{\begin{tabular}[c]{@{}c@{}}Classified as\vspace{-3mm}\\ Galaxy\end{tabular}} & \multicolumn{1}{c|}{\begin{tabular}[c]{@{}c@{}}Classified as\vspace{-3mm}\\ Star\end{tabular}}  & \begin{tabular}[c]{@{}c@{}}Classified as\vspace{-3mm}\\ Quasar\end{tabular}  \\ \hline
\multicolumn{1}{c|}{1,789}                                                          & \multicolumn{1}{c|}{225,834}                                                       & 5,202                                                          \vspace{-3mm} \\
\multicolumn{1}{c|}{(0,77\%)}                                                       & \multicolumn{1}{c|}{(97,00\%)}                                                     & (2,23\%)                                                        \\ \hline\hline
\multicolumn{3}{c}{\textbf{Actual Quasars (N = 53,281)}}                                                                                                                                                                                   \\ \hline
\multicolumn{1}{c|}{\begin{tabular}[c]{@{}c@{}}Classified as\vspace{-3mm}\\ Galaxy\end{tabular}} & \multicolumn{1}{c|}{\begin{tabular}[c]{@{}c@{}}Classified as\vspace{-3mm}\\ Star\end{tabular}}  & \begin{tabular}[c]{@{}c@{}}Classified as\vspace{-3mm}\\ Quasar\end{tabular}  \\ \hline
\multicolumn{1}{c|}{2,048}                                                          & \multicolumn{1}{c|}{1,447}                                                         & 49,760                                                         \vspace{-3mm} \\
\multicolumn{1}{c|}{(3,84\%)}                                                       & \multicolumn{1}{c|}{(2,72\%)}                                                      & (93,44\%)                                                       \\ \hline\hline
\end{tabular}
\caption{Misclassifications according to class (galaxies, stars, and quasars). This table provides a summary of the results depicted in the three panels of Fig.~\ref{fig:conf_r}.}
\label{tab:miss}
\end{table}

\subsection{Stellar locus}

\begin{figure}
\centering
% trim={<left> <lower> <right> <upper>}
\includegraphics[trim={0      .8cm 0  0}, clip, width=\columnwidth]{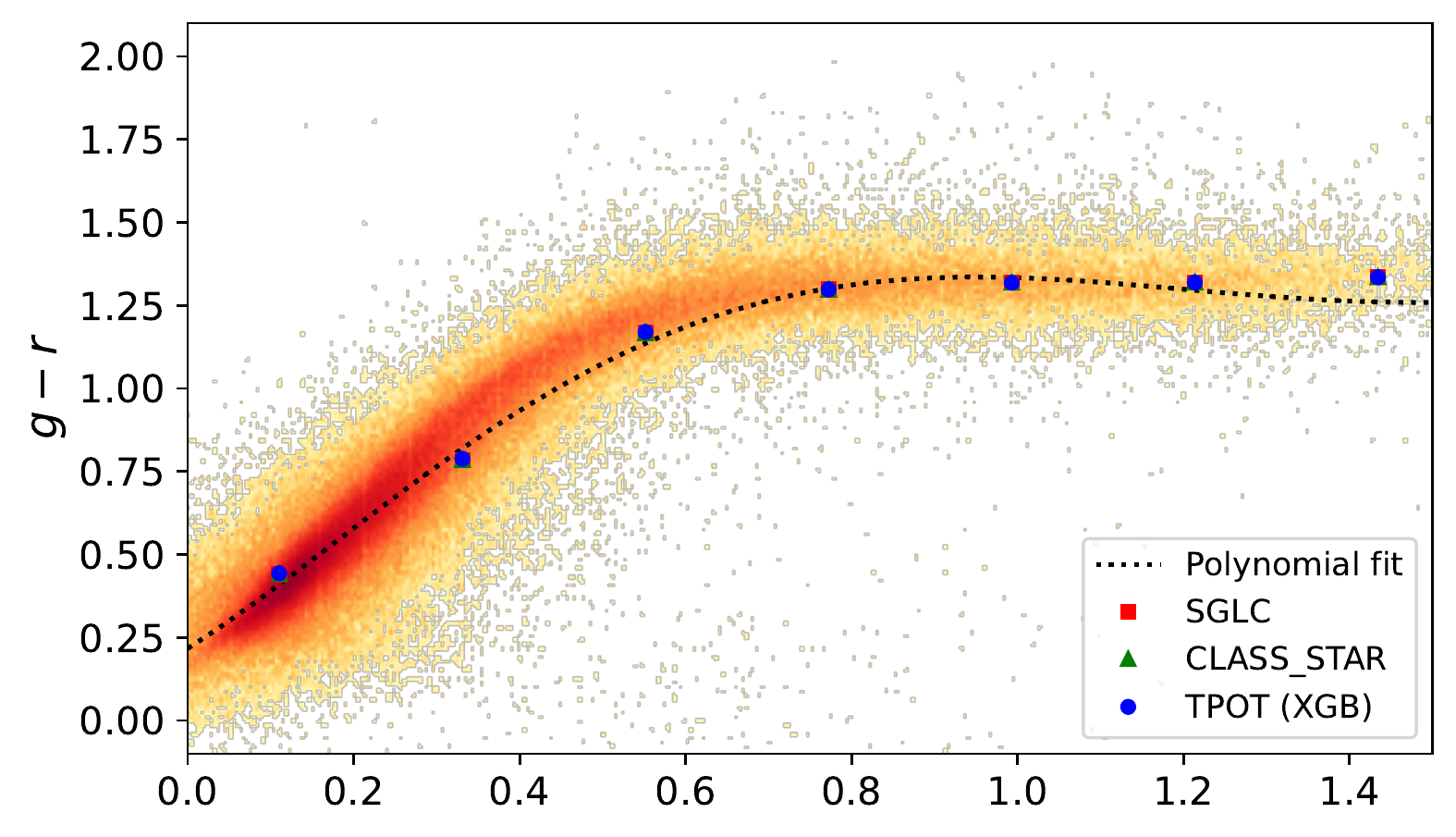}
\includegraphics[trim={0      .8cm 0  0.25cm}, clip, width=\columnwidth]{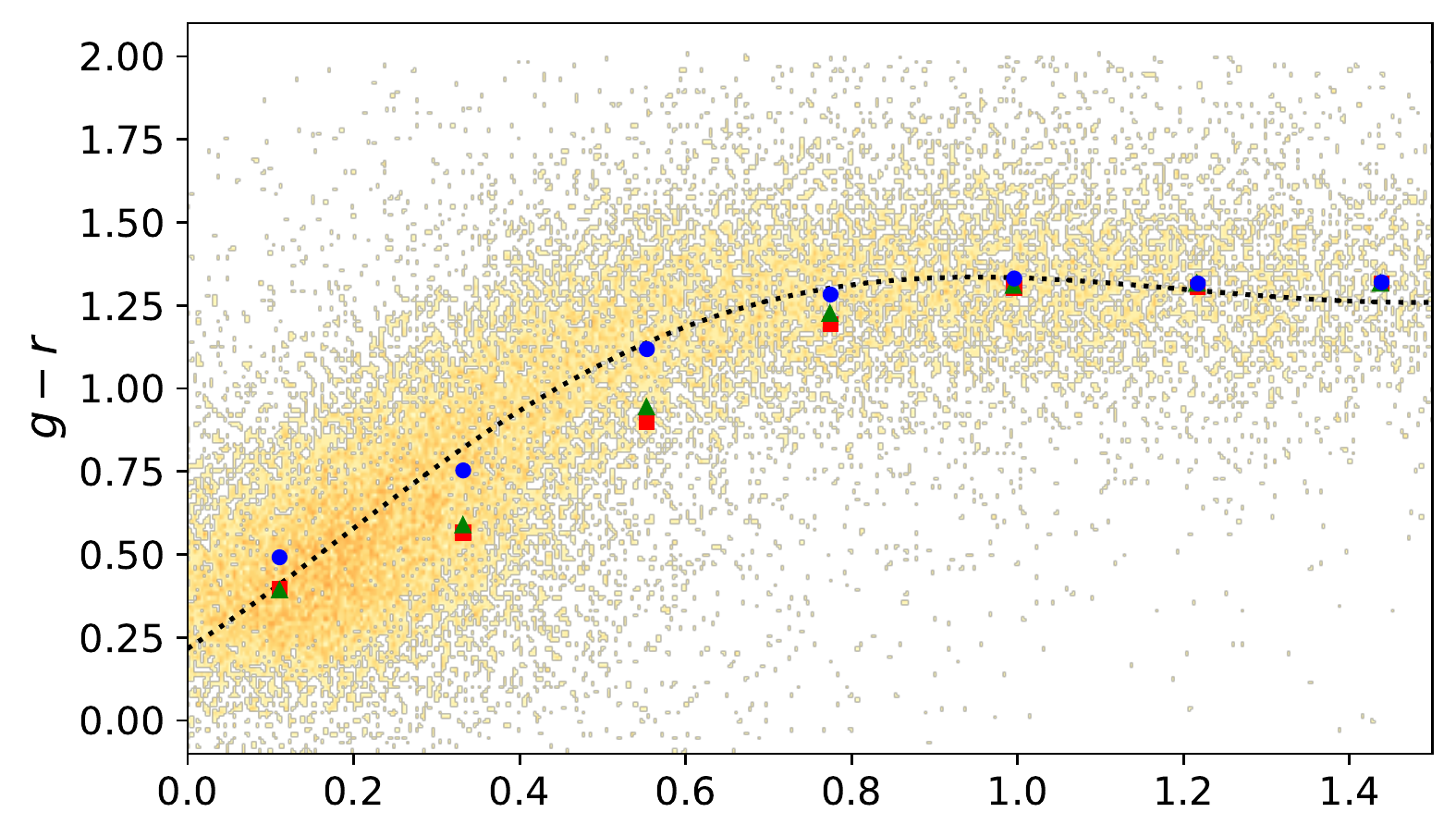}
\includegraphics[trim={0      .3cm 0  0.25cm}, clip, width=\columnwidth]{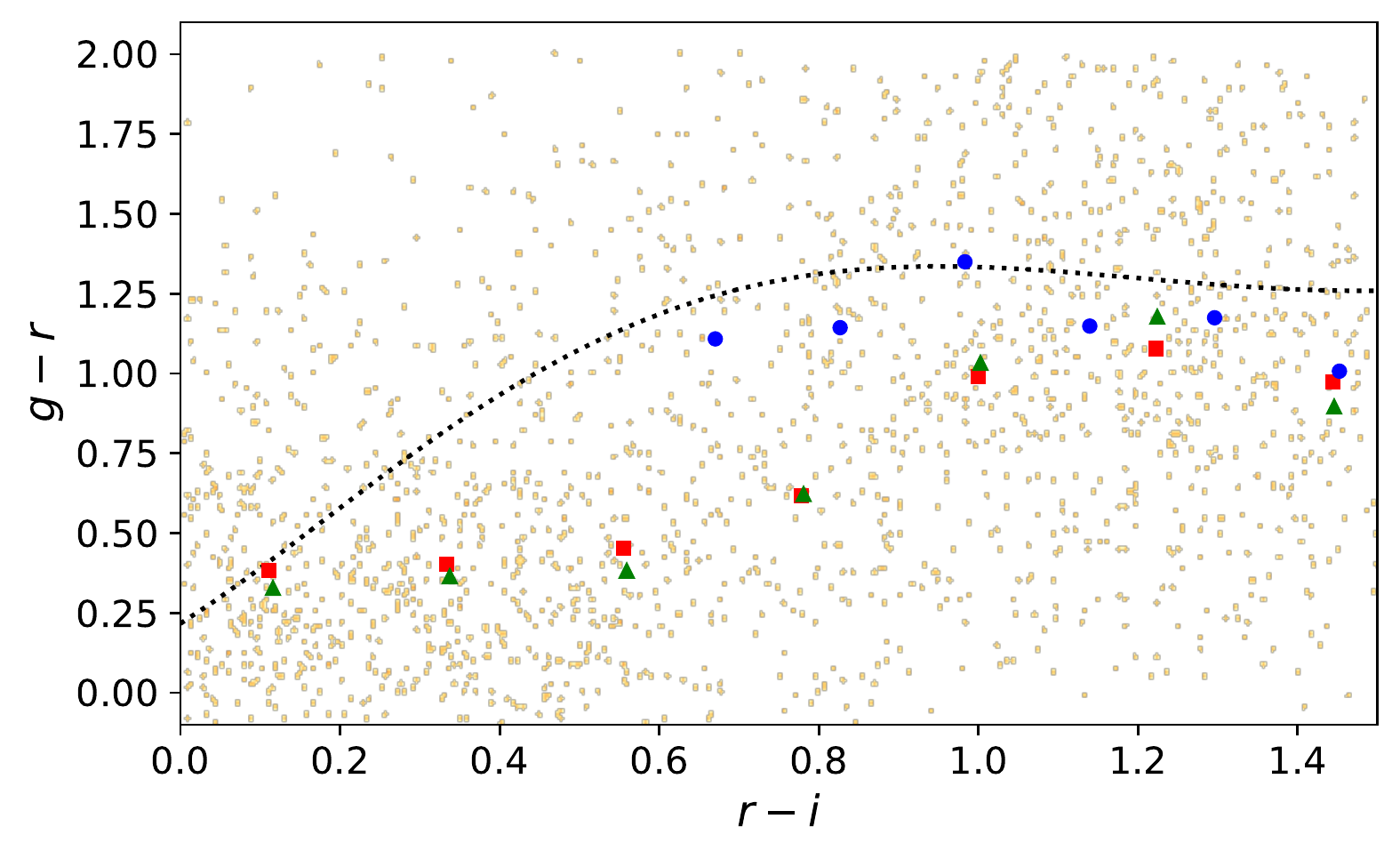}
\caption{Stellar locus for the magnitude bins $r\le 19$ (top), $19<r\le 21$ (middle)  and $r>21$ (bottom). The dashed curve is a polynomial fit to the SDSS DR18, LAMOST DR8 and {\it Gaia} DR3 stars of Section~\ref{sssec:tr} (cloud of points), while the symbols represent the stellar locus as predicted by the classifiers.}
\label{fig:locus}
\end{figure}

Fig.~\ref{fig:locus} shows the stellar locus for the magnitude bins $r\le 19$, $19<r\le 21$  and $r>21$, that is the morphology of the distribution of stars in the color space $g-r$ against $r-i$. Here, stars are  objects with $p_{\rm star}>0.5$.
The performance of all the classifiers is good for the brightest bin, while for the bin $19<r\le 21$ the XGB model outperforms {\tt CLASS\_STAR} and {\tt SGLC}.
%In the fainter bin, all classifiers perform poorly, with XGB comparatively better than  {\tt CLASS\_STAR} and {\tt SGLC}. The poor performance of XGB in the faintest bin is in agreement with the results of Section~\ref{cmf}---the low purity causes a shift in the reconstructed locus.
In the fainter bin, {\tt CLASS\_STAR} and {\tt SGLC} perform poorly. XGB shows a bias and this is in agreement with the results of Section~\ref{cmf}---the lower purity causes a shift in the reconstructed locus.

\subsection{Area under the curve}

\begin{figure*}
\centering
\includegraphics[trim={0      .2cm 0  0}, clip, width=\columnwidth]{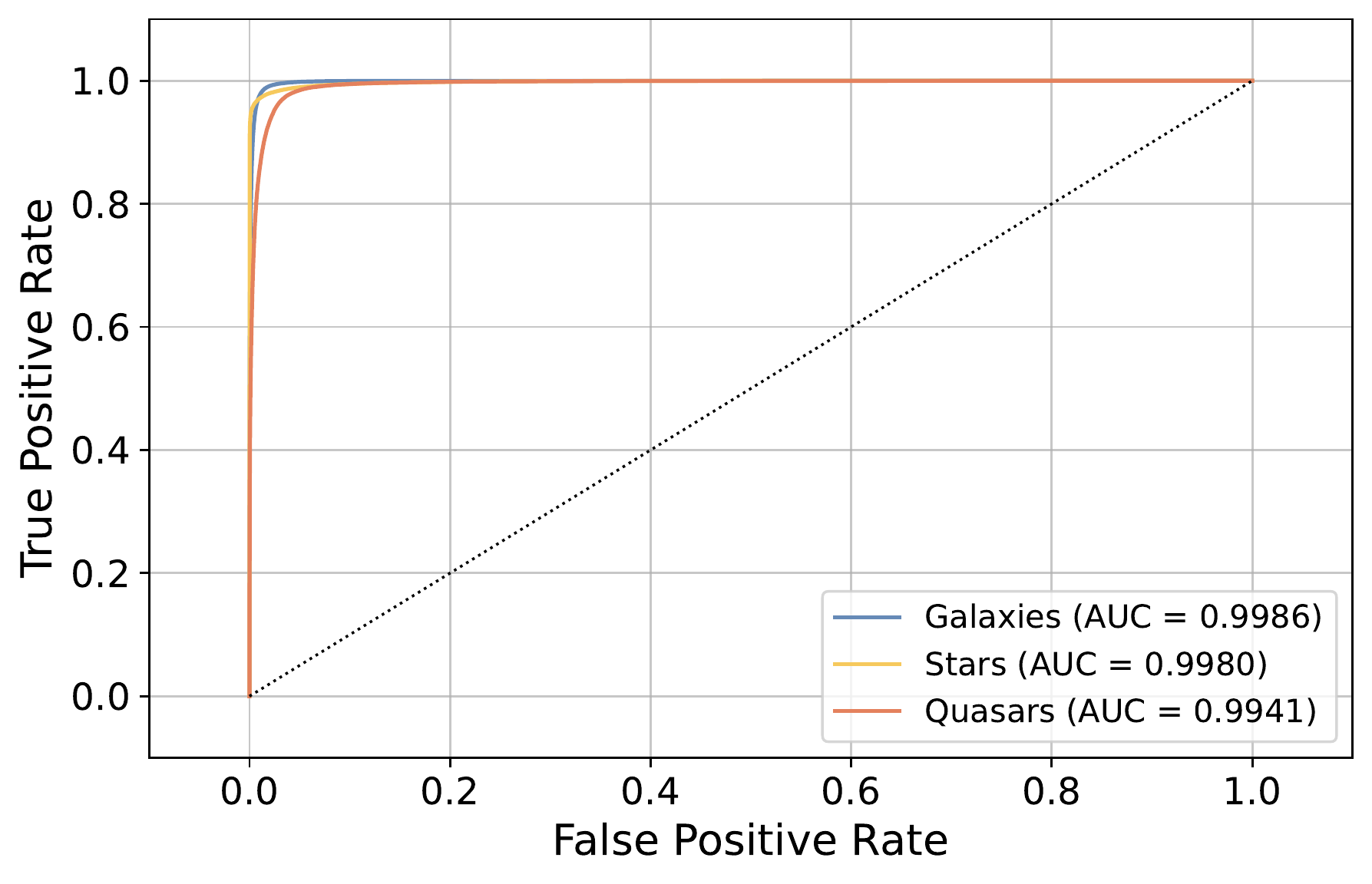}
\includegraphics[trim={0      .2cm 0  0}, clip, width=\columnwidth]{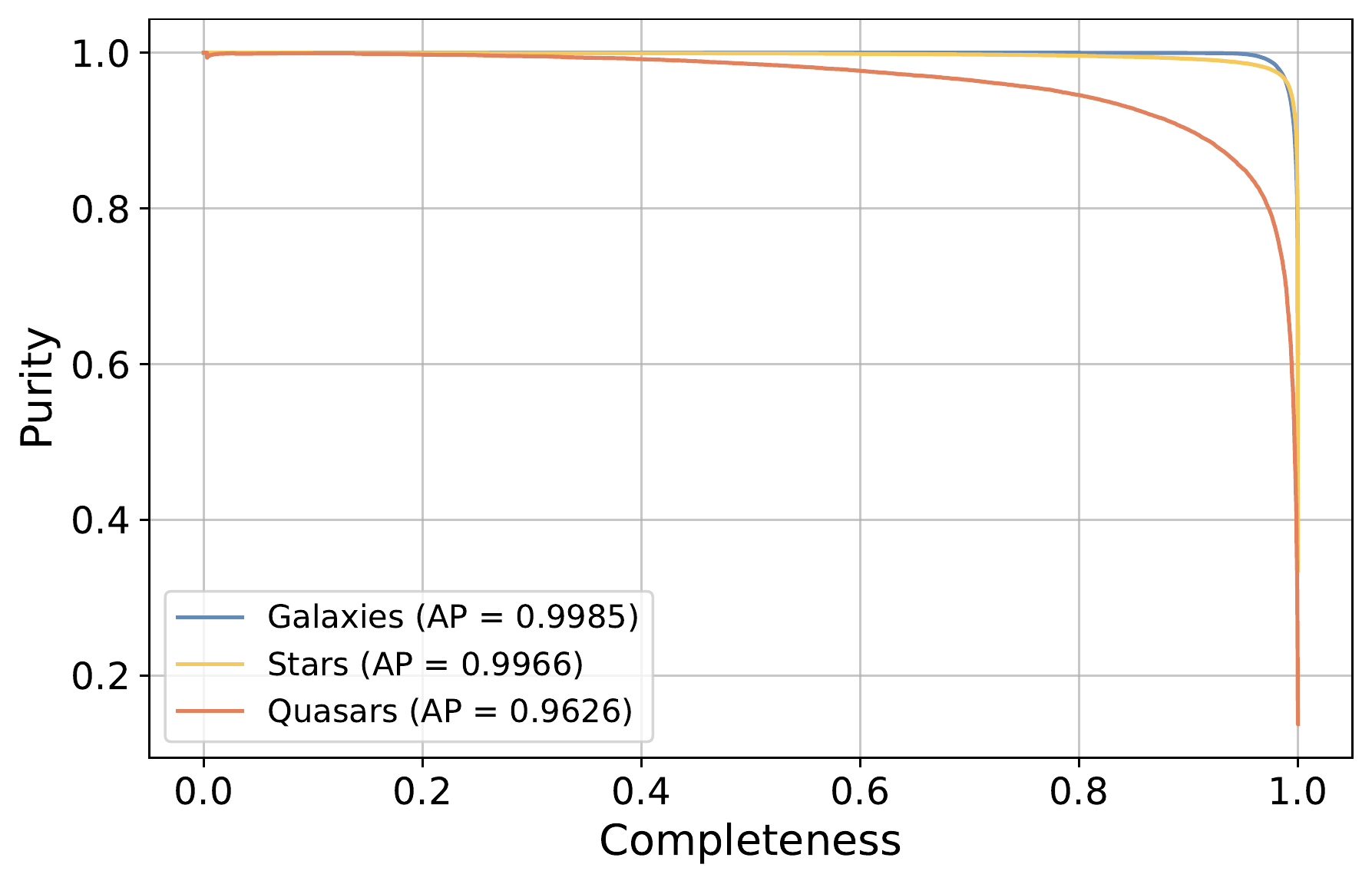}
\caption{ROC and purity-completeness curves for the test set using the three binary classifiers for star, galaxies and quasars.}
\label{fig:roc}
\end{figure*}

As defined in Sec.~\ref{ssec:metrics}, AUC and AP are useful  metrics that summarize the overall performance of the classifier. We used indeed AUC as the metric that is maximized when searching for the best pipeline.
Fig.~\ref{fig:roc} displays the ROC and purity-completeness curves for the binary classifiers for stars, galaxies, and quasars, along with the AUC and AP values. The AUC scores are above 0.99, showing strong discrimination, and the AP scores are above 0.99 for galaxies and stars, and above 0.93 for quasars, reflecting robust classification accuracy. This demonstrates the classifier's overall excellent performance.

\begin{figure*}
\centering
\includegraphics[trim={0      .2cm 0  0}, clip, width=\columnwidth]{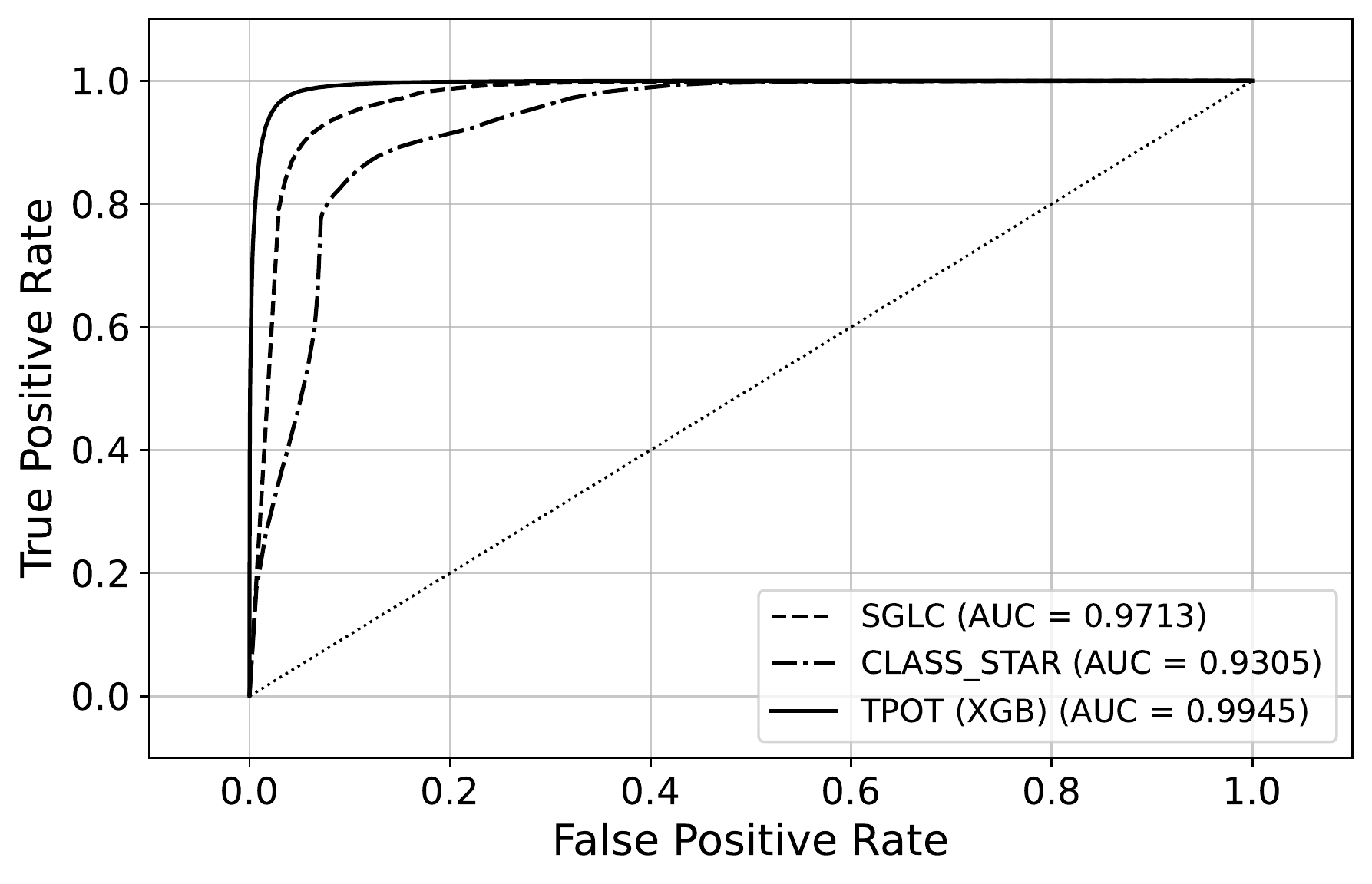}
\includegraphics[trim={0      .2cm 0  0}, clip, width=\columnwidth]{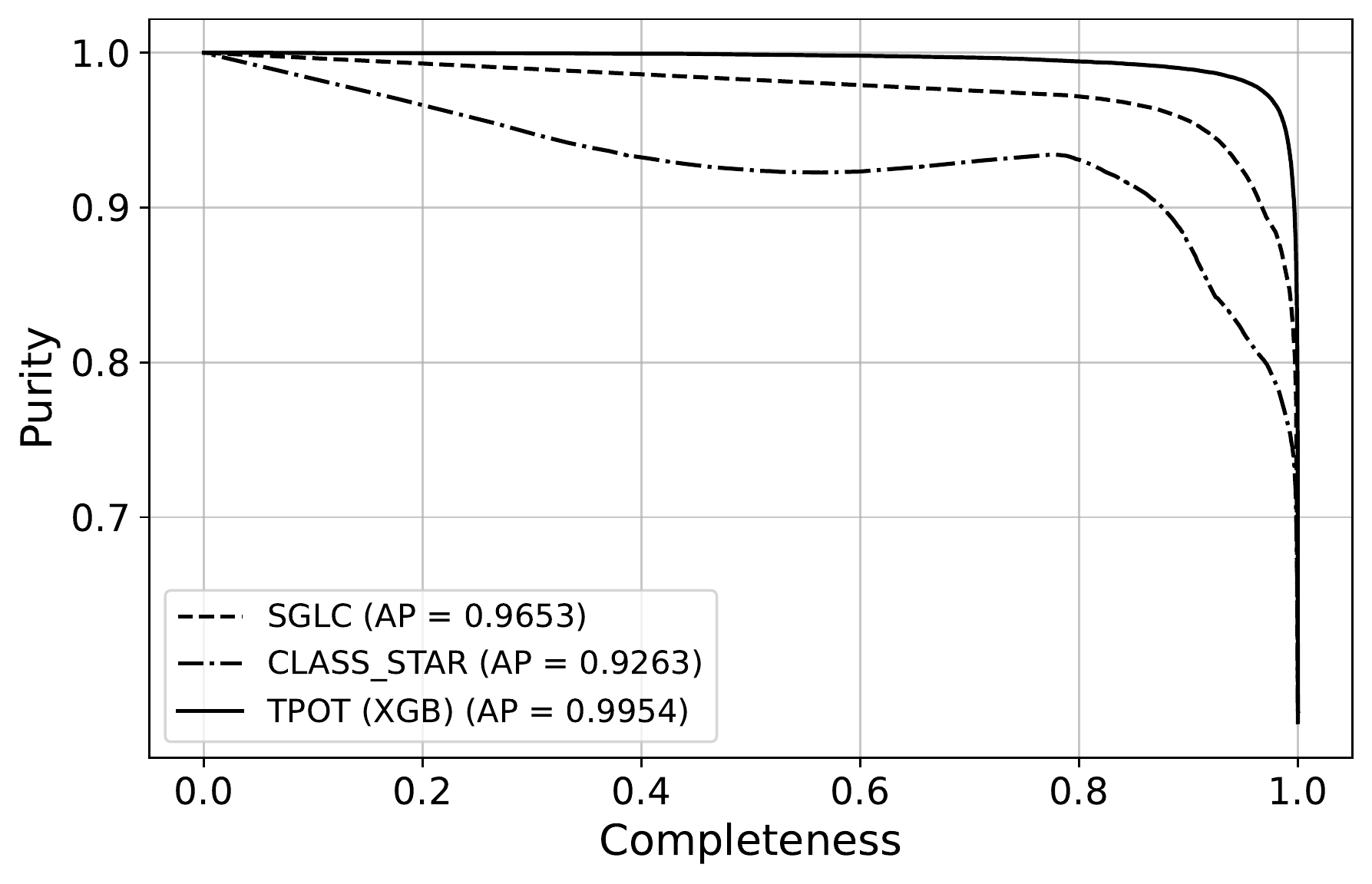}
\caption{ROC and purity-completeness curves for the test set of SDSS DR18 galaxies for XGBoost, {\tt CLASS\_STAR} and {\tt SGLC}.}
\label{fig:roc-sglc}
\end{figure*}

At this point it is important to compare our classifier with the existing classifiers utilized by J-PLUS: {\tt SGLC} and {\tt CLASS\_STAR}. As previously mentioned, our classifier holds an inherent advantage by being able to identify quasars, while {\tt SGLC} and {\tt CLASS\_STAR} are limited to distinguishing between point and extended objects. To ensure a fair comparison, we impose two conditions: ($i$) we exclusively compare galaxies with extended objects, as the existing J-PLUS classifiers can not differentiate between stars and quasars within the point-like objects; ($ii$) we restrict the analysis to the test set data belonging to SDSS DR18, as the probabilities provided by {\tt SGLC} and {\tt CLASS\_STAR} are instrumental in constructing the crossmatch with LAMOST and {\it Gaia} (see Secs.~\ref{sssec:jplusxsdss},~\ref{sssec:jplusxlamost} and~\ref{sssec:jplusxgaia}). In Fig.~\ref{fig:roc-sglc} we show the ROC curve and the purity-completeness curves for galaxies of our best XGBoost model with the curves relative to the J-PLUS classifiers. As can be seen, the XGB classifier outperforms {\tt CLASS\_STAR} and {\tt SGLC}, which, we remind the reader, cannot classify quasars. In this context, we can conclude that our classification model represents a noteworthy advancement in the classification of J-PLUS objects.

\subsection{Feature importance}
\label{subsec:fi}

\begin{figure}
\centering
\includegraphics[width=\columnwidth]{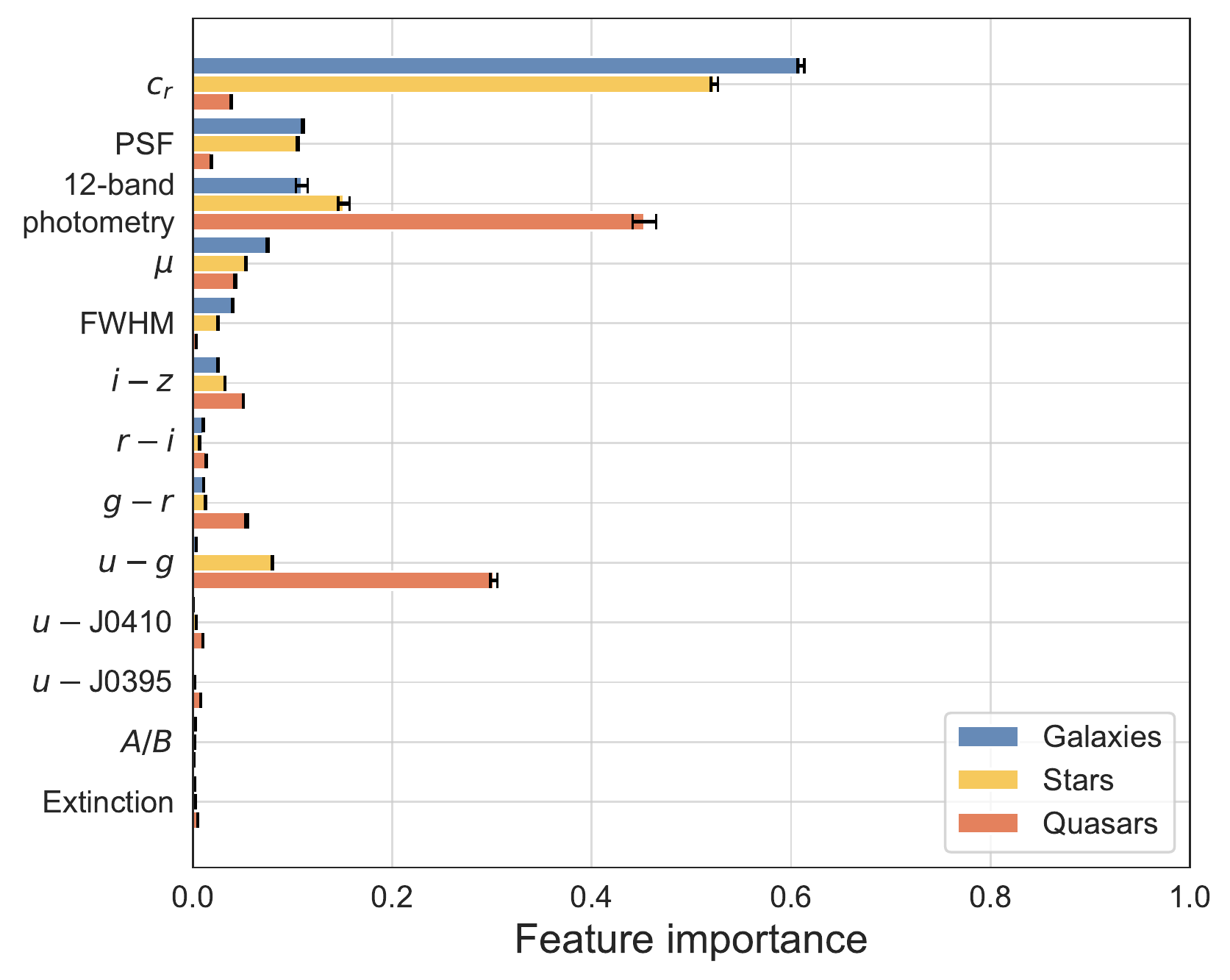}
\caption{Feature importance. See Fig.~\ref{fig:feat_imp_wavelength} for the importance of the individual photometric features.}
\label{fig:feat_imp}
\end{figure}

We estimate feature importance via the permutation method, which evaluates how the prediction error increases when a feature is not available. Also in this case we adopt the AUC of the ROC curve as metric.
In order to present the results, we divide the features in two sets: the photometric bands with errors (24 features) and the rest (13 features).
The latter are shown in Fig.~\ref{fig:feat_imp}, separately for stars, galaxies and quasars.
We also show there the overall importance of the photometric features.
We can see that there are features that similarly characterize star and galaxies (concentration, PSF and photometry) and features that are asymetrically important for quasars (photometry, $u-g$, $g-r$).

\begin{figure*}
\centering
\includegraphics[width=\textwidth]{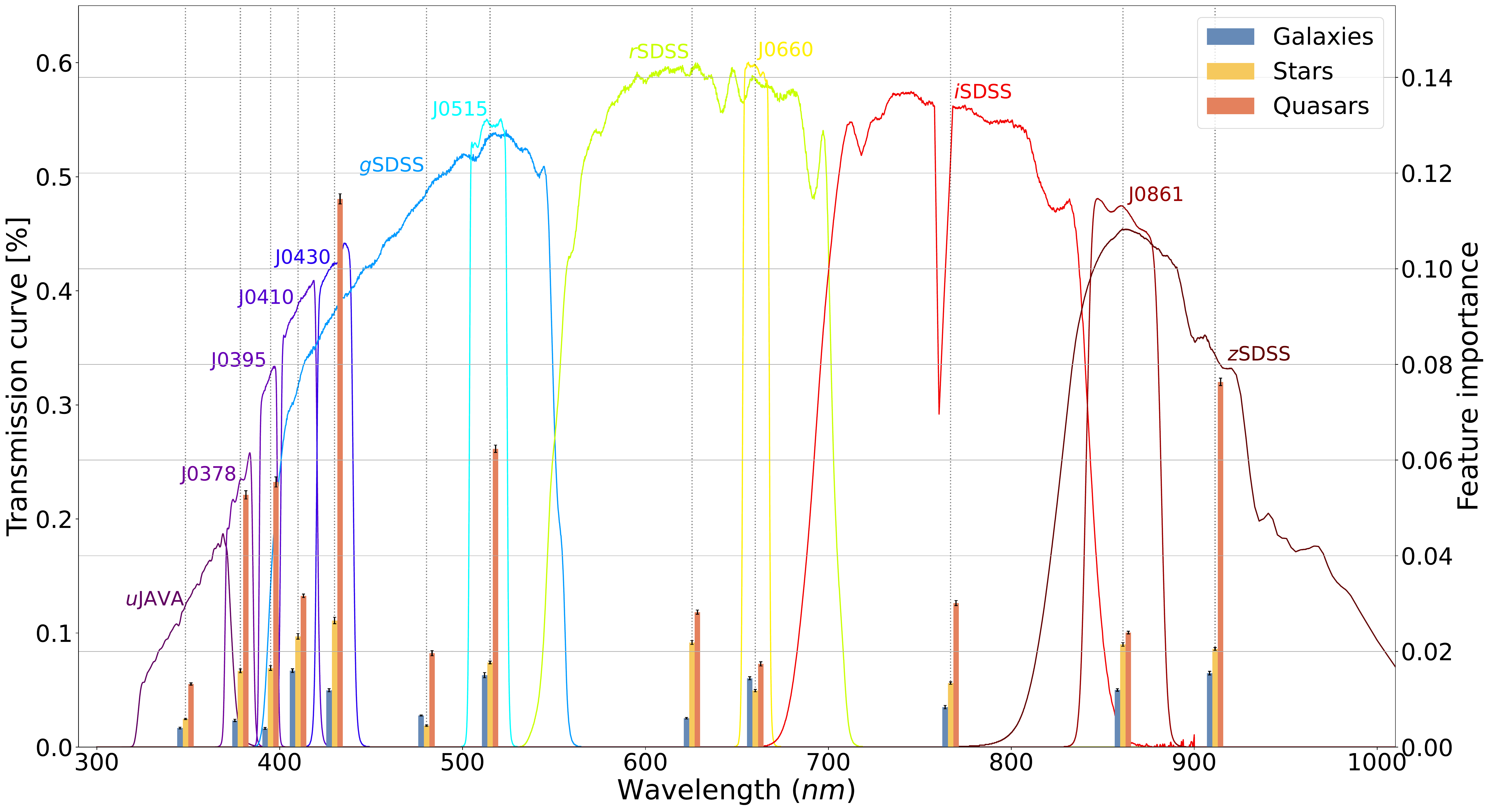}
\caption{Importance for the photometric features as a function of their wavelength,  together with the J-PLUS filter transmission curves. The importance of the photometric bands and the corresponding errors was combined together. A plot of only the transmission curves can be found in \citet{Cenarro:2018uoy}.}
\label{fig:feat_imp_wavelength}
\end{figure*}

The individual photometric features are  shown as a function of their wavelength in Fig.~\ref{fig:feat_imp_wavelength},  together with the filter transmission curves, and separately for galaxies, stars, and quasars. The importance of the photometric bands and the corresponding errors are combined together.

%

%%%%%%%%%%%%%%%%%%%%%%%%%%%%%%%%%%%%%%
%%%%%%%%%%%%%%%%%%%%%%%%%%%%%%%%%%%%%%
\section{Value-Added Catalog}
\label{sec:vac}

%%%%%%%%%%%%%%%%%%%%%%%%%%%%%%%%%%%%%%
%%%%%%%%%%%%%%%%%%%%%%%%%%%%%%%%%%%%%%
%%%%%%%%%%%%%%%%%%%%%%%%%%%%%%%%%%%%%%
\subsection{Training set representativeness of the JPLUS catalog}
\label{sec:represa}

\begin{figure}
\centering
\includegraphics[width=\columnwidth]{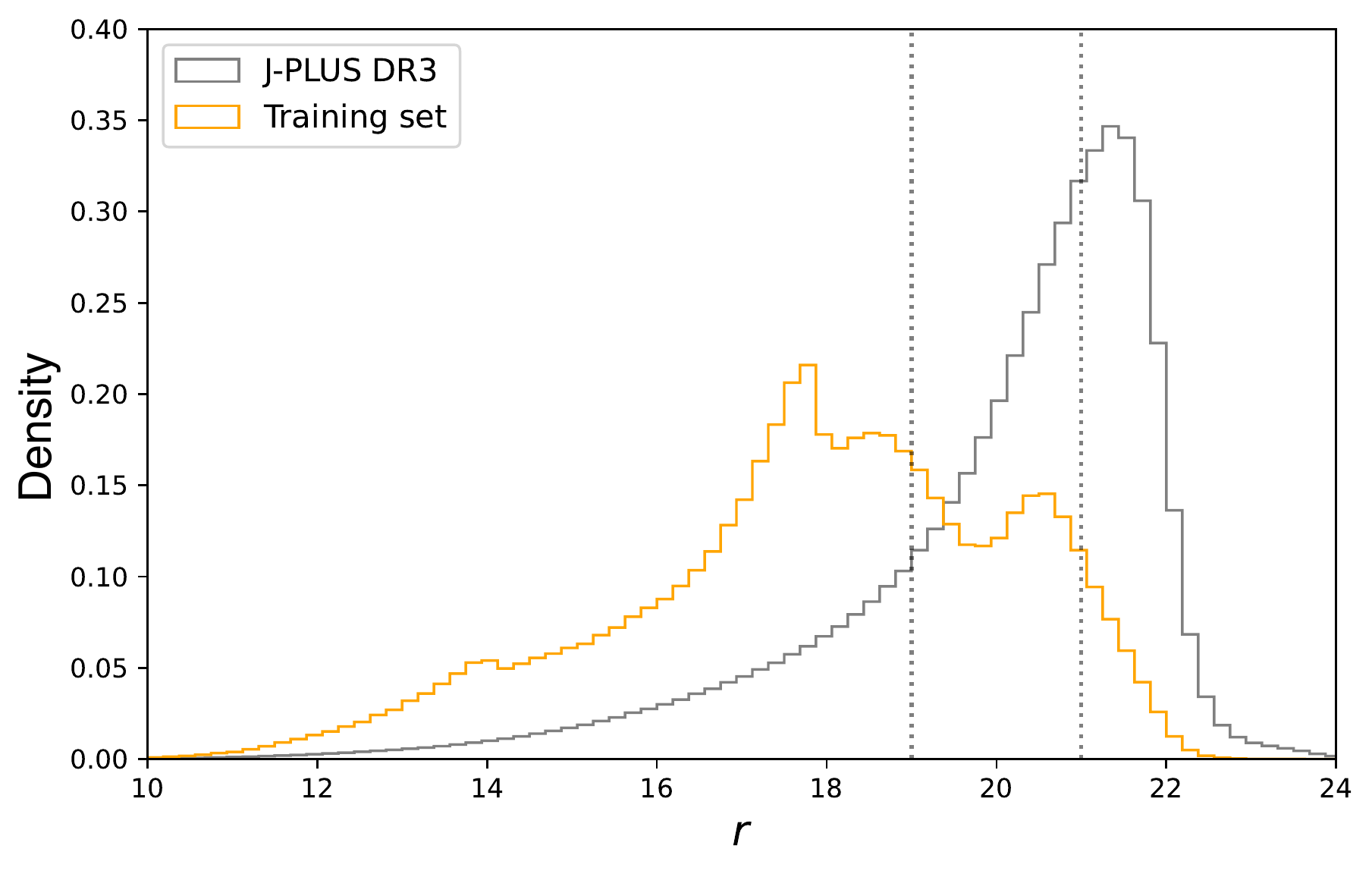}
\caption{Distributions of $r$-band magnitudes for the training set of Section~\ref{sssec:tr} and the full J-PLUS DR3 catalog.}
\label{fig:hist_rband}
\end{figure}

\begin{figure}
\centering
\includegraphics[width=\columnwidth]{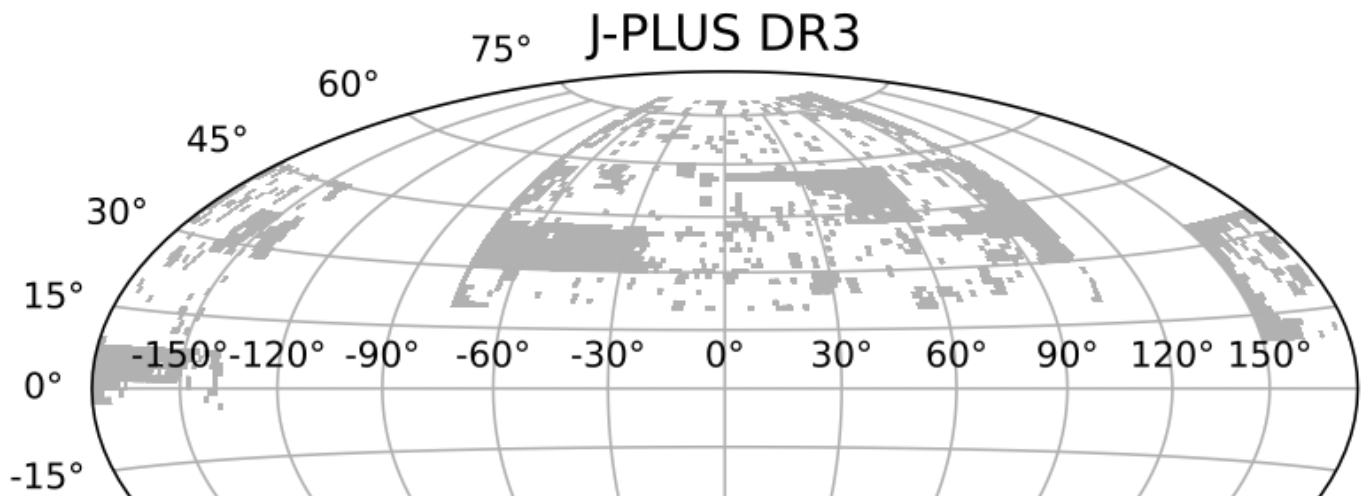}
\includegraphics[width=\columnwidth]{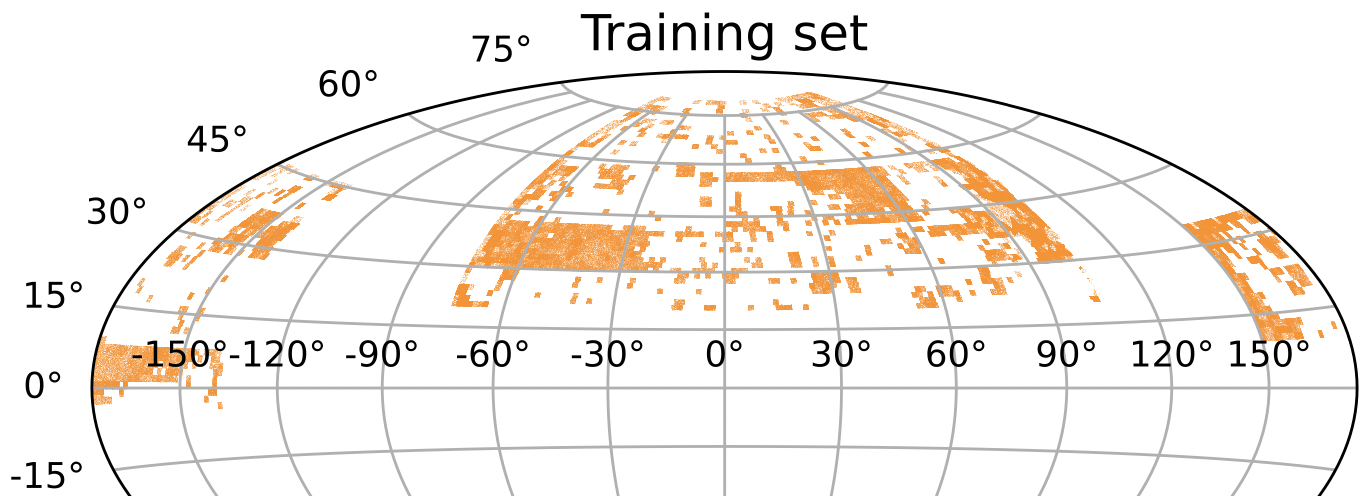}
\caption{Footprints of the  J-PLUS DR3 catalog and of the training set.}
\label{fig:footprint_train}
\end{figure}

Before applying our best model to the full J-PLUS catalog it is important to test the representativeness of the training set that we adopted. This analysis plays a crucial role in assessing the adequacy of our training set in capturing the general characteristics of the entire catalog. In Figs.~\ref{fig:hist_rband} and~\ref{fig:footprint_train} we compare the distributions of the $r$-band magnitudes and angular positions, respectively. Regarding the latter, Fig.~\ref{fig:footprint_train} shows that the training set covers all the J-PLUS footprint, giving us confidence that unknown systematics should be modeled by the machine learning algorithm. Regarding the magnitude distribution, Fig.~\ref{fig:hist_rband} shows that the J-PLUS catalog features objects at $r \gtrsim 20$ which are less represented by the training set. Note that the limiting magnitude of J-PLUS is $\sim$21.8~mag for $S/N\ge 5$. As a consequence (see next Section), our classification is expected to be less reliable for faint objects.

\begin{figure*}
\centering
\includegraphics[trim={0      .3cm 0  0}, clip, width=.9 \textwidth]{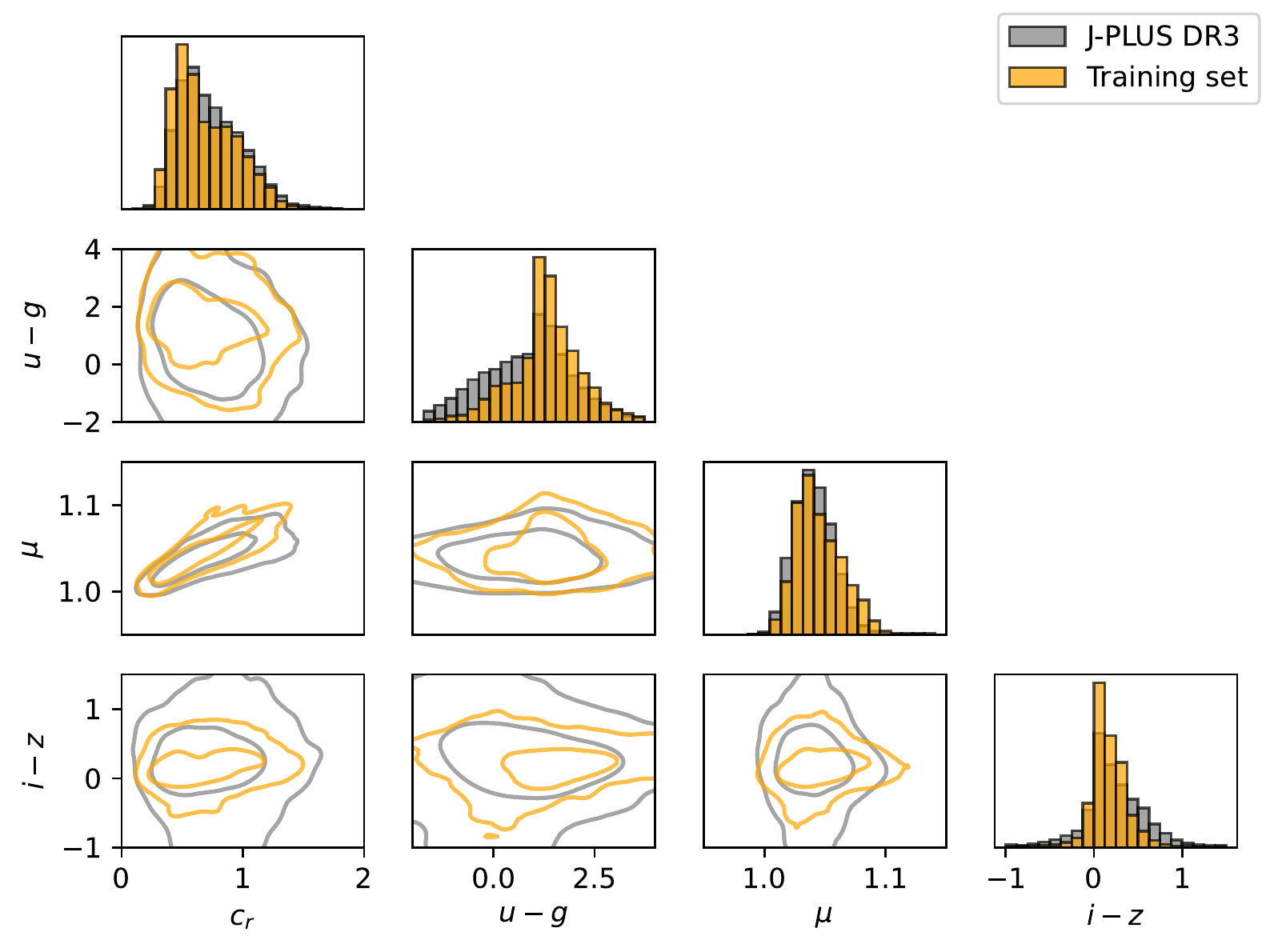}
\caption{Comparison of the 1- and 2-dimensional distributions of the training set and the full J-PLUS catalog relative to four important features. The contours mark 68\% and 95\% of the data. See Section~\ref{sec:represa} for details.}
\label{fig:triplot}
\end{figure*}

Next, we compare the distributions of the most important features.
We select the colors $u-g$ and $i-z$ because, as discussed in Sec.~\ref{subsec:fi} and confirmed in Fig.~\ref{fig:feat_imp}, they are important for all the three sources. Then we select the morphological parameters $c_r$ and $\mu$. We do not select the PSF feature because it characterizes observational conditions, which are not directly connected with representativeness; furthermore it is common to all sources in a given pointing. Finally, the representativeness of the photometry has been studied in Fig.~\ref{fig:hist_rband}.
Fig.~\ref{fig:triplot} compares the 1- and 2-dimensional distributions of the selected features for the training set and the full J-PLUS catalog. We find that the training set well covers the parameter space spanned by the full dataset, except for the $i-z$ color whose distribution a wider.


\begin{table*}
\centering
\setlength{\tabcolsep}{3.5pt}
\small
\begin{tabular}{l|cccc|cccc|cccc}
\hline\hline
\multicolumn{1}{c|}{\multirow{2}{*}{Analysis}} & \multicolumn{4}{c|}{$r\le 19$ (N=10,781,300)}         & \multicolumn{4}{c|}{$19<r\le 21$ (N=19,018,443)}       & \multicolumn{4}{c}{$r>21$ (N=17,631,499)}              \\ \cline{2-13} 
\multicolumn{1}{c|}{}                          & Galaxies  & Stars     & Quasars  & NC                 & Galaxies  & Stars     & Quasars   & NC                 & Galaxies  & Stars     & Quasars   & NC                 \\ \hline \vspace{-3mm}
\multirow{2}{*}{Full catalog}                  & 1,224,291 & 9,511,897 & 45,112   & \multirow{2}{*}{-} & 8,476,280 & 6,629,908 & 3,912,255 & \multirow{2}{*}{-} & 9,192,030 & 1,784,379 & 6,655,090 & \multirow{2}{*}{-} \\ 
                                               & (11.35\%) & (88.23\%) & (0.42\%) &                    & (44.57\%) & (34.86\%) & (20.57\%) &                    & (52.13\%) & (10.12\%) & (37.75\%) &                    \\ \vspace{-3mm}
\multirow{2}{*}{$p_{\rm cut}=0.5$}             & 1,200,941 & 9,473,493 & 34,979   & 71,887             & 7,863,746 & 6,014,965 & 3,109,034 & 2,030,698          & 8,165,084 & 1,369,138 & 5,586,425 & 2,510,852          \\ 
                                               & (11.14\%) & (87.87\%) & (0.32\%) & (0.67\%)           & (41.35\%) & (31.63\%) & (16.35\%) & (10.57\%)          & (46.31\%) & (7.77\%)  & (31.68\%) & (14.24\%)          \\ \vspace{-3mm}
\multirow{2}{*}{$p_{\rm cut}=0.7$}             & 1,171,677 & 9,414,116 & 25,841   & 169,666            & 6,747,967 & 4,812,149 & 1,460,699 & 5,997,628          & 5,658,434 & 872,233   & 2,523,653 & 8,577,179          \\
                                               & (10.87\%) & (87.32\%) & (0.24\%) & (1.57\%)           & (35.48\%) & (25.30\%) & (7.68\%)  & (31,54\%)          & (32.09\%) & (4.95\%)  & (14.31\%) & (48,65\%)          \\ \vspace{-3mm}
\multirow{2}{*}{$p_{\rm cut}=0.9$}             & 1,108,357 & 9,241,969 & 17,426   & 413,548            & 4,517,760 & 3,386,528 & 274,723   & 10,839,432          & 2,108,987 & 444,885   & 132,322   & 14,945,305         \\
                                               & (10.28\%) & (85.72\%) & (0.16\%) & (3.84\%)           & (23.75\%) & (17.81\%) & (1.44\%)  & (56.99\%)          & (11.96\%) & (2.52\%)  & (0.75\%)  & (84.76\%)          \\ \hline\hline 
\end{tabular}
\caption{Number of objects that are classified according to the highest probability criterion as galaxies, stars or quasars, split into three magnitude bins. While the first line shows all sources, the subsequent lines show only sources for which the highest probability is greater than the stated value of $p_{\rm cut}$. The sources that do not pass this criterion are not classified (NC).}
\label{tab:pcut}
\end{table*}

%\begin{table*}
%\centering
%\setlength{\tabcolsep}{3.5pt}
%\small
%\begin{tabular}{l|cccc|cccc|cccc}
%\hline\hline
%\multirow{2}{*}{Analysis} & \multicolumn{4}{c|}{$r\le 19$ (N=10,781,300)}  & \multicolumn{4}{c|}{$19<r\le 21$ (N=19,018,443)} & \multicolumn{4}{c}{$r>21$ (N=17,631,499)}   \\ \cline{2-13} 
%                          & Galaxies & Stars   & Quasars & NC     & Galaxies & Stars   & Quasars & NC      & Galaxies & Stars   & Quasars & NC      \\ \hline
%Full catalog              & 11.35\%  & 88.23\% & 0.42\%  & -      & 44.57\%  & 34.86\% & 20.57\% & -       & 52.13\%  & 10.12\% & 37.75\% & -       \\
%$p_{\rm cut}=0.5$         & 11.14\%  & 87.87\% & 0.32\%  & 0.67\% & 41.35\%  & 31.63\% & 16.35\% & 10,57\% & 46.31\%  &  7.77\% & 31.68\% & 14,24\% \\
%$p_{\rm cut}=0.7$         & 10.87\%  & 87.32\% & 0.24\%  & 1,57\% & 35.48\%  & 25.30\% & 7.68\%  & 31,54\% & 32.09\%  &  4.90\% & 14.31\% & 48,70\% \\
%$p_{\rm cut}=0.9$         & 10.28\%  & 85.72\% & 0.16\%  & 3.84\% & 23.75\%  & 17.81\% & 1.44\%  & 56.99\% & 11.96\%  & 2.52\%  & 0.75\%  & 84.76\% \\ \hline\hline
%\end{tabular}
%\caption{Number of objects that are classified according to the highest probability criterion as galaxies, stars or quasars, split into three magnitude bins. While the first line shows all sources, the subsequent lines show only sources for which the highest probability is greater than the stated value of $p_{\rm cut}$. The sources that do not pass this criterion are not classified (NC).}
%\label{tab:pcut}
%\end{table*}

%%%%%%%%%%%%%%%%%%%%%%%%%%%%%%%%%%%%%%
\subsection{Classification results of JPLUS DR3}

The ultimate goal of this work is to release a value-added catalog with our best galaxy-star-quasar classification. 
We applied our XGB model to all 47,431,242 objects of the J-PLUS DR3 catalog.
The result is given in Table~\ref{tab:pcut}, and is based on the highest probability criterion of Fig.~\ref{fig:3k}.

\begin{figure*}
\centering
\includegraphics[width=0.49\textwidth]{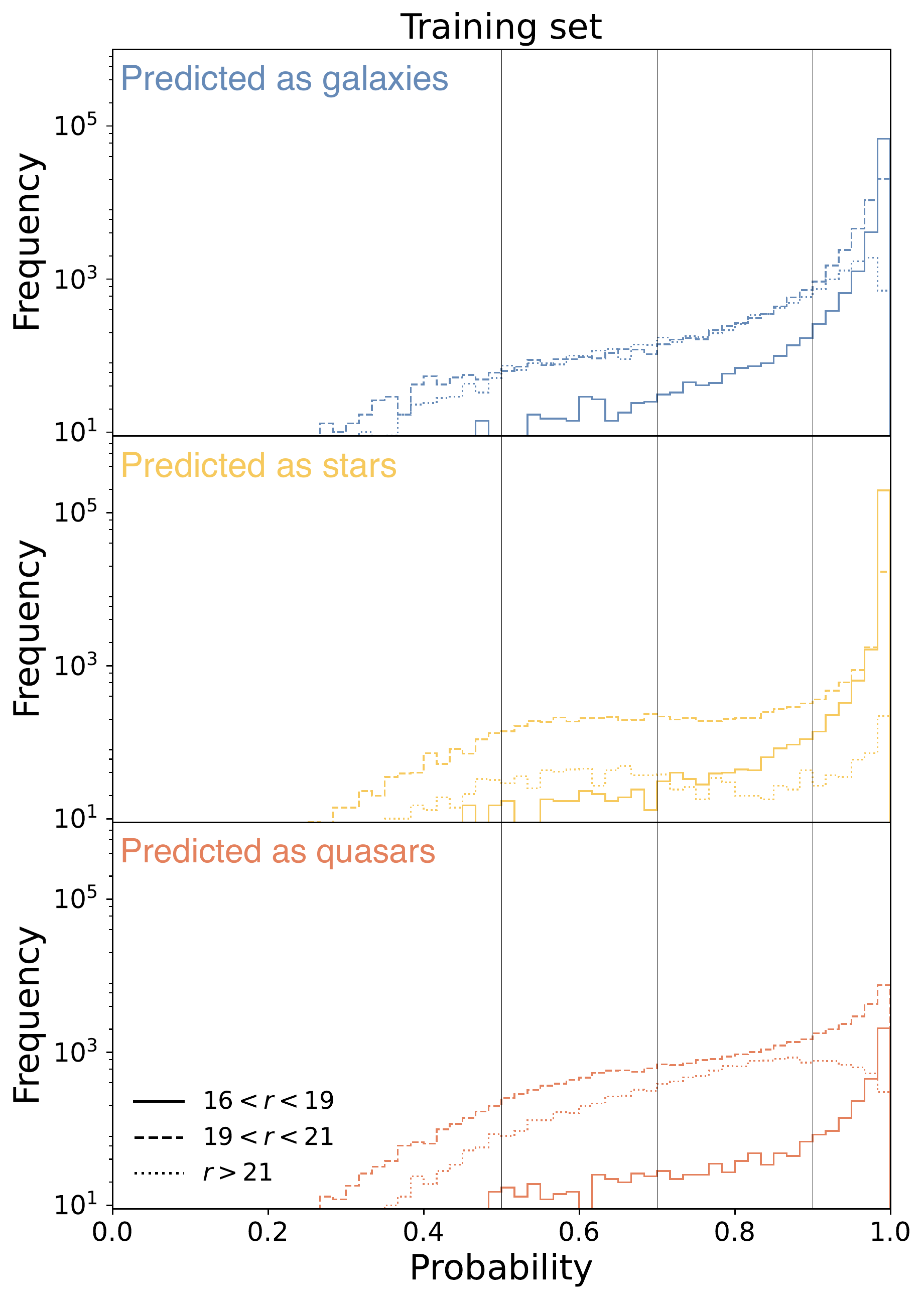}
\includegraphics[width=0.49\textwidth]{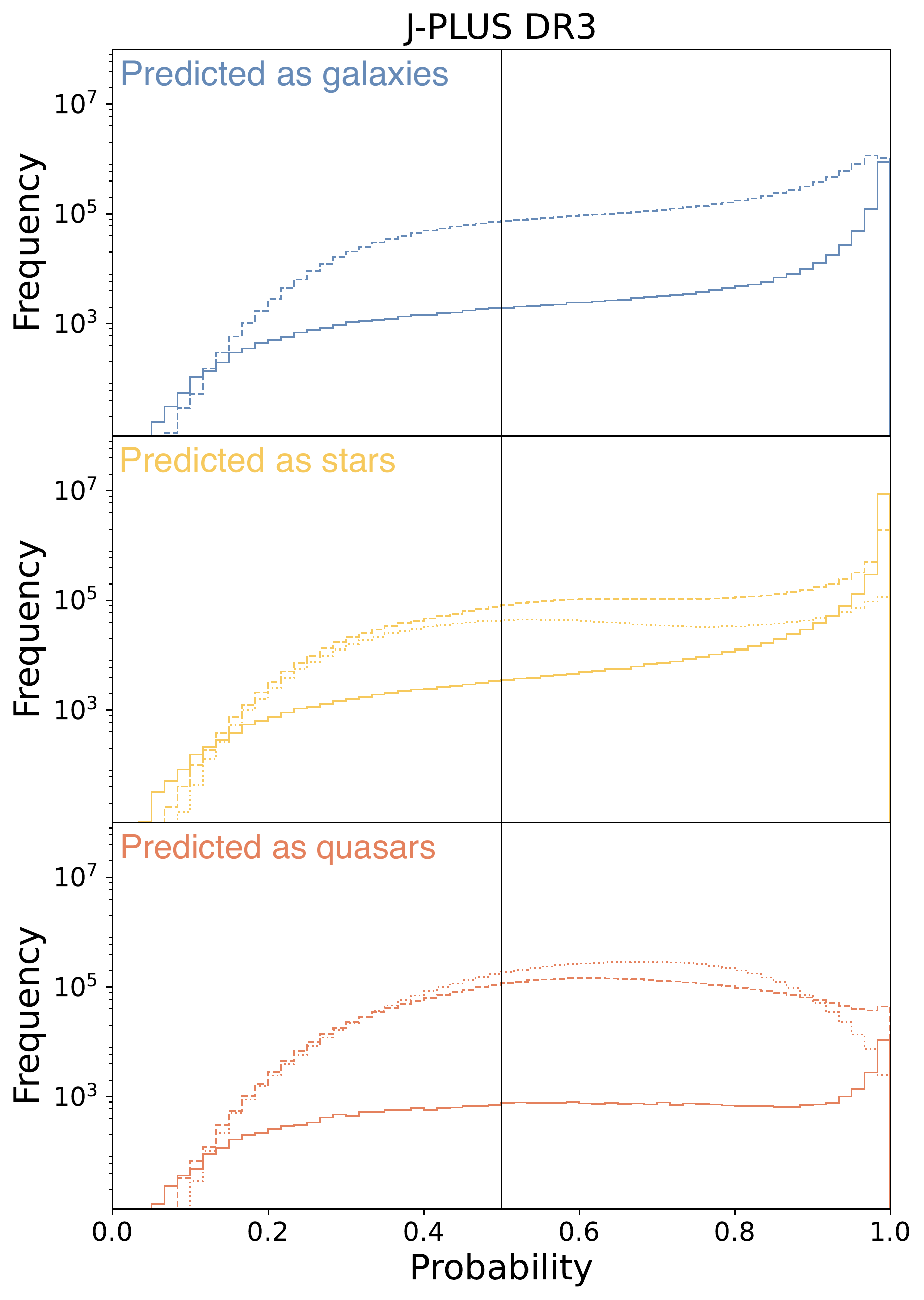}
\caption{Distribution of the probabilities $ p_{\rm gal}$ (top), $ p_{\rm star}$ (middle) and~$ p_{\rm qso}$ (bottom) of the objects that were classified by the XGB model as galaxies, stars and quasars, respectively, for the test set (left) and the full J-PLUS DR3 catalog (right) for three $r$ band bins.
For the sake of clarity we plotted these histograms using a continuous line and 50 bins between 0 and 1. The vertical lines mark the $p_{\rm cut}$ values used in Table~\ref{tab:pcut}.}
\label{fig:prob_full}
\end{figure*}

Fig.~\ref{fig:prob_full} shows the distributions of the probabilities $p_{\rm gal}$, $ p_{\rm star}$ and~$ p_{\rm qso}$ of the objects that were classified by the XGB model as galaxies, stars and quasars, respectively. Also in this case we split the analysis into three magnitude bins.
We can use Fig.~\ref{fig:prob_full} to compare the results relative to the test set (left panels) and the full J-PLUS DR3 catalog (right panels). This test should show if there is a significant number of objects that were underrepresented in the training set, degrading the performance.
Regarding the test set, we can see that the classification of brighter bins produces a distribution of probabilities that is more peaked at the ideal value of unity.%
\footnote{This is a necessary but not sufficient condition for an ideal classifier.}
On the other hand, the fainter bin shows flatter distributions across the classes, featuring an increase around the probability of half, signaling that there are comparatively more objects with indecisive classification. 
If we now look at the results relative to the full dataset, we see a similar but more pronounced trend.
The main differences are that the distributions are a bit flatter and that they reach smaller probability values, as compared to the curves relative to the test set.
Both these features indicate a degradation in performance and can be attributed to the fact that our training set represents less fainter objects at $r\gtrsim 20$, as discussed in Section~\ref{sec:represa}.
These plots allow us to assess the performance for the full set, going beyond the analysis on data representativity. We think they are a useful tool.

In particular, in the case of the full catalog there are significantly more objects with probabilities less than 0.5.
This means, for instance, that an hypothetical object with $p_{\rm gal}=0.08$, $p_{\rm star}=0.01$ and $p_{\rm qso}=0.01$ is classified as galaxy with a probability of 0.08 because the star and quasar binary classifiers returned an even lower value of probability (note that the normalized probability is $\tilde{p}_{\rm gal}=0.8$). However, according to the galaxy classifier, this object is classified with a probability of 0.92 as non-galaxy. It is evident that sources with probabilities below 0.5 lack reliable classification by our model. This observation serves as a motivation to apply a filter to the VAC based on a threshold $p > p_{\rm cut}$, where $p$ represents the highest probability obtained from the binary classifiers. Objects with binary probabilities below $p_{\rm cut}$ are considered unclassified. The minimum value of $p_{\rm cut}$ that is necessary to remove the problem depicted above is 0.5. Besides this value, we also consider the values 0.7 and 0.9 in order to increase the purity of the sample, as shown in Tab~\ref{tab:pcut}.

Table~\ref{tab:pcut} presents an overview encapsulating the key outcomes of our classification. Upon careful examination of the table, some conclusions emerge, aligning with our initial expectations: ($i$) the classification task becomes more challenging for fainter objects, evident from the percentage of unclassified objects and their sensitivity to changes in $p_{cut}$; ($ii$) brighter objects predominantly consist of stars, while galaxies and quasars exhibit higher concentrations among fainter objects, as indicated by the total number of objects per magnitude bin; ($iii$) quasars pose a greater classification challenge, as inferred from the variations in classified quasars across different $p_{cut}$ values, which might imply that only a minority have a probability close to 1.

\begin{figure}
\centering
\includegraphics[width=\columnwidth]{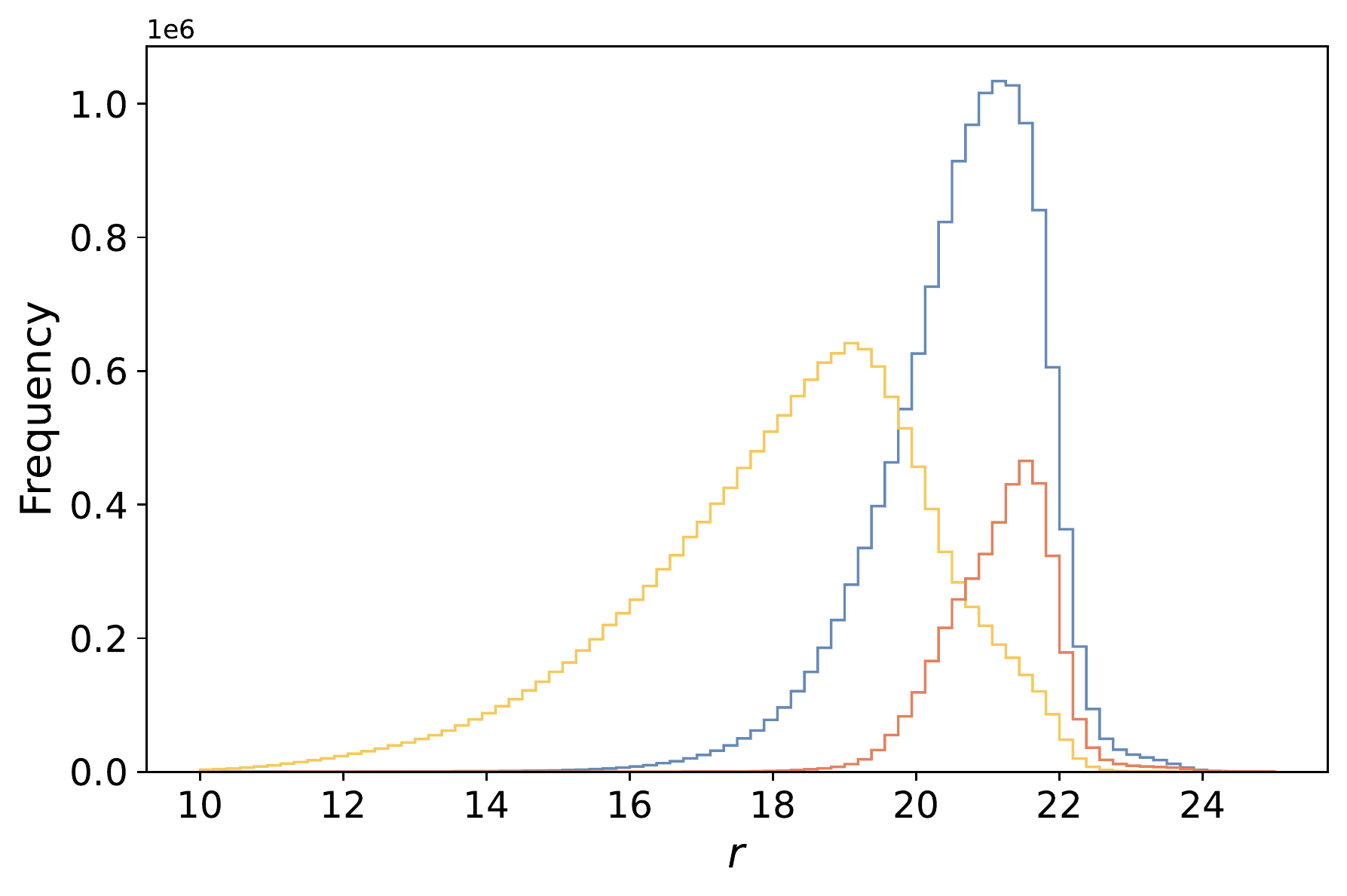}
\includegraphics[width=\columnwidth]{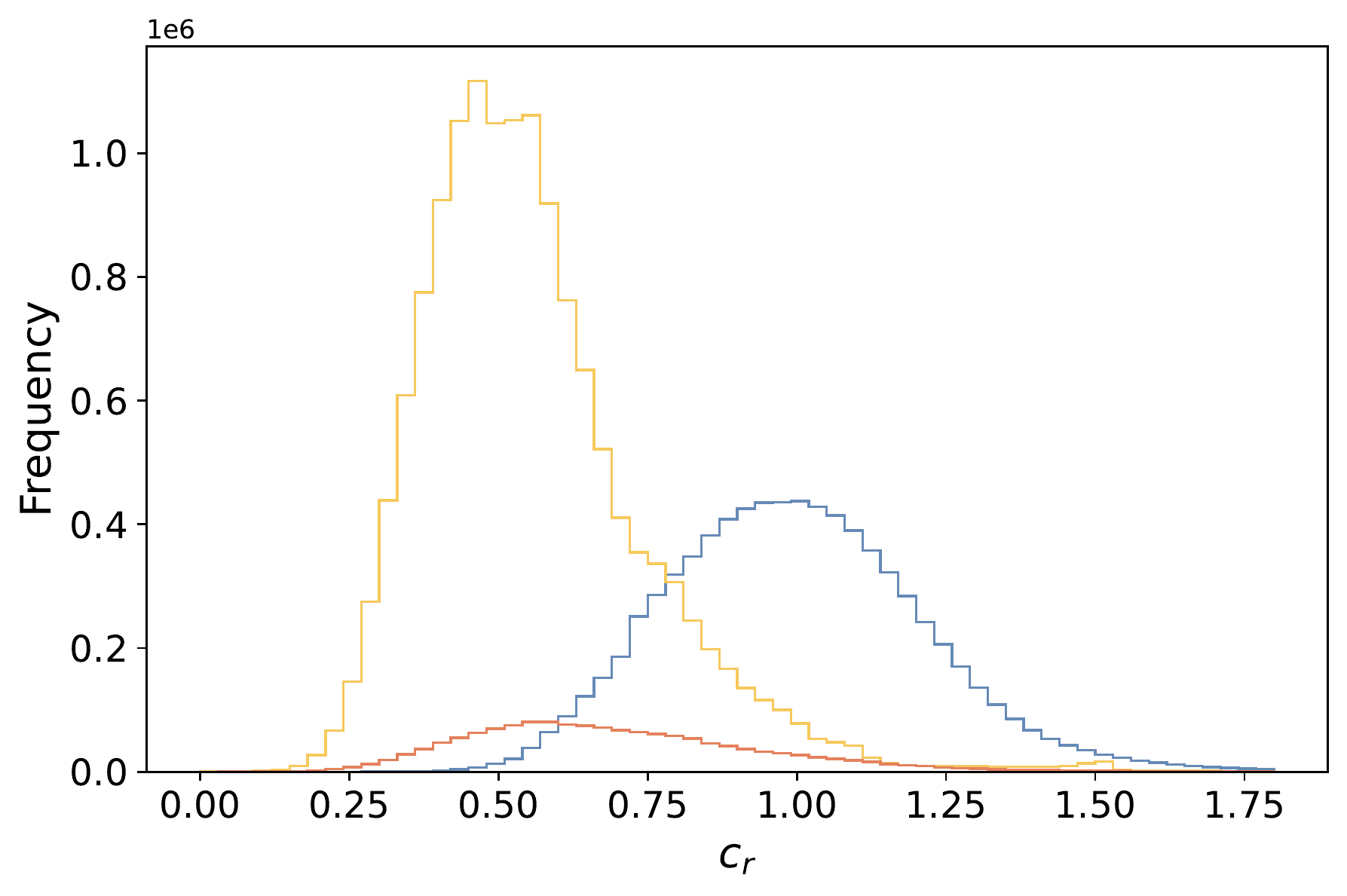}
\includegraphics[width=\columnwidth]{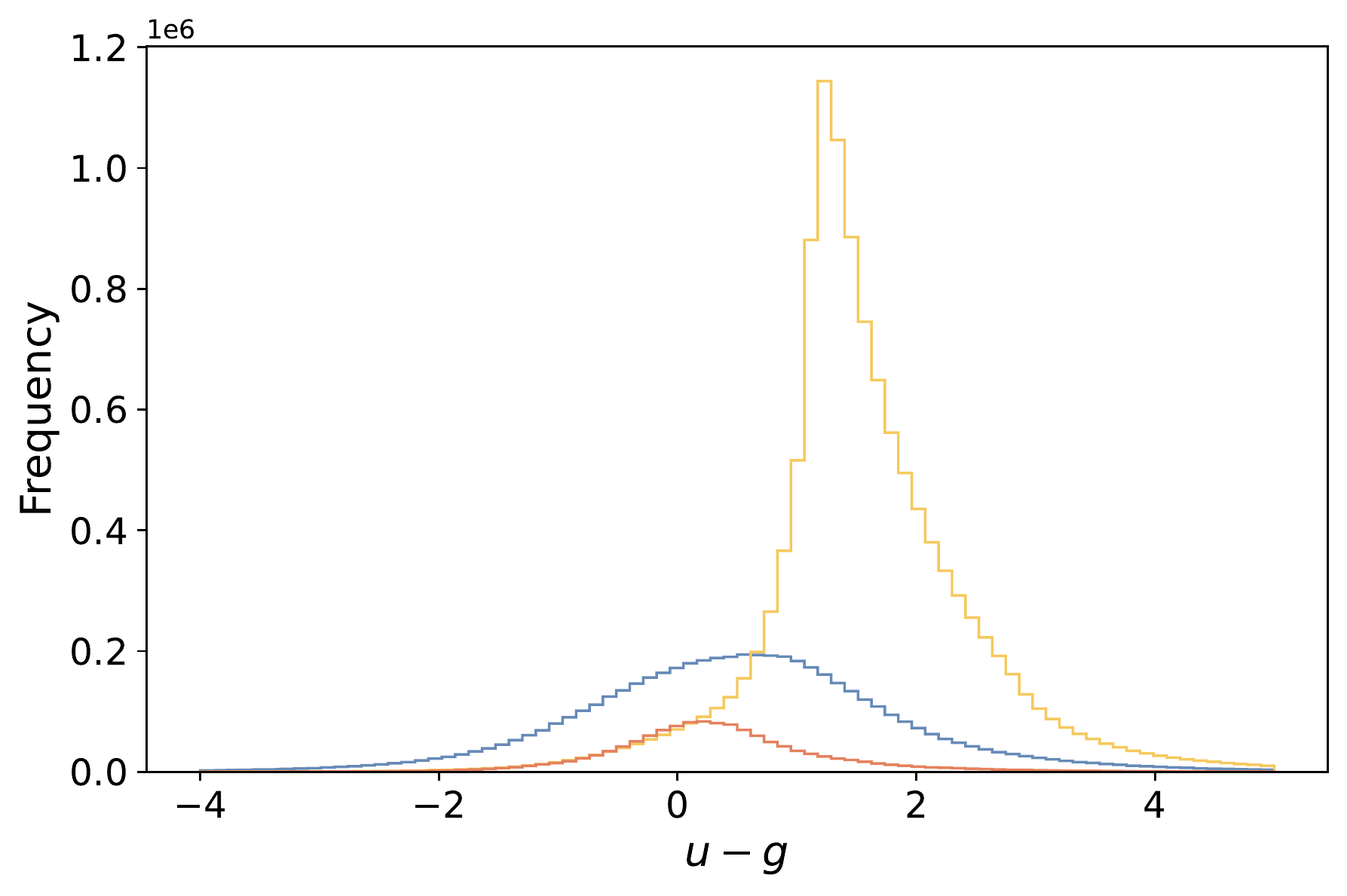}
\caption{Distribution with respect to the $r$ band (top), concentration for $r<21$ (middle) and $u-g$ color for $r<21$ (bottom) of the objects that were classified by the XGB model as galaxies, stars and quasars for the full J-PLUS DR3 catalog. Only objects with $p>p_{\rm cut}=0.7$ are shown, see also Table~\ref{tab:pcut}.}
\label{fig:rband_full}
\end{figure}

Fig.~\ref{fig:rband_full} shows the distribution of the $r$-band magnitude (top), concentration for $r<21$ (middle) and $u-g$ color for $r<21$ (bottom)  of the objects that satisfy the condition $p>p_{\rm cut}=0.7$. For $p>p_{\rm cut}=0.7$ and $r<21$ there are a total of 29,799,743 classified objects (63\% of the full J-PLUS DR3 catalog). As we can see from comparing Fig.~\ref{fig:rband_full} with Fig.~\ref{fig:hist_train}, the classified objects follow the concentration and color trends that we have seen in the training set, i.e., our results reveal that galaxies exhibit higher concentration as extended sources, while stars and quasars have lower concentration as point-like sources. Furthermore, it is noteworthy that quasars tend to have a slightly less positive $u-g$ color compared to stars and galaxies.

Choosing a higher value of $p_{\rm cut}$ has the effect of strongly reducing the number of quasars with respect to the number of stars, in agreement with the estimates from \citet{Palanque-Delabrouille:2015eca, palanque2016extended}, according to which the anticipated count of quasars, for the magnitude range we considered, is a small fraction of the total number of stars, typically a few percent or less.
Note, however, that the chosen value of $p_{\rm cut}$ may give worse completeness and purity for the full catalog than the ones estimated using the test set in Fig.~\ref{fig:prob}. This is due to the fact that, as said earlier, our training set is less representative for fainter objects and the performance is lower.


The catalog is publicly available%
\footnote{\url{https://www.j-plus.es/datareleases/data_release_dr3}}
via the ADQL table \texttt{jplus.StarGalClass}, where one can find the (non-normalized) binary probabilities $p_{\rm gal}$, $p_{\rm star}$ e $p_{\rm qso}$.
As discussed in Sec.~\ref{ssec:onevsall}, one can use directly these probabilities or  build the normalized probabilities of Eq.~\eqref{ptilde}. In the latter case the user may adopt the usual classification, according to the highest-probability criterion, or some other criterion.
See Appendix~\ref{ap:adql} for more information and an ADQL query example.

%%%%%%%%%%%%%%%%%%%%%%%%%%%%%%%%%%%%%%
%%%%%%%%%%%%%%%%%%%%%%%%%%%%%%%%%%%%%%
%%%%%%%%%%%%%%%%%%%%%%%%%%%%%%%%%%%%%%
\section{Conclusions}
\label{sec:conclusions}

By testing a large set of machine learning algorithms and hyperparameters using the TPOT optimization pipeline, we were able to identify the best performing model and classify J-PLUS DR3 sources into galaxies, stars and quasars. The goal is to produce a value-added catalog with a trustworthy classification. In particular, the classification provided in J-PLUS DR3 does not consider quasars.

We adopted supervised machine learning algorithms and built a training set via crossmatches with SDSS DR18, LAMOST DR8 and \textit{Gaia} DR3. We tested that the training set is both representative and pure.
The training set includes 
660 thousand galaxies, 1.2 million stars and 270 thousand quasars.
We considered 37 features: besides the twelve photometric bands with their errors, we adopted six colors, four morphological parameters, extinction with its error and the PSF relative to the corresponding pointing.

After exploring numerous pipeline possibilities through the TPOT genetic algorithm, we found that XGBoost provides the best performance: the AUC for galaxies, stars and quasars is above 0.99 and the average precision is above 0.99 for galaxies and stars and 0.96 for quasars.
XGBoost outperforms the star-galaxy classifiers already provided in J-PLUS DR3 and also efficiently classifies quasars.
Regarding feature importance, we found that photometry was very important in the classification of quasars, showing the relevance of narrow-band photometry.
We also showed which individual bands are most important for the classification.

Finally, the value-added catalog with our best galaxy-star-quasar classification is available via the ADQL table \texttt{jplus.StarGalClass}.
We classified all the 47 million objects of the J-PLUS DR3 catalog and 90\% of the dataset received a meaningful classification ($p>p_{\rm cut}=0.5$).
As discussed in the previous Sections, although at $r\gtrsim 20$ the training set is less representative (see \ref{fig:hist_rband}), we believe that there is merit in classifying also faint objects which may serve as candidates for future surveys.%
\footnote{We remind the reader that the classifier was trained once over all sources.}

As future developments, we stress the importance of increasing the depth of the training set in order to increase representativity. Indeed, by analyzing the distribution of probabilities of classified sources we found hints that the training set is not representing all the relevant types of objects.
This goal may be attainable, for instance, with the forthcoming dataset from the Dark Energy Spectroscopic Instrument \citep{DESI2016}. This will substantially augment the training set, particularly in relation to galaxies and quasars. Furthermore, improvements on the machine learning side could be achieved via the use of deep neural networks.
Finally, it is worth stressing that the methodology developed in this work may be applied to the Javalambre Physics of the Accelerating Universe Astrophysical Survey \citep[J-PAS, ][]{Bonoli:2020ciz}, extending the results of \citet{2021A&A...645A..87B}.

%%%%%%%%%%%%%%%%%%%%%%%%%%%%%%%%%%%%
%%%%%%%%%%%%%%%%%%%%%%%%%%%%%%%%%%%%
\section*{Acknowledgements}

RvM thanks FAPES (Brazil) and the Programa de Capacitaç\~ao Institucional PCI/ON for financial support.
VM thanks CNPq (Brazil) and FAPES (Brazil)  for partial financial support.
MQ is supported by the Brazilian research agencies CNPq, FAPERJ and CAPES. We acknowledge support from the CAPES-DAAD bilateral project  ``Data Analysis and Model Testing in the Era of Precision Cosmology.''
LC acknowledges financial support from CNPq.
AAC acknowledges support from the State Agency for Research of the Spanish MCIU through the ``Center of Excellence Severo Ochoa'' award to the Instituto de Astrofísica de Andalucía (SEV-2017-0709).
JAFO acknowledges the financial support from the Spanish Ministry of Science and Innovation and the European Union -- NextGenerationEU through the Recovery and Resilience Facility project ICTS-MRR-2021-03-CEFCA.
LADG~acknowledges financial support from the State Agency for Research of the Spanish MCIU through the `Center of Excellence Severo Ochoa' award to the Instituto de Astrof\'isica de Andaluc\'ia (SEV-2017-0709), and to the PID2019-109067-GB100.
LSJ acknowledges the support from CNPq (308994/2021-3)  and FAPESP (2011/51680-6).
This work made use of the Virgo Cluster at Cosmo-Ufes/UFES, which is funded by FAPES (Fundação de Amparo à Pesquisa e Inovação do Espírito Santo) and administered by Renan Alves de Oliveira.\\
%J-PLUS
Based on observations made with the JAST80 telescope and T80Cam camera for the J-PLUS project at the Observatorio Astrof\'{\i}sico de Javalambre (OAJ), in Teruel, owned, managed, and operated by the Centro de Estudios de F\'{\i}sica del  Cosmos de Arag\'on (CEFCA). We acknowledge the OAJ Data Processing and Archiving Unit (UPAD) for reducing and calibrating the OAJ data used in this work.
Funding for the J-PLUS project has been provided by the Governments of Spain and Arag\'on through the Fondo de Inversiones de Teruel; the Aragonese Government through the Research Groups E96, E103, E16\_17R, and E16\_20R; the Spanish Ministry of Science, Innovation and Universities (MCIU/AEI/FEDER, UE) with grants PGC2018-097585-B-C21 and PGC2018-097585-B-C22; the Spanish Ministry of Economy and Competitiveness (MINECO/FEDER, UE) under AYA2015-66211-C2-1-P, AYA2015-66211-C2-2, AYA2012-30789, and ICTS-2009-14; and European FEDER funding (FCDD10-4E-867, FCDD13-4E-2685). The Brazilian agencies FAPERJ and FAPESP as well as the National Observatory of Brazil have also contributed to this project.\\
%SDSS
Funding for SDSS-III has been provided by the Alfred P.~Sloan Foundation, the Participating Institutions, the National Science Foundation, and the U.S.~Department of Energy Office of Science. The SDSS-III web site is \href{http://www.sdss3.org/}{sdss3.org}.
SDSS-III is managed by the Astrophysical Research Consortium for the Participating Institutions of the SDSS-III Collaboration including the University of Arizona, the Brazilian Participation Group, Brookhaven National Laboratory, Carnegie Mellon University, University of Florida, the French Participation Group, the German Participation Group, Harvard University, the Instituto de Astrofisica de Canarias, the Michigan State/Notre Dame/JINA Participation Group, Johns Hopkins University, Lawrence Berkeley National Laboratory, Max Planck Institute for Astrophysics, Max Planck Institute for Extraterrestrial Physics, New Mexico State University, New York University, Ohio State University, Pennsylvania State University, University of Portsmouth, Princeton University, the Spanish Participation Group, University of Tokyo, University of Utah, Vanderbilt University, University of Virginia, University of Washington, and Yale University.\\
%Gaia
This work has made use of data from the European Space Agency (ESA) mission {\it Gaia} (\href{https://www.cosmos.esa.int/gaia}{www.cosmos.esa.int/gaia}), processed by the {\it Gaia} Data Processing and Analysis Consortium (DPAC, \href{https://www.cosmos.esa.int/web/gaia/dpac/consortium}{www.cosmos.esa.int/web/gaia/dpac/consortium}). Funding for the DPAC has been provided by national institutions, in particular the institutions participating in the {\it Gaia} Multilateral Agreement.\\
%LAMOST
Guoshoujing Telescope (the Large Sky Area Multi-Object Fiber Spectroscopic Telescope LAMOST) is a National Major Scientific Project built by the Chinese Academy of Sciences. Funding for the project has been provided by the National Development and Reform Commission. LAMOST is operated and managed by the National Astronomical Observatories, Chinese Academy of Sciences.

%%%%%%%%%%%%%%%%%%%%%%%%%%%%%%%%%%%%
%%%%%%%%%%%%%%%%%%%%%%%%%%%%%%%%%%%%

\section*{Data availability}

The data underlying this article (training and test sets) will be shared on reasonable request to the corresponding author.
The classifier configuration file as well as the validation set analyses are publicly available at \url{https://github.com/rodrigovonmarttens/J-PLUS-classification}.
The VAC will be made publicly accessible in compliance with the rules of the J-PLUS collaboration.

%%%%%%%%%%%%%%%%%%%%%%%%%%%%%%%%%%%%
%%%%%%%%%%%%% REFERENCES %%%%%%%%%%%%%%%
\bibliographystyle{mnrasArxiv}
\bibliography{biblio}

%%%%%%%%%%%%%%%%%%%%%%%%%%%%%%%%%%%%
%%%%%%%%%%%%%%%%%%%%%%%%%%%%%%%%%%%%

\appendix

%%%%%%%%%%%%%%%%%%%%%%%%%%%%%%%%%%%%%%
%%%%%%%%%%%%%%%%%%%%%%%%%%%%%%%%%%%%%%
%%%%%%%%%%%%%%%%%%%%%%%%%%%%%%%%%%%%%%
\section{Pipeline}
\label{ap:pipeline}

As discussed in Sec.~\ref{ssec:tpotxgb}, the resulting pipeline from the TPOT analysis is simply the use of  eXtreme Gradient Boosting (XGBoost). However, beyond the ML method,  TPOT also analyzed the hyperparameters. In our analysis, the result of the optimization of the hyperparameters is the following:
\begin{itemize}
    \item \texttt{base\_score=None} %\vspace{-3mm}
    \item \texttt{booster='gbtree'} %\vspace{-3mm}
    \item \texttt{bootstrap=False} %\vspace{-3mm}
    \item \texttt{class\_weight='balanced\_subsample'} %\vspace{-3mm}
    \item \texttt{colsample\_bylevel=None} %\vspace{-3mm}
    \item \texttt{colsample\_bynode=None} %\vspace{-3mm}
    \item \texttt{colsample\_bytree=None} %\vspace{-3mm}
    \item \texttt{criterion='gini'} %\vspace{-3mm}
    \item \texttt{eta=0.1} %\vspace{-3mm}
    \item \texttt{gamma=None} %\vspace{-3mm}
    \item \texttt{importance\_type='gain'} %\vspace{-3mm}
    \item \texttt{interaction\_constraints=None} %\vspace{-3mm}
    \item \texttt{max\_delta\_step=None} %\vspace{-3mm}
    \item \texttt{max\_depth=10} %\vspace{-3mm}
    \item \texttt{max\_features=None} %\vspace{-3mm}
    \item \texttt{min\_child\_weight=None} %\vspace{-3mm}
    \item \texttt{monotone\_constraints=None} %\vspace{-3mm}
    \item \texttt{n\_estimators=200} %\vspace{-3mm}
    \item \texttt{num\_parallel\_tree=None} %\vspace{-3mm}
    \item \texttt{random\_state=5} %\vspace{-3mm}
    \item \texttt{reg\_alpha=None} %\vspace{-3mm}
    \item \texttt{reg\_lambda=None} %\vspace{-3mm}
    \item \texttt{scale\_pos\_weight=None} %\vspace{-3mm}
    \item \texttt{subsample=None} %\vspace{-3mm}
    \item \texttt{tree\_method=None} %\vspace{-3mm}
    \item \texttt{validate\_parameters=None} %\vspace{-3mm}
\end{itemize}

%%%%%%%%%%%%%%%%%%%%%%%%%%%%%%%%%%%%%%
%%%%%%%%%%%%%%%%%%%%%%%%%%%%%%%%%%%%%%
%%%%%%%%%%%%%%%%%%%%%%%%%%%%%%%%%%%%%%
\section{Classification distribution}
\label{ap:goodies}

\begin{figure}
\centering
\includegraphics[trim={0      .2cm 0  0}, clip, width=\columnwidth]{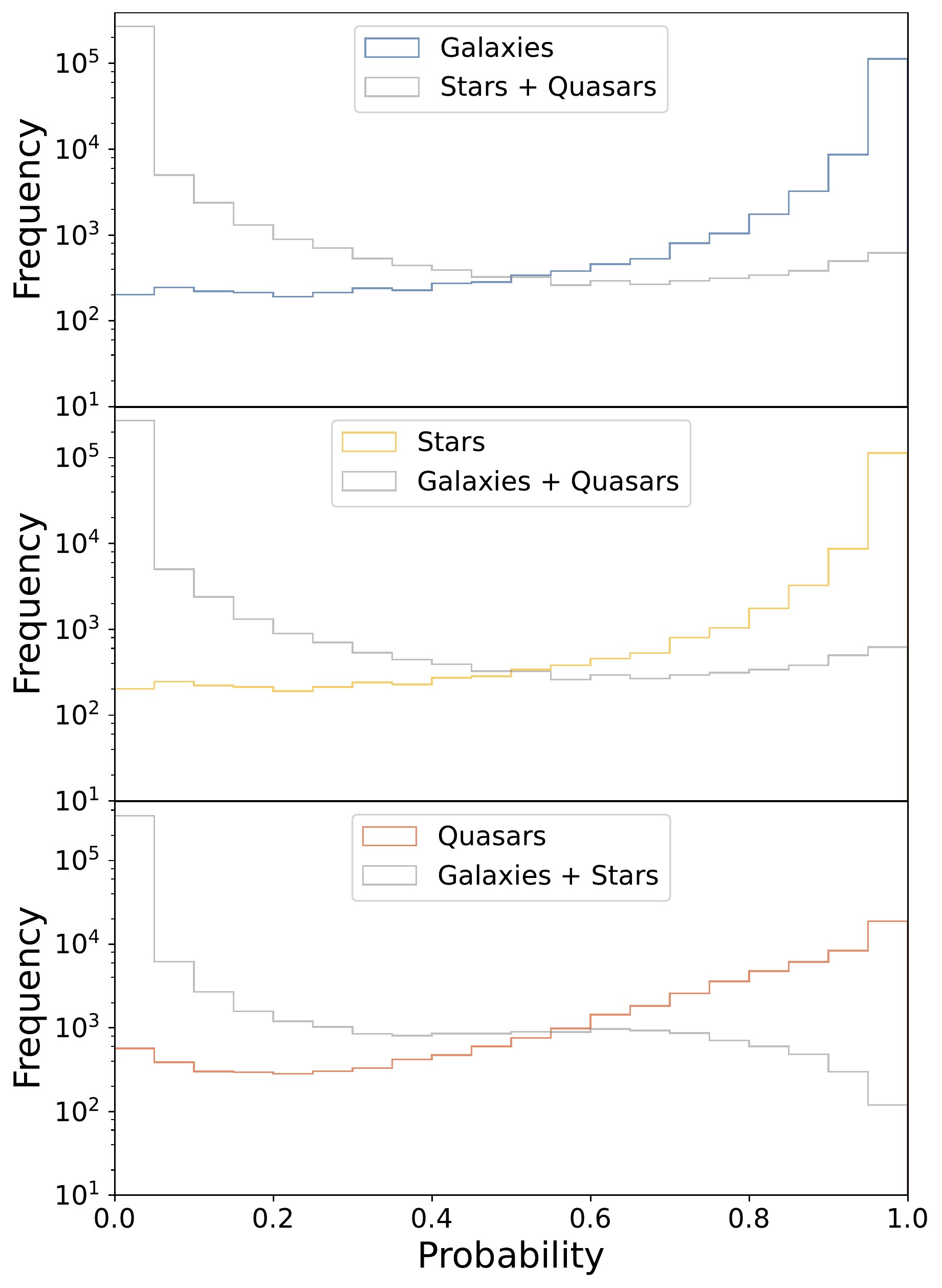}
\caption{Distribution of probabilities $p_{\rm gal}$, $p_{\rm star}$ and~$p_{\rm qso}$ for the test set, see App.~\ref{ap:goodies}.}
\label{fig:n-prob}
\end{figure}

The top panel of Fig.~\ref{fig:n-prob} shows the distribution of probabilities $p_{\rm gal}$ that XGB assigns to the actual galaxies in the test set. Also shown for completeness are the probabilities of the remaining sources. The other two panels show the same for stars and quasars.
We see that most of the actual objects can be recovered by selecting, for instance, $p_{\rm gal}>0.5$ in the case of galaxies.
We also see that the distribution of $p_{\rm qso}$ is flatter as compared to stars and galaxies, showing that a larger fraction is classified with lower values of $p_{\rm qso}$. For an ideal classifier the distribution would be a Dirac delta at unity.

%

%%%%%%%%%%%%%%%%%%%%%%%%%%%%%%%%%%%%%%
%%%%%%%%%%%%%%%%%%%%%%%%%%%%%%%%%%%%%%
%%%%%%%%%%%%%%%%%%%%%%%%%%%%%%%%%%%%%%
\section{Sky positions of misclassifications}
\label{ap:sky}

\begin{figure*}
\centering
% trim={<left> <lower> <right> <upper>}
\includegraphics[trim={0  0 0  0}, clip, width=0.49\textwidth]{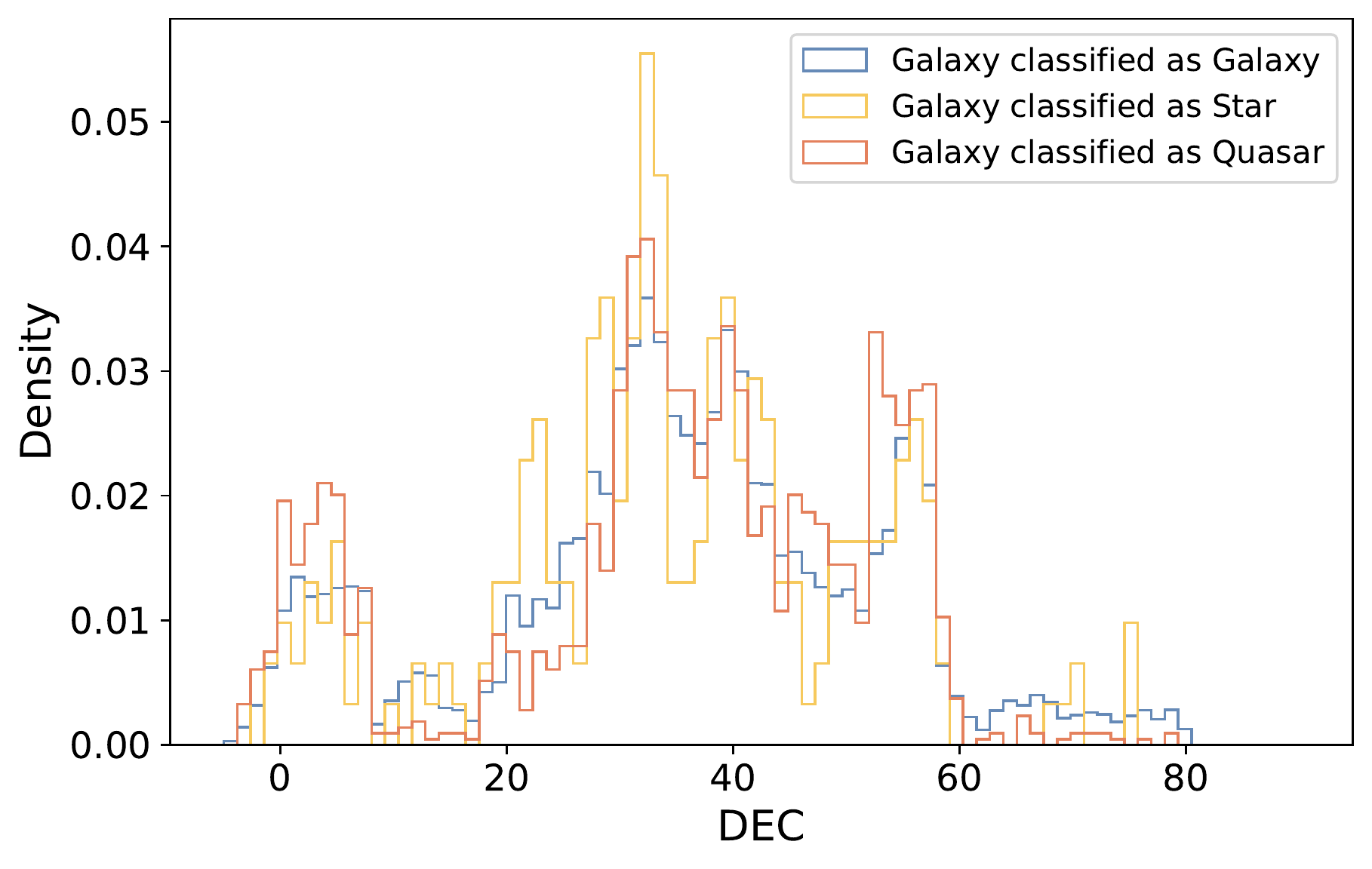}
\includegraphics[trim={0  0 0  0}, clip, width=0.49\textwidth]{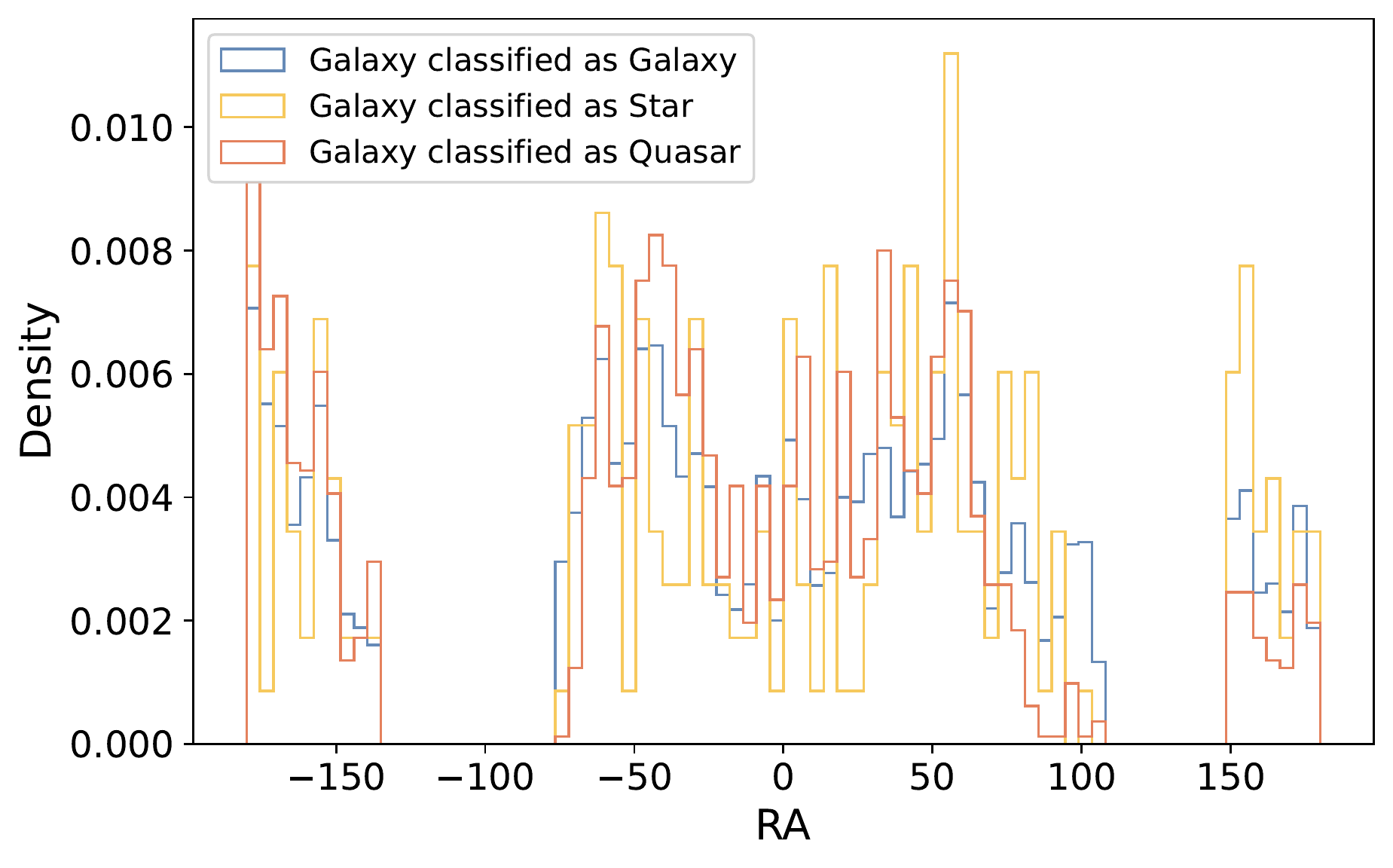}
\includegraphics[trim={0  0 0  0}, clip, width=0.49\textwidth]{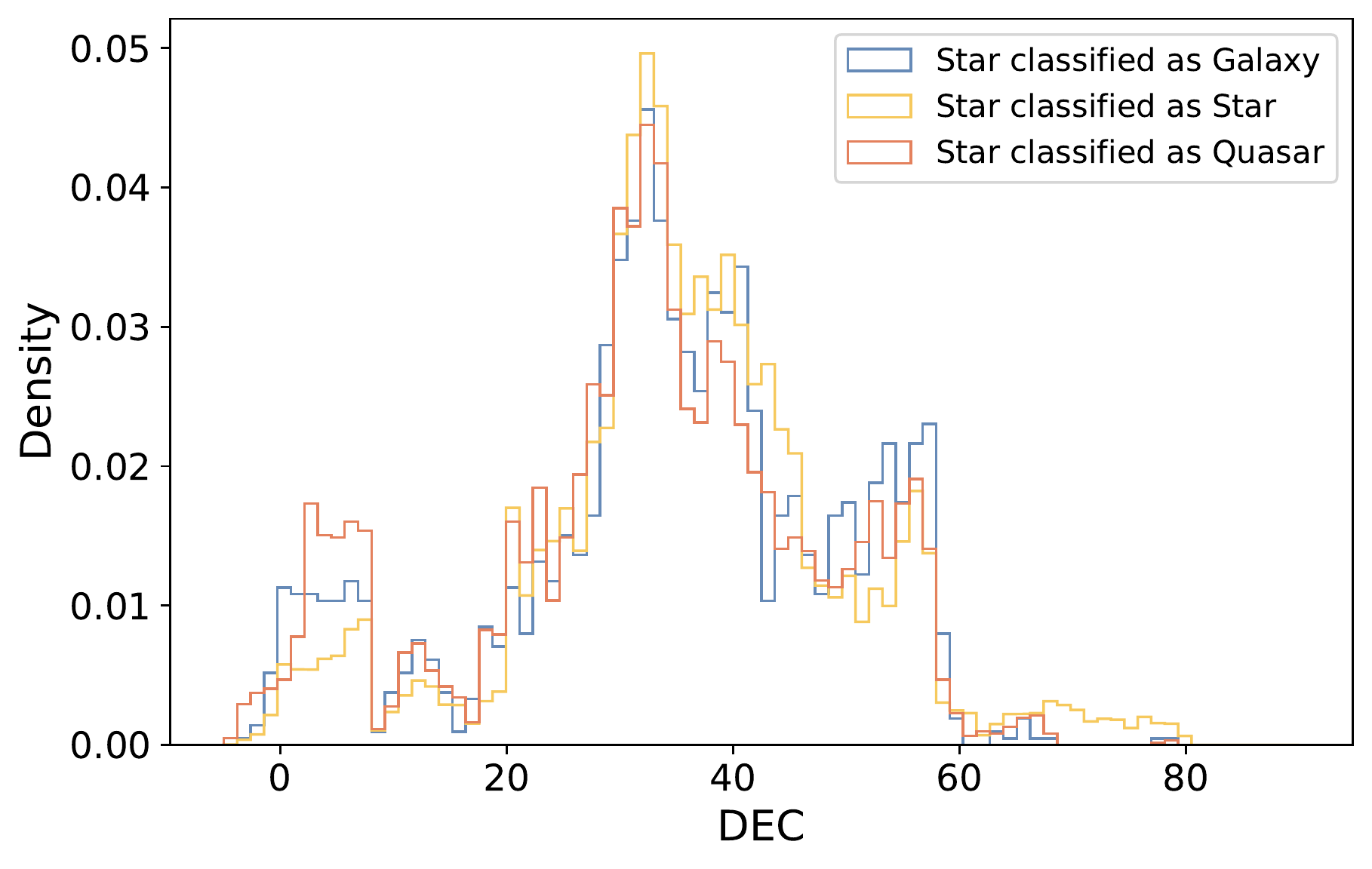}
\includegraphics[trim={0  0 0  0}, clip, width=0.49\textwidth]{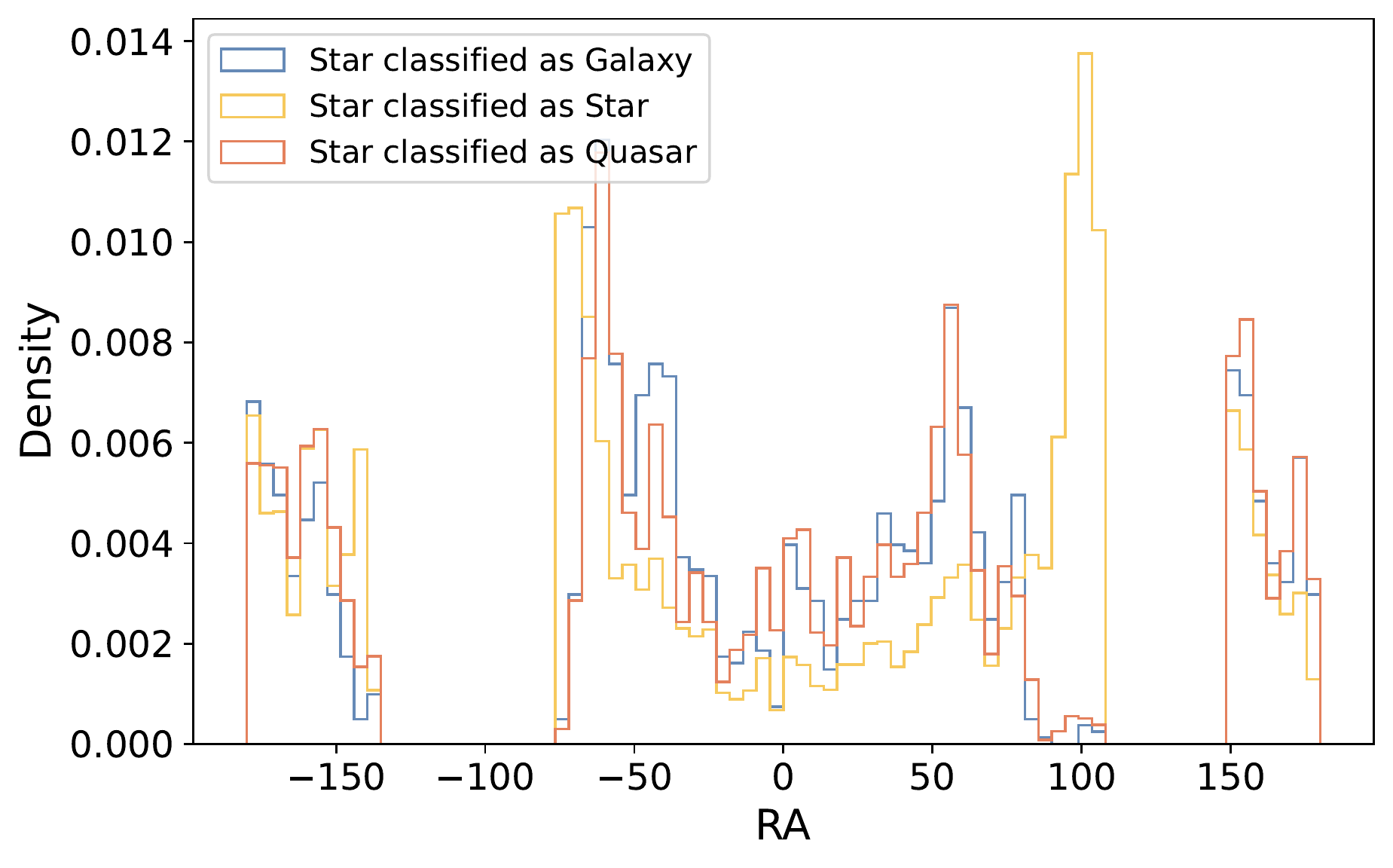}
\includegraphics[trim={0  0 0  0}, clip, width=0.49\textwidth]{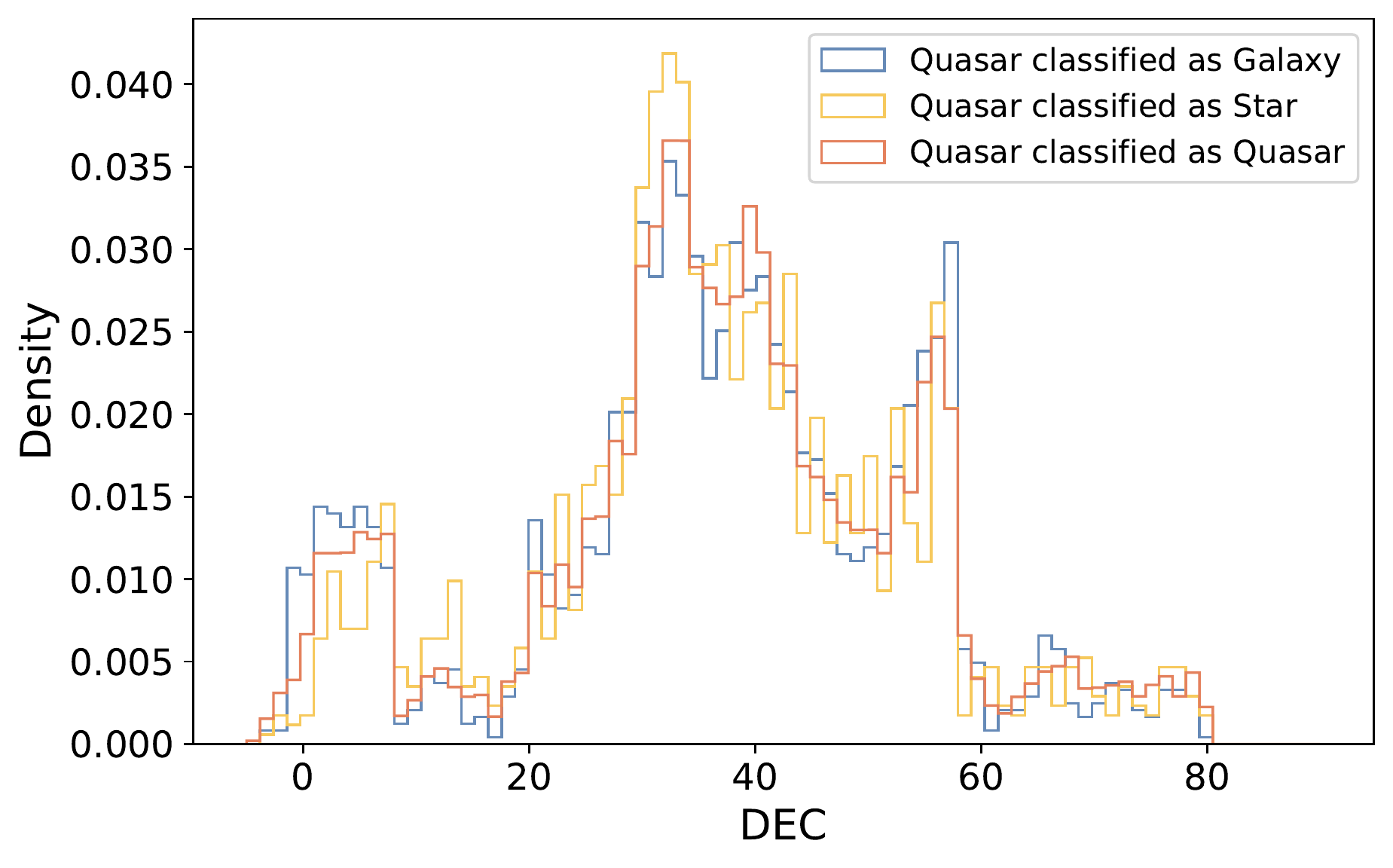}
\includegraphics[trim={0  0 0  0}, clip, width=0.49\textwidth]{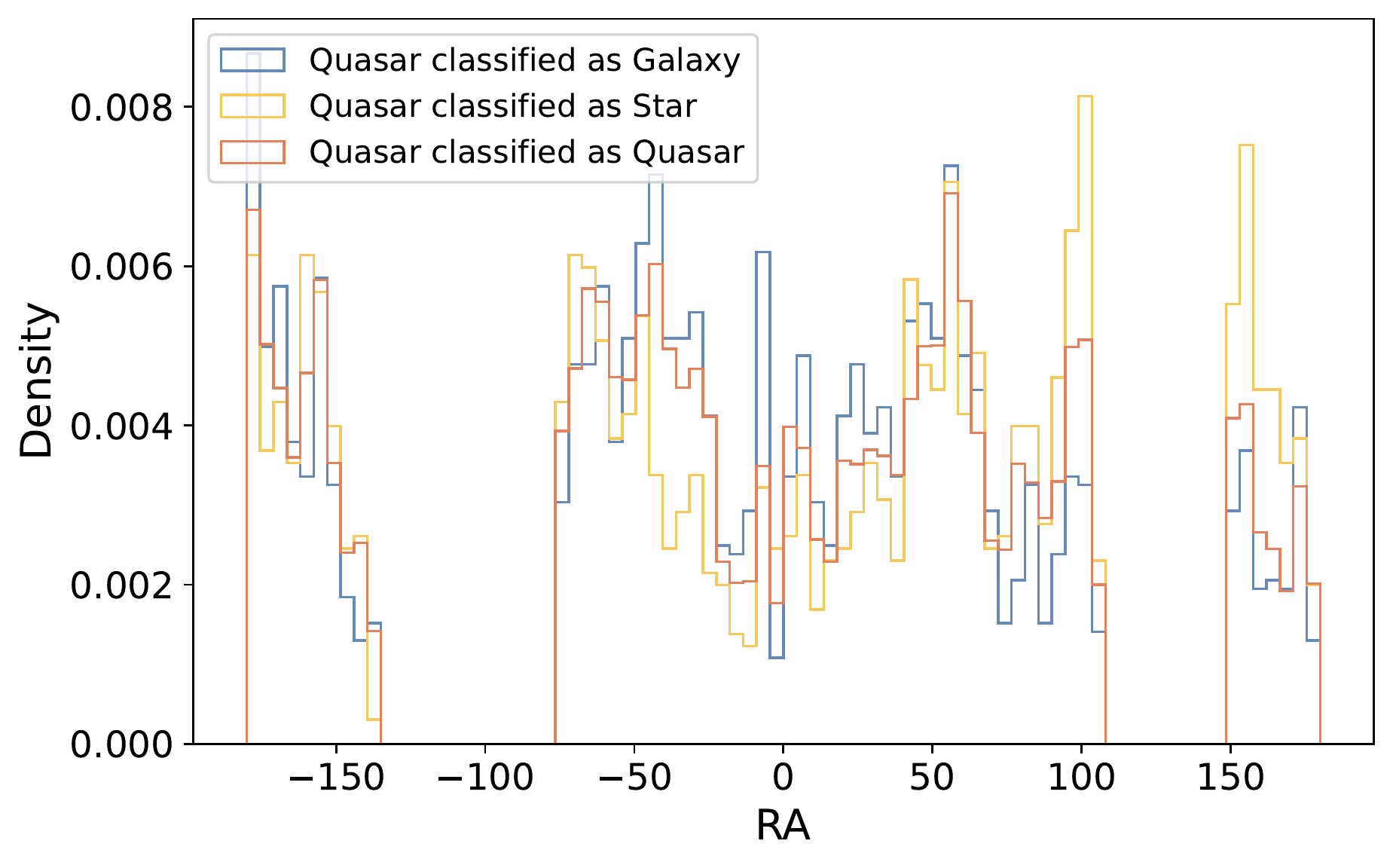}
\caption{Sky positions of misclassifications.}
\label{fig:sky}
\end{figure*}

Fig.~\ref{fig:sky} shows the sky position, divided by classes and by correct or incorrect classifications. We did not find significant misclassification trends with respect to the sky positions of the sources nor crowding in specific sky regions, except an excess of objects that are classified as stars in the region close to the galactic plane (see Fig.~\ref{fig:footprint_train}) as highlighted in the RA histograms.

%

%%%%%%%%%%%%%%%%%%%%%%%%%%%%%%%%%%%%%%
%%%%%%%%%%%%%%%%%%%%%%%%%%%%%%%%%%%%%%
%%%%%%%%%%%%%%%%%%%%%%%%%%%%%%%%%%%%%%
\section{ADQL query}
\label{ap:adql}

The value added catalog with the XGBoost classifications is publicly available via the ADQL table \texttt{jplus.StarGalClass} from the J-PLUS webpage.
\footnote{\url{https://www.j-plus.es/datareleases/data_release_dr3}}
The columns \texttt{xgb\_prob\_gal}, \texttt{xgb\_prob\_star} and \texttt{xgb\_prob\_qso} give $p_{\rm gal}$, $p_{\rm star}$ and $p_{\rm qso}$, respectively, which are the unnormalized (OVA) probabilities, see Section~\ref{ssec:onevsall}.

In order to facilitate access to our results we now report a simple query example that allows one to access the XGBoost classifications  along with the J-PLUS photometric bands with flag and mask quality cuts:
\begin{verbatim}
SELECT

t1.MAG_AUTO[jplus::uJAVA] as uJAVA,
t1.MAG_AUTO[jplus::rSDSS] as rSDSS,
t1.MAG_AUTO[jplus::J0395] as J0395,
t1.MAG_AUTO[jplus::J0861] as J0861,
t2.xgb_prob_gal,
t2.xgb_prob_star,
t2.xgb_prob_qso

FROM

jplus.MagABDualObj t1

JOIN

jplus.StarGalClass t2

ON

t1.tile_id = t2.tile_id AND
t1.number=t2.number

WHERE

t1.flags[jplus::rSDSS]=0 AND
t1.mask_flags[jplus::rSDSS]=0

\end{verbatim}

%%%%%%%%%%%%%%%%%%%%%%%%%%%%%%%%%%%%%%%%%%%%%%%%%%

% Don't change these lines
\bsp	% typesetting comment
\label{lastpage}
\end{document}